\documentclass[12pt]{article}

\usepackage{graphicx}
\usepackage{amsmath}
\usepackage{amssymb}
\usepackage{setspace}
\usepackage{booktabs,dcolumn}
\usepackage{amssymb}
\usepackage{lineno}

\usepackage{changebar}

\newcommand{\ee}{$\mathrm{EE}$}
\newcommand{\eeinit}{$\mathrm{EE}_{0}$}
\newcommand{\eelast}{$\mathrm{EE}_{44}$}
\newcommand{\ei}{$\mathrm{EI}$}

\pagewiselinenumbers

\begin{document}

\title{ Comparison of Image Registration Based Measures\\ 
       of Regional Lung Ventilation from Dynamic Spiral CT with Xe-CT}
\date{}
\author{}

\maketitle
\begin{center}
{\large Kai Ding,$^{1,*}$ Kunlin Cao,$^2$ Matthew K. Fuld,$^{1,3}$\\
Kaifang Du,$^1$ Gary E. Christensen,$^{2,4}$ Eric A. Hoffman,$^{1,3,**}$\\ and Joseph M. Reinhardt$^{1,**}$}\\
\vspace{0.25in}

$^1$Department of Biomedical Engineering,\\
$^2$Department of Electrical and Computer Engineering,\\
$^3$Department of Radiology,\\
$^4$Department of Radiation Oncology,\\
The University of Iowa, Iowa City, IA 52242\\

\end{center}

\vspace{0.25 in}

\vfill $^*$Dr. Ding is currently with the Department of Radiation
Oncology, University of Virginia, Charlottesville, VA 22908\\
$^{**}$Drs. Reinhardt and Hoffman are founders and shareholder of
VIDA Diagnostics, Inc.

\newpage
\noindent
{\bf Correspondence:}\\
Joseph M. Reinhardt, PhD\\
Chair, Department of Biomedical Engineering\\
The University of Iowa\\
1402 SC \\
Iowa City, IA 52242\\
Tel. (319) 335-5634\\
FAX: (319) 335-5631\\
E-mail: joe-reinhardt@uiowa.edu\\

\newpage

\begin{abstract}
\noindent\textbf{Purpose:}  Regional lung volume change as a
function of lung inflation serves as an index of parenchymal and
airway status as well as an index of regional ventilation and can be
used to detect pathologic changes over time. In this article, we
propose a new regional measure of lung mechanics --- the specific
air volume change by corrected Jacobian. We compare this new
measure, along with two existing registration-based measures of lung
ventilation, to a regional ventilation measurement
derived from xenon-CT (Xe-CT) imaging.
\\\textbf{Methods:} 4DCT and Xe-CT data sets from four adult sheep are used in this study. Nonlinear, 3D image
registration is applied to register an image acquired near end
inspiration to an image acquired near end expiration. Approximately
200 annotated anatomical points are used as landmarks to evaluate
registration accuracy. Three different registration-based measures
of  regional lung mechanics  are derived and compared:
the specific air volume change calculated from the Jacobian (SAJ);
the specific air volume change calculated by the corrected Jacobian
(SACJ); and the specific air volume change by intensity change
(SAI). We show that the commonly-used SAI measure can be derived
from the direct SAJ measure by using the air-tissue mixture model
and assuming there is no tissue volume change between the end
inspiration and end expiration data sets. All three ventilation
measures are evaluated by comparing to Xe-CT estimates of regional
ventilation.
\\\textbf{Results:} After registration, the mean registration error is on the
order of 1 mm. For cubical ROIs in cubes with size 20 mm $\times$ 20
mm $\times$ 20 mm, the SAJ and SACJ measures show significantly
higher correlation (linear regression, average $r^2=0.75$ and
$r^2=0.82$) with the Xe-CT based measure of specific ventilation
(sV) than the SAI measure. For ROIs in slabs along the
ventral-dorsal vertical direction with size of 150 mm $\times$ 8 mm
$\times$ 40 mm, the SAJ, SACJ, and SAI all show high correlation
(linear regression, average $r^2=0.88$, $r^2=0.92$ and $r^2=0.87$)
with the Xe-CT based sV without significant differences when
comparing between the three methods. We demonstrate a linear
relationship between the difference of specific air volume change
(DSA) and difference of tissue volume (DT) in all four animals
(linear regression, average $r^2=0.86$).
\\\textbf{Conclusion:}  Given a deformation field by an image
registration algorithm, significant differences between the SAJ,
SACJ, and SAI measures were found at a regional level compared to
the Xe-CT sV in four sheep that were studied. The SACJ introduced
here, provides better correlations with Xe-CT based sV than the SAJ
and SAI measures, thus providing an improved surrogate for regional
ventilation. 

\end{abstract}

\newpage
\section{Introduction}
\label{sec:introduction}  %
 Regional ventilation is the term used to characterize the
volume of fresh gas per unit time that enters or exits the lung at
the acinar (gas exchange) level. Disruption of regional ventilation
can reflect alterations to airways (physiological or pathological),
alterations in parenchymal mechanics, changes to the muscles of
respiration, body posture effects and inhaled gas properties. Thus,
measures of regional lung mechanics can serve as a sensitive test of
the status of the respiratory system and should be considerably more
sensitive and informative than global pulmonary function test.
Recent advances in multi-detector-row CT (MDCT), %
4DCT respiratory gating methods, and image processing
techniques enable us to study pulmonary function at the regional
level with high resolution anatomical information compared to other
methods.  MDCT can be used to acquire multiple static breath-hold CT
images of the lung taken at different lung volumes, or 4DCT images
of the lung acquired during spiral scanning using a low pitch and
retrospectively reconstructed at different respiratory phases with
proper respiratory control~\cite{Keall2002,low:1254,pan:627}. Image
registration can be applied to these data to estimate a deformation
field that transforms the lung from one volume configuration to the
other. This deformation field can be analyzed to estimate local lung
tissue expansion, calculate voxel-by-voxel intensity change, and
make biomechanical measurements.  When combined with image
segmentation
algorithms~\cite{Reinhardt2008752,ding2008b,ding2009,ding2009b},
functional and biomechanical measurements can be reported on a lung,
lobe, and sublobar basis, and can be used to interpret regional lung
function relative to specific segments of bronchial tree. Such
measurements of pulmonary function have proven useful as a planning
tool during RT planning~\cite{Yaremko2007562, Yamamoto2010} and may be useful for
tracking the progression of toxicity to nearby normal tissue during
RT and can be used to evaluate the effectiveness of a treatment
post-therapy~\cite{ding:1261}.

Early studies using CT to study regional air volume changes have
proved to enhance our understanding of normal lung function. 
Several groups have proposed methods that couple image registration
and CT imaging to study regional lung function. Guerrero et al.\ have
used optical-flow registration to compute lung ventilation from
4DCT~\cite{Guerrero2005630,guerrero2006a} with an intensity-based
ventilation measure. Christensen et al.\ used image registration to
match images across cine-CT sequences and estimate rates of local
tissue expansion and contraction~\cite{christensen2007a} using a
Jacobian-based ventilation measure. While they were able to show that
their accumulated measurements matched well with the global
measurements, they were not able to compare the registration-based
measurements to local measures of regional tissue
ventilation. Recently, Castillo et al.\ compared the intensity-based
and Jacobian-based calculations of ventilation from 4DCT with the
ventilation from $^{99m}Tc$-labeled aerosol
SPECT/CT~\cite{Castillo2010}. A statistically higher correlation to
the SPECT/CT based ventilation was found for intensity-based based
calculation over the Jacobian-based calculation. However, the
comparison of the two techniques was based on the Dice similarity
coefficient between the thresholded masks within 20\% variation from
the 4DCT and from SPECT/CT. Though their experiment is novel and
important, since the average mask size is about 490.5 mL (with average
subject exhale volume 2452.7 mL, and 5 sub-masks per subject), the
comparison is more global than regional. In addition, as shown in
Section~\ref{sect:meth_regvent}, both the intensity-based and
Jacobian-based ventilation measures are based on the assumption that
regional lung volume change is due solely to air content change,
which may not always be a valid assumption.   Other factors,
such as blood volume change, may also introduce the regional lung
volume change. 

The physiologic significance of these registration-based measures of
respiratory function can be established by comparing to more
conventional measurements, such as nuclear medicine or contrast
wash-in/wash-out studies with CT or MR\@.  Xenon-enhanced CT
(Xe-CT) measures regional ventilation by observing the gas wash-in
or wash-out rate on serial CT
images~\cite{Marcucci02012001,tajik2002,Chon200565}
Xe-CT imaging has the advantage of high temporal
resolution and spatial resolution and reflects a measure of fresh
gas delivery to the gas exchange units of the lung.  Although the
dynamic Xe-CT method is limited in Z-axis coverage,
requires expensive Xe gas, and is technically challenging, it serves as
the gold standard of regional ventilation and can be used to compare
with registration-based measures of regional lung function in animal
studies for validation purposes. 

This paper describes three measures to estimate regional ventilation
from image registration of CT images: the specific air volume change
calculated from the Jacobian (SAJ); the specific air volume change
calculated by the corrected Jacobian (SACJ); and the specific air
volume change by intensity change (SAI). We show that the SAI
ventilation measure can be derived from the SAJ measure by making the
assumption that there is no tissue volume change between
registration volumes.  We evaluate these three measures by comparing
them with a Xe-CT measure of ventilation in a regional basis (20 mm $\times$ 20
mm $\times$ 20 mm cube, or 8 mL). Among these three
registration based measures, we show that the corrected
Jacobian-based measure, SACJ, has the best correlation with the
Xe-CT derived measure of specific ventilation.

\section{Material and methods}
\label{sect:meth}

\subsection{Method Overview}
\label{sect:meth_overview}

\begin{figure}[tpb]
  \centering
  \includegraphics[width=\textwidth]{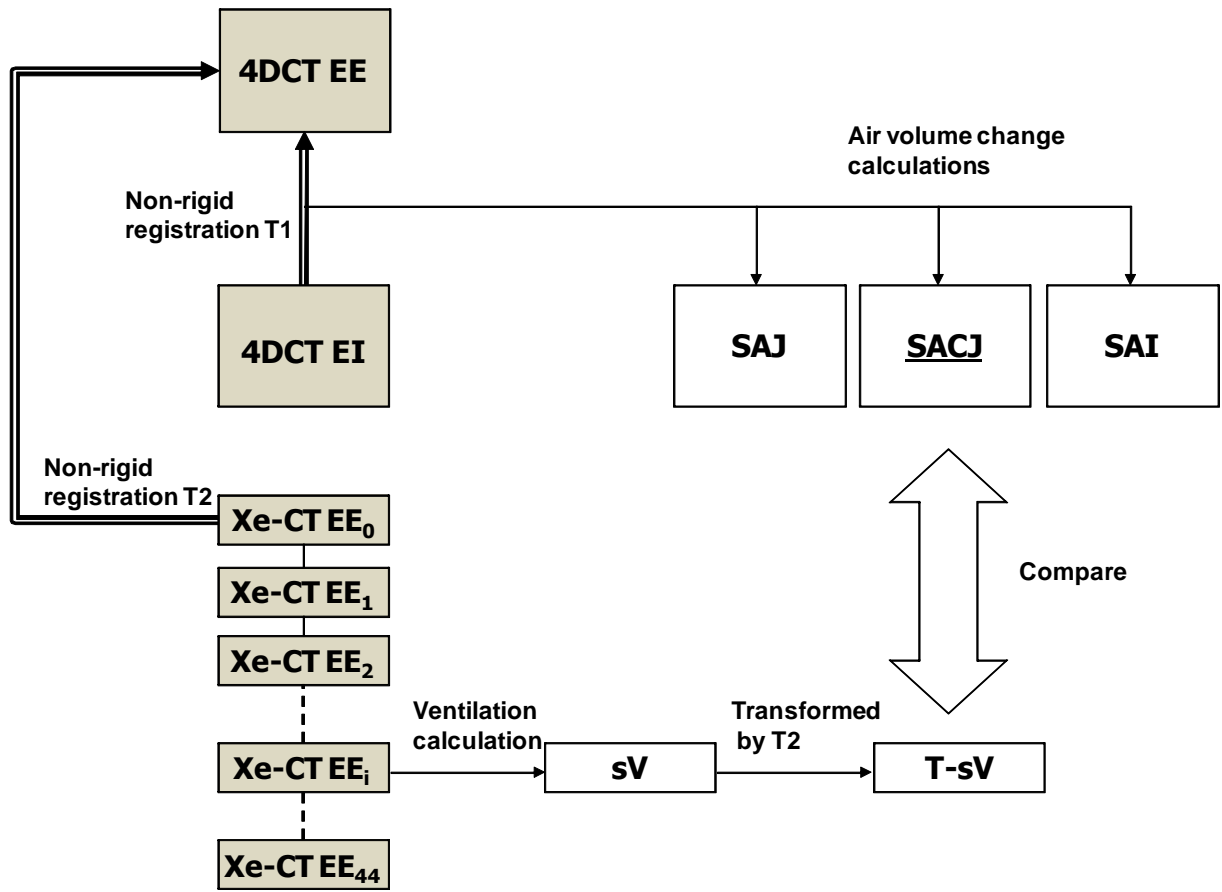}
  \caption{Figure shows the two types of images, a image pair of full lung volumetric phases \ee{} and \ei{} from a 4DCT scan and a Xe-CT scan acquired at end of expiration over 45 respiratory cycles (\eeinit{} to \eelast{}), which are analyzed during the
    processing.  Transformation T1 registers end inspiration
    (EI) to end expiration (EE) data and can be used to assess local
    lung function via calculations of three ventilation measures: specific air volume change
by specific volume change (SAJ), specific air volume change by
corrected Jacobian (SACJ), and specific air volume change by
intensity (SAI). The 45 distinctive partial lung volumetric Xe-CT
scans \eeinit{} to \eelast{} are used to calculate Xe-CT-based
measure of specific ventilation (sV). Transformation T2 maps the sV
data into the coordinate system of the
    \ee{} image~(end expiration phase of the 4DCT scan) to allow direct comparison with
    the 4DCT and registration-based measures of ventilation. Both \ee{} and \eeinit{} are at volumes near end inspiration.  (Shaded boxes indicate CT image data; white boxes
    indicated derived or calculated data; thick arrows indicate image
    registration transformations being calculated; thin solid lines
    indicate other operations.)}
 \label{fig:ventmeasureflowchart}
\end{figure}

Our goal is to validate and compare the measures used to estimate
regional lung ventilation from image registration by comparing them
to Xe-CT estimated ventilation.
Figure~\ref{fig:ventmeasureflowchart} shows a block diagram of the
entire process. Two types of data were acquired for each animal: a
4DCT scan and a Xe-CT scan. In order to make our comparisons under
the same physiological conditions, each animal was scanned and
mechanically ventilated with the same respiratory rate, tidal volume
(TV) and positive end-expiratory pressure (PEEP) during the two
types of scans. The data sets from the 4DCT scan were reconstructed
in volumes at eight phases of the respiratory cycle. For this study
we focus on the data sets from two of the phases, a volume near end
expiration (\ee{}) and a volume near end inspiration (\ei{}). For
the Xe-CT scan, 45 distinctive partial lung volumetric scans were
performed at volume near end expiration, or the initial end
expiration scan (\eeinit{}) to the last expiration scan (\eelast{}).

The nonlinear image registration is used to define the
transformation T1 between the \ee{} and \ei{} in order to measure
the regional lung ventilation from the 4DCT scan. The Xe-CT-based
estimated regional lung ventilation is computed on the \eeinit{} by
using Pulmonary Analysis Software Suite 11.0 (PASS) software by
finding the constant of the exponential rise of the density from
xenon gas wash-in over multiple breaths~\cite{guo2008}. The same
nonlinear image registration is also applied to define the
transformation T2 which maps the \eeinit{} to the \ee{} so that the
Xe-CT based estimate of ventilation can be mapped into the same
coordinate system as that of the registration-based estimate of
ventilation. Additional details on the registration algorithm and
other processing steps are given below.

\subsection{Image Data Sets}
\label{sect:meth_data}

Appropriate animal ethics approval was obtained for these protocols
from the University of Iowa Animal Care and Use Committee and the
study adhered to NIH guidelines for animal experimentation. Four
adult male sheep A, B, C, and D (with weights 44.0, 37.8, 40.4, and
46.7 kg) were used for this study.  The sheep were
anesthetized using intravenous pentobarbital and pancuronium to
ensure adequate sedation and to prevent spontaneous breathing.
Animals were positive pressure ventilated during experiments using a
custom built dual Harvard apparatus piston ventilator designed for
computer control. The 4DCT images were acquired with the animals in
the supine position using the dynamic imaging protocol with a pitch
of 0.1, slice collimation of 0.6 mm, rotation time of 0.5 sec, slice
thickness of 0.75 mm, slice increment of 0.5 mm, 120 kV, 400 mAs,
and kernel B30f.
The airway pressure signal was simultaneously recorded with
the X-ray projections and images were reconstructed retrospectively
at 0, 25, 50, 75, and 100\% of the inspiration duration and 75, 50
and 25\% of the expiration duration. The 0\% (\ee{}) and 100\%
(\ei{}) inspiration phases were used for later ventilation
measurements. A slab of twelve contiguous axial slices were imaged
over 45 breaths for Xe-CT scans. Images were acquired using
respiratory gating by triggering the scan during end-expiration with
80 keV energy (for higher density resolution, approximately 2 HU per
1\% Xe), 160 mAs tube current, a 360$^\circ$ rotation, a 0.33 sec
scan time, and 2.4 mm slice thickness. Respiratory gating is
achieved using a custom built LabVIEW program which controls the
ventilators and triggers the CT scanner. The respiratory rate (RR)
for four animals ranged from 15 to 18 breaths per minute with an
inspiratory-expiratory ratio of 1:1 which was sufficient to maintain
a normocapnic state. Both of the two types of images were acquired
with a matrix of 512 by 512 and without moving the animal between
scans. 

\subsection{Image Registration}
\label{sect:meth_reg} A tissue volume and vesselness measure
preserving nonrigid registration (TVP)
algorithm~\cite{cao:762309,caowbir2010} is used to estimate the
transformations \ei{} to \ee{} and \eeinit{} to \ee{}. The TVP
algorithm minimizes the sum of squared tissue volume difference
(SSTVD)~\cite{Yin2009,yin:4213,cao2009a,Yin20102159} and vesselness
measure difference (SSVMD), utilizing the rich image intensity
information and natural anatomic landmarks provided by the vessels.
This method has been shown to be effective at registering across
lung CT images with high accuracy~\cite{cao:762309,caowbir2010}.

Let $I_1$ and $I_2$ represent two 3D image volumes to be registered.
The vector $\mathbf{x}=(x_1, x_2, x_3)^T$ defines the voxel
coordinate within an image. The algorithm finds the optimal
transformation $\mathbf{h}$ that maps the template image $I_1$ to
the target image $I_2$ by minimizing the cost function

\begin{eqnarray}
  C_{\text{TOTAL}}&=&\rho \int_\mathbf{\Omega}
 \left[V_2(\mathbf{x})-V_1(\mathbf{h(x)})\right]^2 d\mathbf{x}
 +\chi \int_\mathbf{\Omega}
  \left[F_2(\mathbf{x})-F_1(\mathbf{h(x)})\right]^2d\mathbf{x}
  \label{eq:total_cost}.
\end{eqnarray}
 where $\Omega$ is the union domain of the lung regions in
images $I_1$ and $I_2$. $V_2$ and  $V_1$ are the tissue volumes as
defined in Equation~\ref{eq:TissueVolume}. $F_2$ and $F_1$ are the
vesselness measures as defined in Equation~\ref{eq:fangi_fcn}. The
transformation $\mathbf{h}$ is a $(3 \times 1)$ vector-valued
function that maps a point $\mathbf{h(x)}$ in the target image to
its corresponding location in the template image.  The first
integral of the cost function defines the SSTVD cost and the second
integral of the cost function defines the SSVMD cost.

The SSTVD cost assumes that the measured Hounsfield units (HU) in
the lung CT images is a function of tissue and air content.
Following the air-tissue mixture model by Hoffman et
al.~\cite{Hoffman1985}, from the CT value of a given voxel, the
tissue volume can be estimated as
\begin{equation}
\label{eq:TissueVolume} V(\mathbf{x}) = \nu(\mathbf{x})
\frac{I(\mathbf{x})-HU_{air}
}{HU_{tissue}-HU_{air}}=\nu(\mathbf{x})\beta(I(\mathbf{x})),
\end{equation}
and the air volume can be estimated as
\begin{equation}
\label{eq:AirVolume} V'(\mathbf{x}) = \nu(\mathbf{x})
\frac{HU_{tissue}-I(\mathbf{x})
}{HU_{tissue}-HU_{air}}=\nu(\mathbf{x})\alpha(I(\mathbf{x})),
\end{equation}
\noindent where $\nu (\mathbf{x})$ denotes the volume of voxel
$\mathbf{x}$ and $I(\mathbf{x})$ is the intensity of a voxel at
position $\mathbf{x}$.  $HU_{air}$ and $HU_{tissue}$ refer to the
intensity of air and tissue, respectively. In this work, we assume
that air is -1000 HU and tissue is 0 HU. 
$\alpha(I(\mathbf{x}))=\frac{HU_{tissue}-I(\mathbf{x})
}{HU_{tissue}-HU_{air}}$ and
$\beta(I(\mathbf{x}))=\frac{I(\mathbf{x})-HU_{air}
}{HU_{tissue}-HU_{air}}$ are introduced for notational simplicity.
Notice that $\alpha(I(\mathbf{x}))+\beta(I(\mathbf{x}))=1$.

Given~(\ref{eq:TissueVolume}), we can then define the SSTVD cost:
\begin{eqnarray}
  C_{\text{SSTVD}}
  &=&\int_\mathbf{\Omega}\left[V_2(\mathbf{x})-V_1(\mathbf{h(x)})\right]^2d\mathbf{x}\\
  &=&\int_\mathbf{\Omega}\left[ \nu_2(\mathbf{x})\beta(I_2(\mathbf{x})) - \nu_1(\mathbf{h(x)})\beta(I_1(\mathbf{h(x)})) \right]^2 d\mathbf{x}
 \label{eq:sstvd_cost_1},
\end{eqnarray}
 The notation $I_1(\mathbf{h(x)})$ is interpreted as the
image $I_1(\mathbf{x})$ deformed by the transformation
$\mathbf{h(x)}$ and is computed using trilinear interpolation. The
deformed volume element $\nu _1(\mathbf{h(x)})$ is  calculated using
the Jacobian $J(\mathbf{x})$ times the volume element $\nu _2
(\mathbf{x})$, i.e., $\nu _1(\mathbf{h(x)})=J(\mathbf{x}) \nu _2
(\mathbf{x})$. 

\begin{figure}[tbp]
  \centering
  \includegraphics[width=0.8\textwidth]{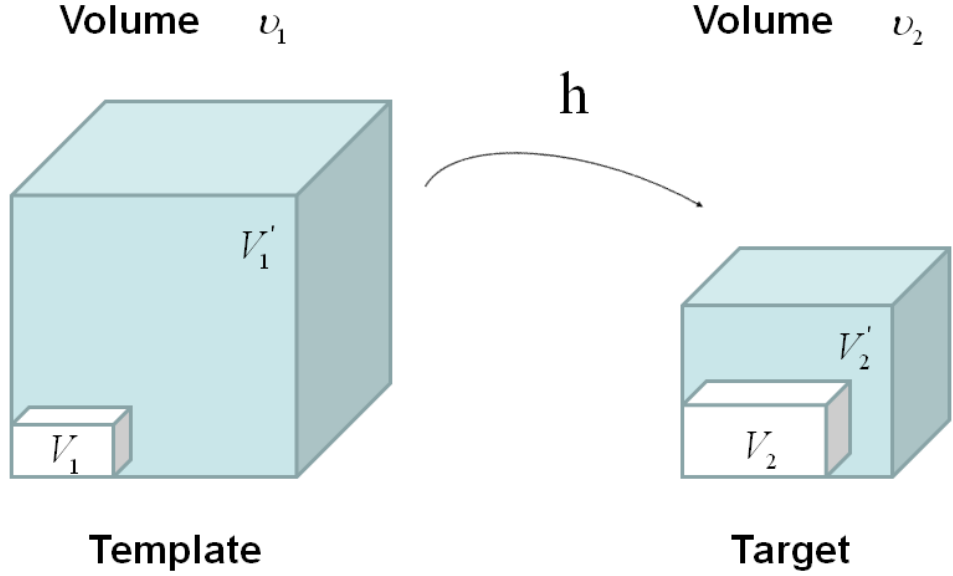}
  \caption{Example of a region under deformation
  $\mathbf{h(x)}$ from template image to target image. $V_1$ and
  $V_2$ are tissue volumes in the regions. $V'_1$ and
$V'_2$ are air volumes in the regions. Region volumes $\nu_1 = V_1 +
V'_1$ and $\nu_2 = V_2 + V'_2$.}
  \label{fig:airtissuevolume}
\end{figure}

Figure~\ref{fig:airtissuevolume} shows an example of a cubic shaped
region under deformation $\mathbf{h}$ from template image to target
image. The region volumes are $\nu_1$ and $\nu_2$. The volumes can
be decomposed into the tissue volume fraction and air volume
fraction based on the mean voxel intensity within the cube. The
small white sub volumes inside the cubes represent the tissue volume
$V_1$ and $V_2$. Air volumes are represented by $V'_1$ and $V'_2$
(in blue). As the ratio of air to tissue decreases, the CT intensity
of a voxel increases. The mean cube voxel intensities for the
template, $I_1$, and target images, $I_2$, are functions of the
ratios of air to tissue volumes within the cubes.

As the blood vessels branch to smaller and smaller diameters, the
raw grayscale information from vessel voxels provide very little
contribution to guide the intensity-based registration. To better
utilize the information of blood vessel locations, we use the
vesselness measure based on the eigenvalues of the Hessian matrix of
image intensity. Frangi's vesselness
function~\cite{Frangi98multiscalevessel} is defined as
\begin{equation}
  F(\lambda)=\left\{\begin{array}{ll}
    (1-e^{-\frac{R_A^2}{2\alpha^2}})\cdot e^{\frac{-R_B^2}{2\beta^2}} \cdot
    (1-e^{-\frac{S^2}{2\gamma^2}})& \text{if}~\lambda_2<0~\text{and}~\lambda_3<0\\
    0& \text{otherwise}
  \end{array}\right.
\label{eq:fangi_fcn}
\end{equation}
with
\begin{equation}
  R_A=\frac{|\lambda_2|}{|\lambda_3|}, ~ ~ ~ ~ ~
  R_B=\frac{|\lambda_1|}{\sqrt{|\lambda_2\lambda_3|}}, ~ ~ ~ ~ ~
  S=\sqrt{\lambda_1^2+\lambda_2^2+\lambda_3^2},
\label{eq:fangi_ra}
\end{equation}
where $R_A$ distinguishes between plate-like and tubular structures,
$R_B$ accounts for the deviation from a blob-like structure, and $S$
differentiates between tubular structure and noise. The vesselness
function has been previously widely used in vessel segmentations in
lungs~\cite{shikata2004, Shikata2009} and in
retinas~\cite{joshi2010a}. $\alpha$, $\beta$, $\gamma$ control the
sensitivity of the vesselness measure. The vesselness measure is
rescaled to [0, 1] and can be considered as a probability-like
estimate of vesselness features. For this study, $\alpha=0.5$,
$\beta=0.5$, and $\gamma=5$ and the weighting constants in the total
cost were set as $\rho=1$ and $\chi=0.2$. These parameters are
similar to those used in our previous
work~\cite{cao:762309,caowbir2010}.

The transformation $\mathbf{h(x)}$ is a cubic B-splines transform:
\begin{equation}
  \mathbf{h(x)}=\mathbf{x}+\sum_{i\in G}\phi_i \beta^{(3)}(\mathbf{x}-\mathbf{x}_i)
  \label{eq:param},
\end{equation}
where $\mathbf{\phi}_i$ describes the displacements of the control
nodes and $\beta^{(3)}(\mathbf{x})$ is a three-dimensional tensor
product of basis functions of cubic B-Spline. A spatial
multiresolution procedure from coarse to fine is used in the
registration in order to improve speed, accuracy and robustness. The
total cost in Equation~\ref{eq:total_cost} is optimized using a
limited-memory, quasi-Newton minimization method with bounds
(L-BFGS-B)~\cite{lbfgsb} algorithm. The B-Splines coefficients are
constrained so that the transformation maintains the topology using
the sufficient conditions that guarantee the local injectivity of
functions parameterized by uniform cubic B-Splines proposed by Choi
and Lee~\cite{Choi00injectivityconditions}.

\subsection{Regional Ventilation Measures from Image Registration}
\label{sect:meth_regvent} 
\begin{figure}[tbp]
  \centering
  \includegraphics[width=0.8\textwidth]{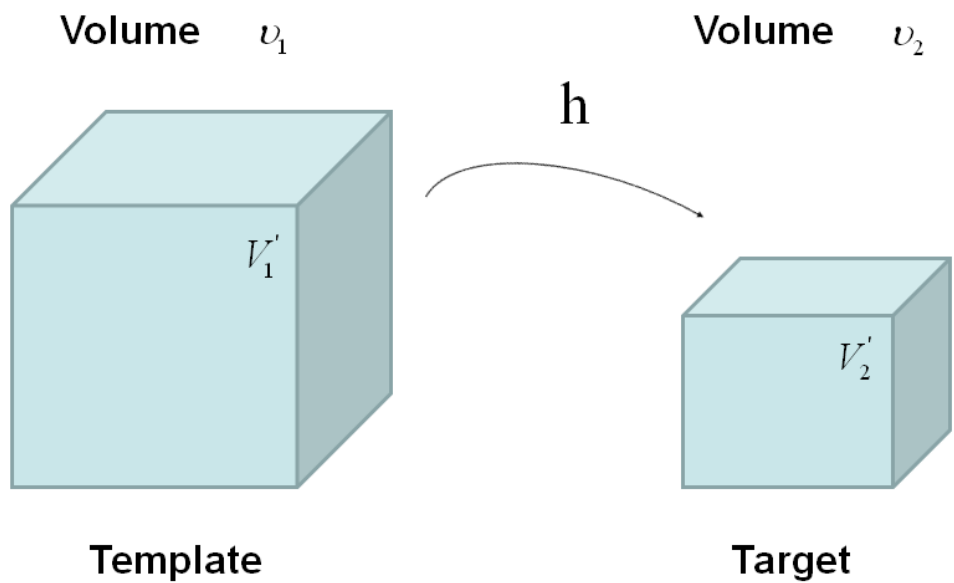}
  \caption{Example of a given region under deformation
  $\mathbf{h(x)}$ from template image to target image, with the
  assumption of no tissue volume ($V_1 = V_2 = 0$). $V'_1$ and
$V'_2$ are air volumes.}
  \label{fig:airtissuevolume_saj}
\end{figure}

After we obtain the optimal warping
function $\mathbf{h(x)}$, we can calculate the regional ventilation,
which is equal to the difference in local air volume change per unit
time. The commonly-used ventilation measure is the specific
ventilation sV which takes the initial air volume into account. The
sV is equal to the specific air volume change sVol per unit time. Or
in other words, in a unit time,
\begin{equation}
\label{eq:sVol} sV=
sVol=\frac{V'_1(\mathbf{h(x)})-V'_2(\mathbf{x})}{V'_2(\mathbf{x})}.
\end{equation}
Three different approaches for estimating~(\ref{eq:sVol}) are
described below:

\paragraph{Specific air volume change by specific volume change (SAJ):}
The SAJ regional ventilation measure is based on the assumption that
local volume change is due to air volume change alone, and thus, any
regional volume change is due only to local air volume change.  Or,
equivalently, the SAJ measure assumes that there is no tissue volume
within the template or target volumes.
Figure~\ref{fig:airtissuevolume_saj} illustrates such an assumption.
Compared with the general condition in
Figure~\ref{fig:airtissuevolume}, the region volume now is pure air
volume, or equivalently,  $\nu_1=V'_1$ and $\nu_2=V'_2$. In this
case, the specific air volume change is equal to specific volume
change. Since the Jacobian tells us the local volume expansion (or
contraction), the regional ventilation can be measured by:
\begin{equation}
\label{eq:SAJ} SAJ=\frac{\nu _1(\mathbf{h(x)})-\nu
_2(\mathbf{x})}{\nu _2(\mathbf{x})}
=J(\mathbf{x})-1.
\end{equation}

Previously, SAJ has been used as an index of the regional function
and was compared with Xe-CT estimates of regional lung
function~\cite{Reinhardt2008752}. Regional lung expansion, as
estimated from the Jacobian of the image registration
transformations, was well correlated with xenon CT specific
ventilation~\cite{Reinhardt2008752,ding2009} (linear regression,
average $r^2=0.73$).

\paragraph{Specific air volume change by corrected Jacobian (SACJ):}
Starting with~(\ref{eq:SAJ}) and expressing the air volumes
$V'_1(\mathbf{h(x)})$ and $V'_2(\mathbf{x})$ using the air-tissue
mixture model (\ref{eq:TissueVolume}) and~(\ref{eq:AirVolume}), we
obtain the corrected Jacobian measure of region air volume change,
SACJ,
\begin{eqnarray}
SACJ&=&\frac{V'_1(\mathbf{h(x)})-V'_2(\mathbf{x})}{V'_2(\mathbf{x})}\\
&=&\frac{V'_1(\mathbf{h(x)})}{V'_2(\mathbf{x})}-1\\
&=&\frac{\nu_1(\mathbf{h(x)})\alpha(I_1(\mathbf{h(x)}))}{\nu_2(\mathbf{x})\alpha(I_2(\mathbf{x}))}-1
\end{eqnarray}

As $\nu_1(\mathbf{h(x)})=J(\mathbf{x}) \nu_2 (\mathbf{x})$, the
specific air volume change is then

\begin{eqnarray}
SACJ
&=&J(\mathbf{x})\frac{\alpha(I_1(\mathbf{h(x)}))}{\alpha(I_2(\mathbf{x}))}-1\\
\label{eq:General_SACJ}
&=&J(\mathbf{x})\frac{HU_{tissue}-I_1(\mathbf{h(x)})}{HU_{tissue}-I_2(\mathbf{x})}-1
\end{eqnarray}
If we assume that pure air is -1000 HU and pure tissue is 0 HU, then
specific air volume change is
\begin{eqnarray}
\label{eq:SACJ}
SACJ&=&J(\mathbf{x})\frac{I_1(\mathbf{h(x)})}{I_2(\mathbf{x})}-1.
\end{eqnarray}
Compared to Equation~\ref{eq:SAJ}, the term
$\frac{I_1(\mathbf{h(x)})}{I_2(\mathbf{x})}$ is a correction factor
that depends on the voxel intensities in the template and target
images.  The SACJ measure is illustrated in
Figure~\ref{fig:airtissuevolume}, and represents the most general
case of where there is both tissue volume and air volume change
within the region.

\paragraph{Specific air volume change by intensity change (SAI):}
The intensity-based measure of regional air volume change SAI can be
derived from the SACJ by assuming that tissue volume is preserved
during deformation, or equivalently, that the tissue volume
difference
${\Delta}V(\mathbf{x})=V_1(\mathbf{h(x)})-V_2(\mathbf{x})=0$. Under
this assumption, $V_1(\mathbf{h(x)})=V_2(\mathbf{x})$ and we have
\begin{equation}
\nu_1(\mathbf{h(x)})\beta(I_1(\mathbf{h(x)}))=\nu_2(\mathbf{x})\beta(I_2(\mathbf{x})),
\end{equation}
and
\begin{equation}
\nu_1(\mathbf{h(x)})=\nu_2(\mathbf{x})\frac{\beta(I_2(\mathbf{x}))}{\beta(I_1(\mathbf{h(x)}))},
\end{equation} Since $\nu_1(\mathbf{h(x)})=J(\mathbf{x}) \nu _2 (\mathbf{x})$, with
above equation, we have
\begin{eqnarray}
\label{eq:SAI-assumption} J(\mathbf{x})
&=&\frac{\beta(I_2(\mathbf{x}))}{\beta(I_1(\mathbf{h(x)}))}\\
&=&\frac{I_2(\mathbf{x})-HU_{air}}{I_1(\mathbf{h(x)})-HU_{air}}.
\end{eqnarray}

\noindent Substituting the above equation into
equation~\ref{eq:General_SACJ}, yields
\begin{equation}
SAI
=\frac{I_2(\mathbf{x})-HU_{air}}{I_1(\mathbf{h(x)})-HU_{air}}\frac{HU_{tissue}-I_1(\mathbf{h(x)})}{HU_{tissue}-I_2(\mathbf{x})}-1\\
\end{equation}

\begin{equation}
\label{eq:General_SAI}
=\frac{I_2(\mathbf{x})HU_{tissue}+HU_{air}I_1(\mathbf{h(x)})-I_1(\mathbf{h(x)})HU_{tissue}-HU_{air}I_2(\mathbf{x})}{(I_1(\mathbf{h(x)})-HU_{air})(HU_{tissue}-I_2(\mathbf{x}))}
\end{equation}

\noindent Finally, if we assume that pure air is -1000 HU and pure
tissue is 0 HU, then
\begin{equation}
\label{eq:SAI} SAI
=1000\frac{I_1(\mathbf{h(x)})-I_2(\mathbf{x})}{I_2(\mathbf{x})(I_1(\mathbf{h(x)})+1000)}
\end{equation}

\noindent which is exactly the result from Simon~\cite{Simon2000},
Guerrero et al.~\cite{Guerrero2005630}, and Fuld et
al.~\cite{Fuld04012008}.

\begin{figure}[tbp]
  \centering
  \includegraphics[width=0.8\textwidth]{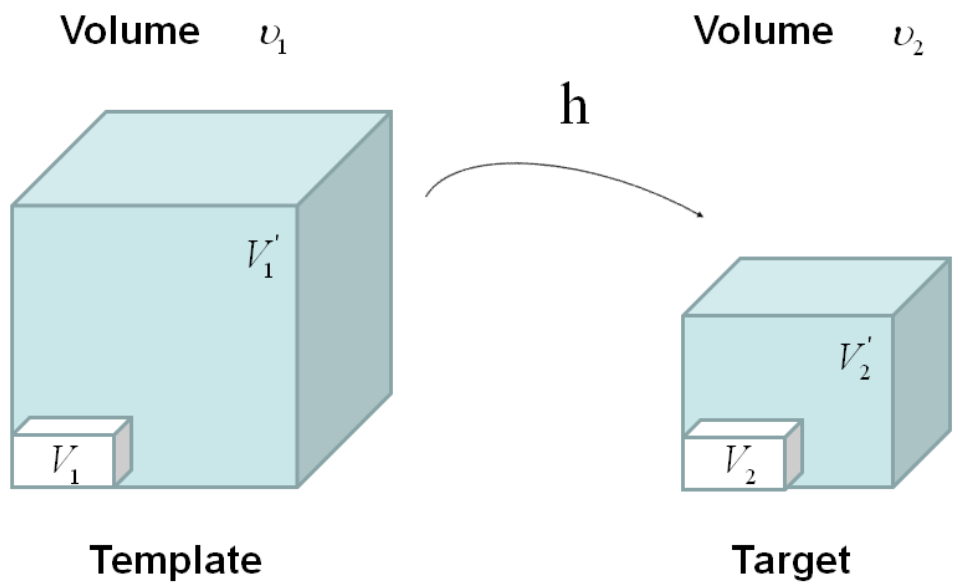}
  \caption{Example of a given voxel under deformation
  $\mathbf{h(x)}$ from template image to target image, with the
  assumption of no tissue volume change. Notice the tissue volume
  $V_1=V_2$ under this assumption. $V'_1$ and
$V'_2$ are air volumes.}
  \label{fig:airtissuevolume_sai}
\end{figure}

Figure~\ref{fig:airtissuevolume_sai} illustrates the assumption with
no tissue volume change in SAI\@. In
Figure~\ref{fig:airtissuevolume_sai} as the region volume changes
from $\nu_1$ to $\nu_2$, the tissue volume inside the cube remains
the same ($V_1=V_2$).

\paragraph{Difference of specific air volume change (DSA) and difference of tissue volume (DT):}
To investigate the relationship between the measurements of specific
air volume changes and the tissue volume change, we can calculate
the difference between equation~(\ref{eq:General_SACJ}) and
equation~(\ref{eq:General_SAI}) and define the difference of
specific air volume change (DSA) between SACJ and SAI, and the
difference of tissue volume (DT) as:

\begin{equation}
DSA = |SACJ - SAI|
\end{equation}

\begin{eqnarray}
DT &=& |V_1(\mathbf{h(x)})-V_2(\mathbf{x})|\\
&=& |\nu_1(\mathbf{h(x)})\beta(I_1(\mathbf{h(x)}))-\nu_2(\mathbf{x})\beta(I_2(\mathbf{x}))|\\
&=& |J(\mathbf{x})\nu_2(\mathbf{x})\beta(I_1(\mathbf{h(x)}))-\nu_2(\mathbf{x})\beta(I_2(\mathbf{x}))|\\
&=&
|\nu_2(\mathbf{x})\frac{J(\mathbf{x})(I_1(\mathbf{h(x)})-HU_{air})-(I_2(\mathbf{x})-HU_{air})}{HU_{tissue}-HU_{air}}|
\end{eqnarray}

\noindent Again, if we assume that air is -1000 HU and tissue is 0
HU, then the tissue volume difference is:
\begin{equation}
DT
=|\nu_2(\mathbf{x})\frac{J(\mathbf{x})(I_1(\mathbf{h(x)})+1000)-(I_2(\mathbf{x})+1000)}{1000}|
\end{equation}

\subsection{Computational Setup}
Processing starts by identifying the lung regions in all images
using the Pulmonary Workstation 2.0 (VIDA Diagnostics, Inc., Iowa
City, IA). The Xe-CT estimate of sV is computed in the coordinates
of the \eeinit{} using Pulmonary Analysis Software Suite 11.0
(PASS)~\cite{guo2008} at the original image size of 0.5 mm $\times$
0.5 mm $\times$ 2.4 mm voxels. Overlapping 1 $\times$ 8 regions of
interest (ROI) are defined in the lung region on each 2D slice. All
images, including the \ee{}, \ei{}, \eeinit{} and their
corresponding lung region masks or sV map, are then resampled to a
voxel size of 1 mm $\times$ 1 mm $\times$ 1 mm. After preprocessing,
\ei{} is registered to \ee{} using the TVP registration for measuring the
regional ventilation from these two phases in a 4DCT scan. The
resulting transformation is used to estimate the SAJ, SACJ and SAI.
Then \eeinit{} is registered to \ee{} using TVP registration as well to map the
sV to the same coordinate system as that of the SAJ, SACJ and SAI.
Due to the fact that the denominator of equation
(\ref{eq:General_SACJ}) and (\ref{eq:General_SAI}) may become zero,
we eliminated from consideration any points that have the absolute
value of the denominator become less than $0.001$. For the TVP registration, the
multiresolution strategy is used in the processing and it proceeds
from low to high image resolution starting at one-eighth the spatial
resolution and increases by a factor of two until the full
resolution is reached. Meanwhile, a hierarchy of B-spline grid
spaces from large to small is used. The finest B-spline grid space
used in the experiments is 4~mm. The images and image grid space are
refined alternatively.

\subsection{Assessment of image registration accuracy}
A semi-automatic landmark system is used for landmark selection and
matching~\cite{murphy2008}. This system first uses an automatic
landmark detection algorithm to find the landmarks in the \ee{}
image. The algorithm automatically detects ``distinctive'' points in
the target image as the landmarks based on a distinctiveness value
$D(p)$. Around each point $p$, 45 points, $q_1, \dots, q_{45}$ are
uniformly distributed on a spherical surface. A region of interest
$ROI(q_i)$ is compared with the corresponding region of interest
$ROI(p)$ around the original point, and then combined with its
gradient value to calculate the distinctiveness value $D(p)$.

The same system is then applied to guide the observer to match
landmarks in the target image with corresponding landmarks in the
template image. Each landmark-pair manually annotated by the
observer is added to a thin-plate-spline to warp the template image.
The system utilizes the warped image to estimate where the anatomic
match will be located for a new landmark point presented to the
observer, therefore the observer can start the matching from a
system estimated location. Thus, as the warped image becomes more
accurate by the new added landmarks, the task of the observer
becomes easier.

For each animal, after 200 anatomic landmarks are identified in the
\ee{} image, the observer marks the locations of the voxels corresponding
to the anatomic locations of the landmarks in the \ei{} image. For each
landmark, the actual landmark position is compared to the
registration-derived estimate of landmark position and the error is
calculated. With the evaluated accuracy of transformation from the
lung image registration algorithm, the resulting regional
ventilation measures estimated using the transformation can be then
compared to Xe-CT estimated ventilation.

\subsection{Compare Registration Regional Ventilation \\Measures to Xe-CT Estimated Ventilation}
In our previous work~\cite{Reinhardt2008752,ding2009}, regional lung
expansion, as estimated from the Jacobian of the image registration
transformations, was compared with Xe-CT-based sV.  The
analysis was conducted by evaluating Jacobian value between a pair
of lung volumes from static (multiple airway pressures) and dynamic
image data sets, and comparing the Jacobian along the y
(ventral-dorsal) axis.  While the correlation between the
Jacobian value and sV reflect the fact that regional expansion
estimated from image registration can be used as an index as
regional lung function, the spatial resolution of the analysis
method employed were likely not sufficient to distinguish the
differences between regional ventilation measures as we have
described in Section~\ref{sect:meth_regvent}. Therefore, to locally
compare the regional ventilation measures, the corresponding region
of Xe-CT image \eeinit{} in the \ee{} is divided into approximately
100 non-overlapping cubes with size of 20 mm $\times$ 20 mm $\times$
20 mm. We compare the average regional ventilation measures (SAJ,
SACJ and SAI) to the corresponding average sV measurement from Xe-CT
images within each cube. The correlation coefficients between any
two estimates (SAJ-sV, SACJ-sV or SAI-sV) are calculated by linear
regression. To compare two correlation coefficients, the Fisher
Z-transform of the $r$ values is used and the level of significance
is determined~\cite{papoulis1990}. The relationship between the
specific air volume change and difference of tissue volume is also
studied in four animals by linear regression analysis.

\section{Results} \label{sec:results}
\subsection{Registration Accuracy}
\begin{figure}[htbp]
  \centering
  \begin{tabular}{cc}
  \includegraphics[width=0.52\textwidth]{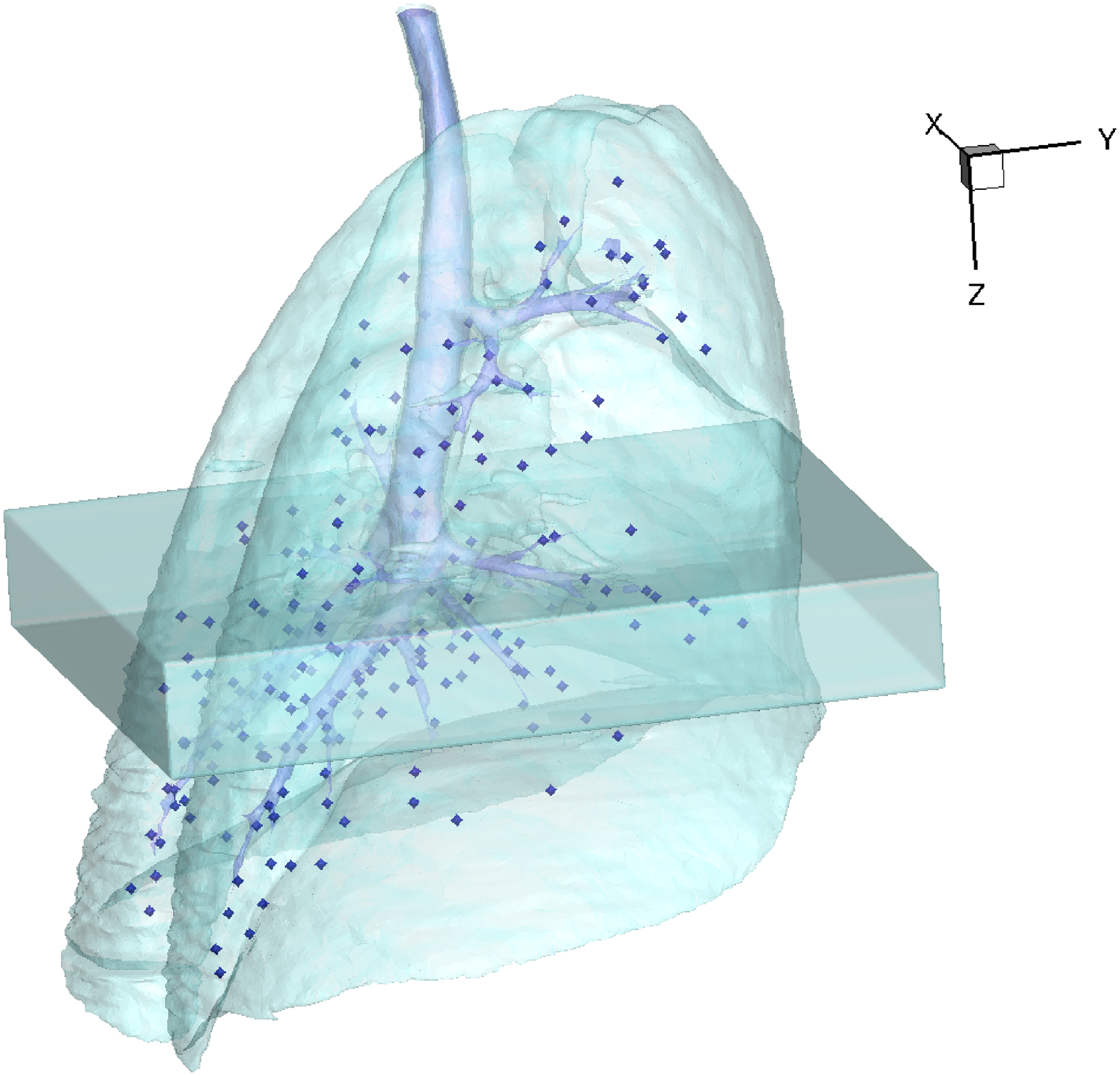}
  &
  \includegraphics[width=0.52\textwidth]{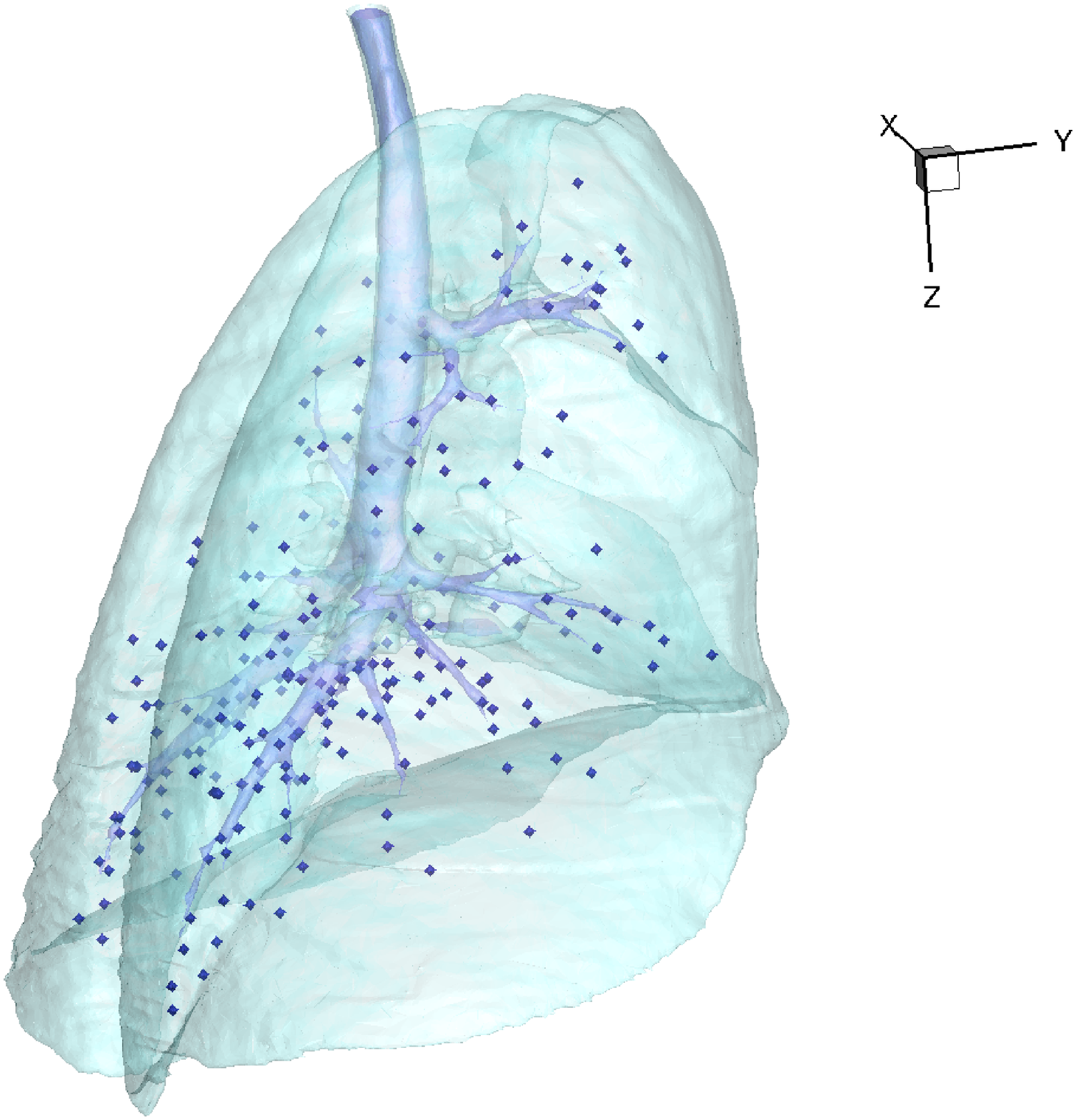}\\
  (a)&(b)\\
  \end{tabular}
  \caption{3D view of the landmarks in: (a) \ee{} with \eeinit{} and (b) \ei{}. The dark region below the carina in (a) is the \eeinit{} and the spheres are the automatically defined landmarks.}
  \label{fig:xe3d}
\end{figure}

\begin{figure}[htbp]
  \centering
  \begin{tabular}{c}
    \includegraphics[width=0.8\textwidth]{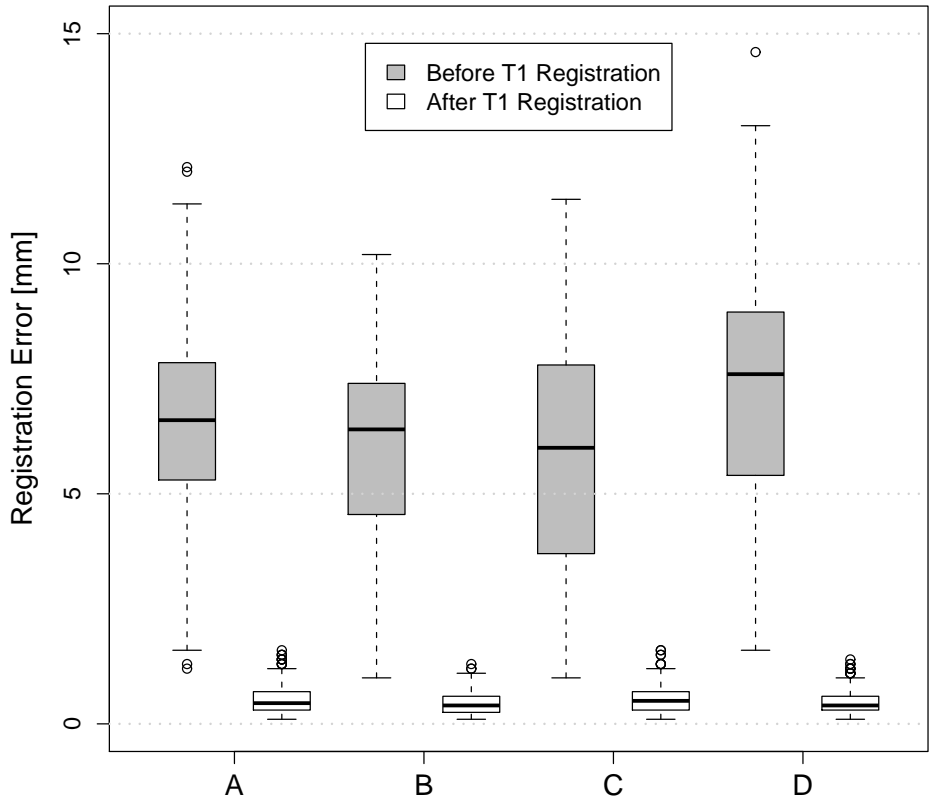}
  \end{tabular}
  \caption{Landmarks distances of the registration pair \ei{} to \ee{} for all four animals.  Boxplot lower extreme is first quartile,
  boxplot upper extreme is third quartile.  Median is shown with solid horizontal
line.   Whiskers show either the minimum (maximum) value or extend
1.5 times the first to third quartile range beyond the lower (upper)
extreme of the box, whichever is smaller (larger). Outliers are
marked with circles.}
  \label{fig:lmkerror}
\end{figure}

For each animal, approximately 200 automatically identified
landmarks within the lungs are used to compute registration
accuracy. The landmarks are widely distributed throughout the lung
regions. Figure~\ref{fig:xe3d} shows an example of the distribution
of the landmarks in animal D for both the \ee{} and \ei{} images.
The coordinate of each landmark location is recorded for each image
data set before and after registration for all four animals.
Figure~\ref{fig:lmkerror} shows the landmark distance before and
after registration for four animals. The grey boxes show the
magnitude of respiratory motion during the tidal breathing. For all
four animals, before registration, the average landmark distance is
6.6 mm with a minimum distance of 1.0 mm, maximum distance of 14.6
mm, and standard deviation of 2.42 mm. After registration, the
average landmark distance is 0.4 mm with a minimum distance of 0.1
mm, a maximum distance of 1.6 mm, and a standard deviation 0.29 mm.
The trends for all animals are consistent and the results
demonstrate that the registrations produced good anatomic
correspondences. All registrations were examined and it was
confirmed that all Jacobian values had positive values.

\begin{figure}[p]
  \centering
  \begin{tabular}{cc}
    \includegraphics[width=0.38\textwidth]{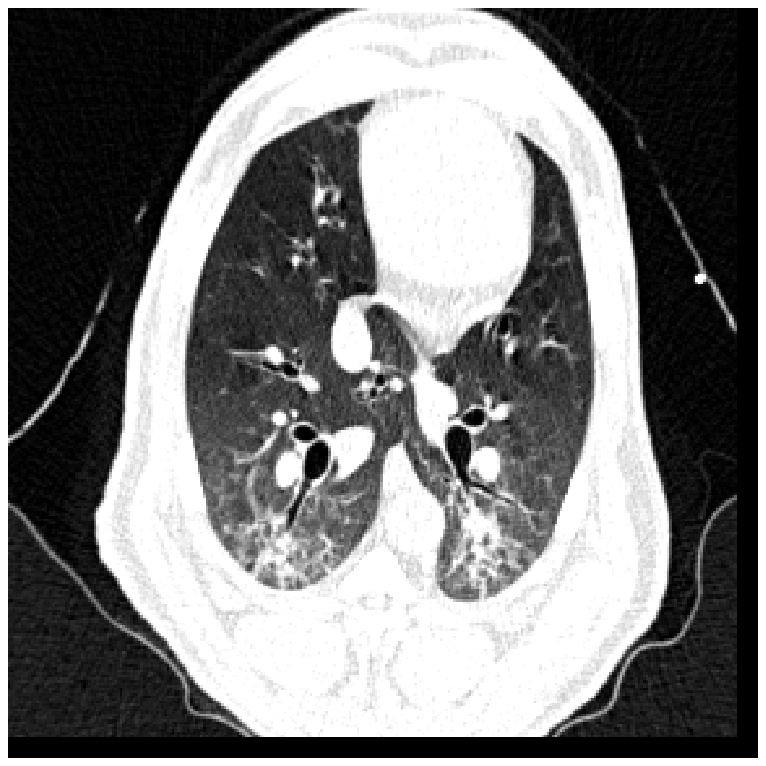}\\
    (a)\\
    \includegraphics[width=0.38\textwidth]{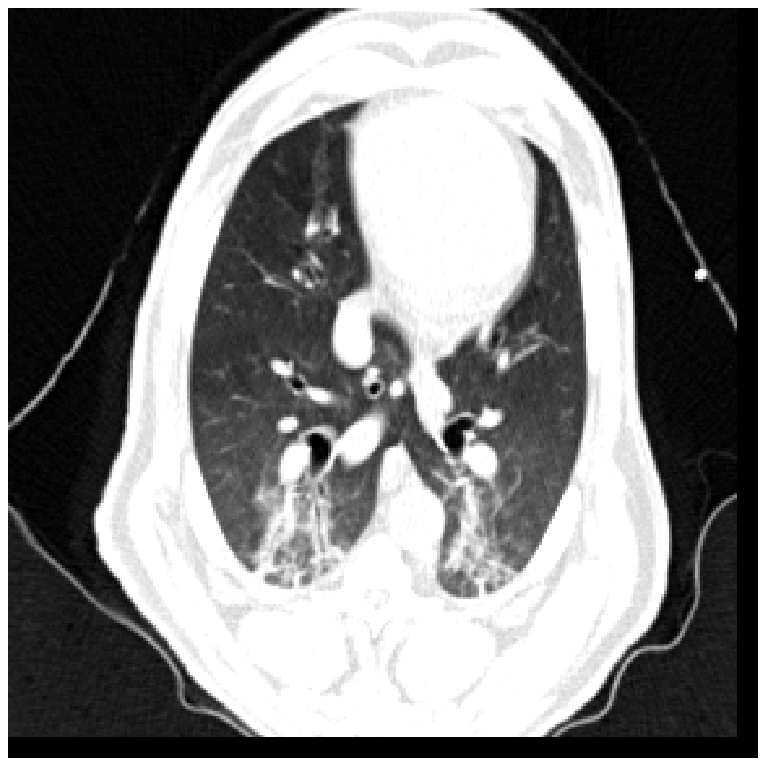}
    &
    \includegraphics[width=0.38\textwidth]{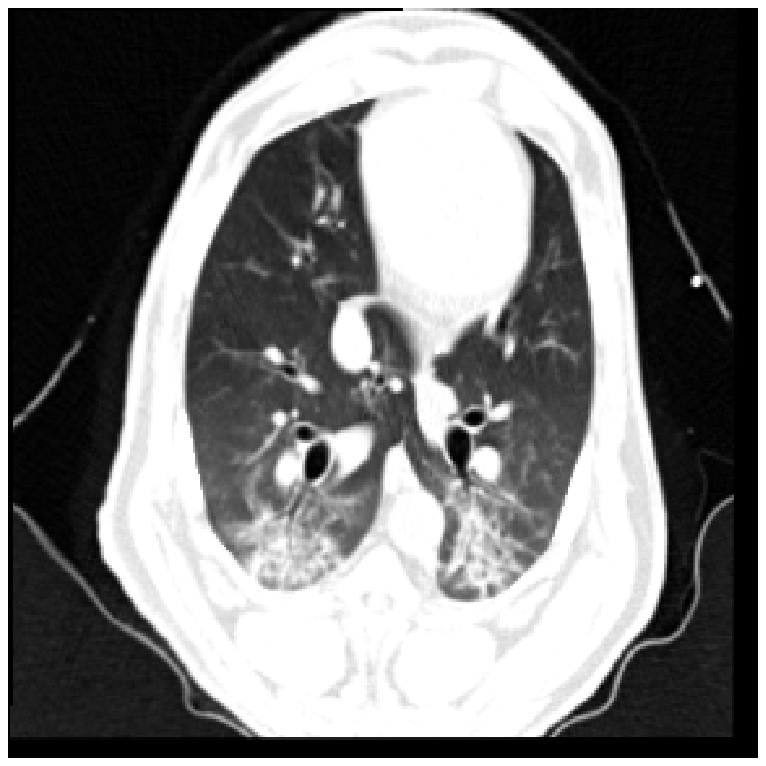}
    \\
    (b)&(c)\\
    \includegraphics[width=0.38\textwidth]{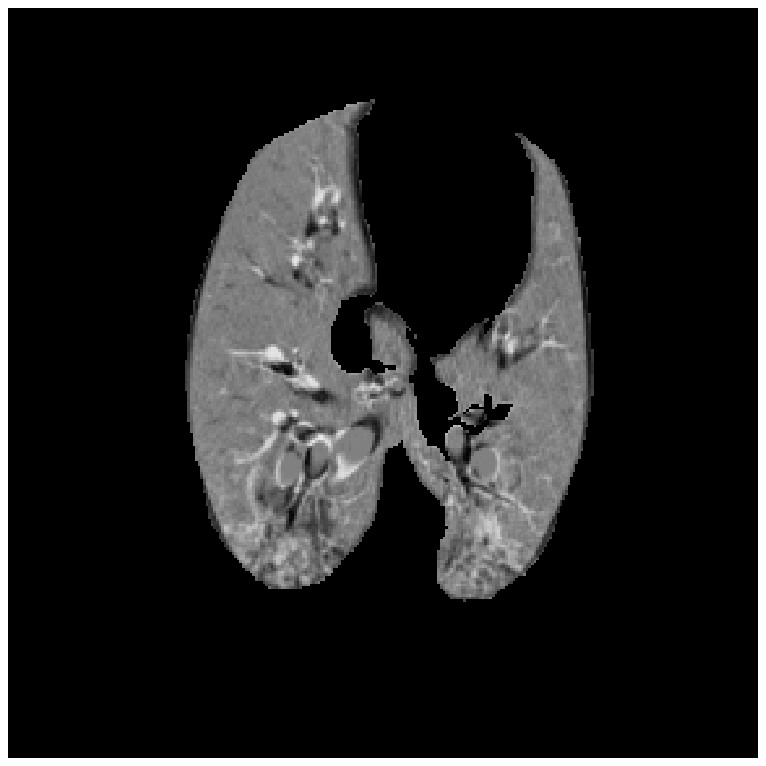}
    &
    \includegraphics[width=0.38\textwidth]{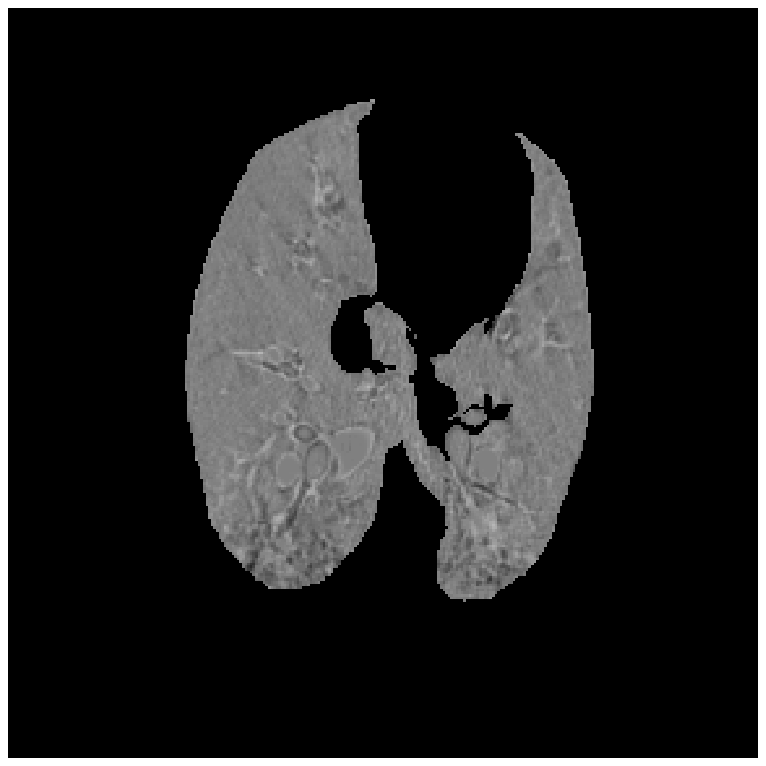}
    \\
    (d)&(e)\\
  \end{tabular}
  \caption{Visualization of the result of the transformation that
  maps the Xe-CT estimated ventilation sV to the \ee{} coordinate system:
  (a) \ee{}, (b) \eeinit{}, (c) deformed \eeinit{} after registration,
  (d) intensity difference between \ee{} and \eeinit{}
  before registration, (e) intensity difference
  between \ee{} and \eeinit{} after
registration.}
   \label{fig:xe-0in-error}
\end{figure}

Figure~\ref{fig:xe3d}(a) shows the location of the \eeinit{} (Xe-CT)
slab overlaid on the \ee{} image. Figure~\ref{fig:xe-0in-error}
shows an example of the image registration result from the \eeinit{}
image to the \ee{} image. The first row shows the misalignment
between the images before image registration. Though the images were
acquired without moving the animal between the scans, there is still
non-rigid deformation between scans as shown in
Fig~\ref{fig:xe-0in-error}(d), as the black and white regions
represent the large intensity difference between
Fig.~\ref{fig:xe-0in-error}(a) and (b).  In addition, the
slice thicknesses were quite different which causes partial volume
artifacts. After image registration, the \eeinit{} image is
aligned to the \ee{} image, and the resulting difference image
(shown in Fig.~\ref{fig:xe-0in-error}(e)) is near zero. The
transformation from the \eeinit{} to the \ee{} image allows us to
map the Xe-CT sV into the coordinate system of \ee{} image. Note
that since the regions outside the lung are not included in the
registration process, the mediastinum and other body tissues are not
aligned. Also note that the dorsal region of the lung shows a
intensity difference after registration. This is due mainly to the
gradual progression of atelectasis and tissue edema during the
course of the experiment.

\subsection{Registration Estimated Ventilation Compared \\to Xe-CT Estimated Ventilation}
\begin{figure}[p]
  \centering
  \begin{tabular}{cc}
    \includegraphics[width=0.35\textwidth]{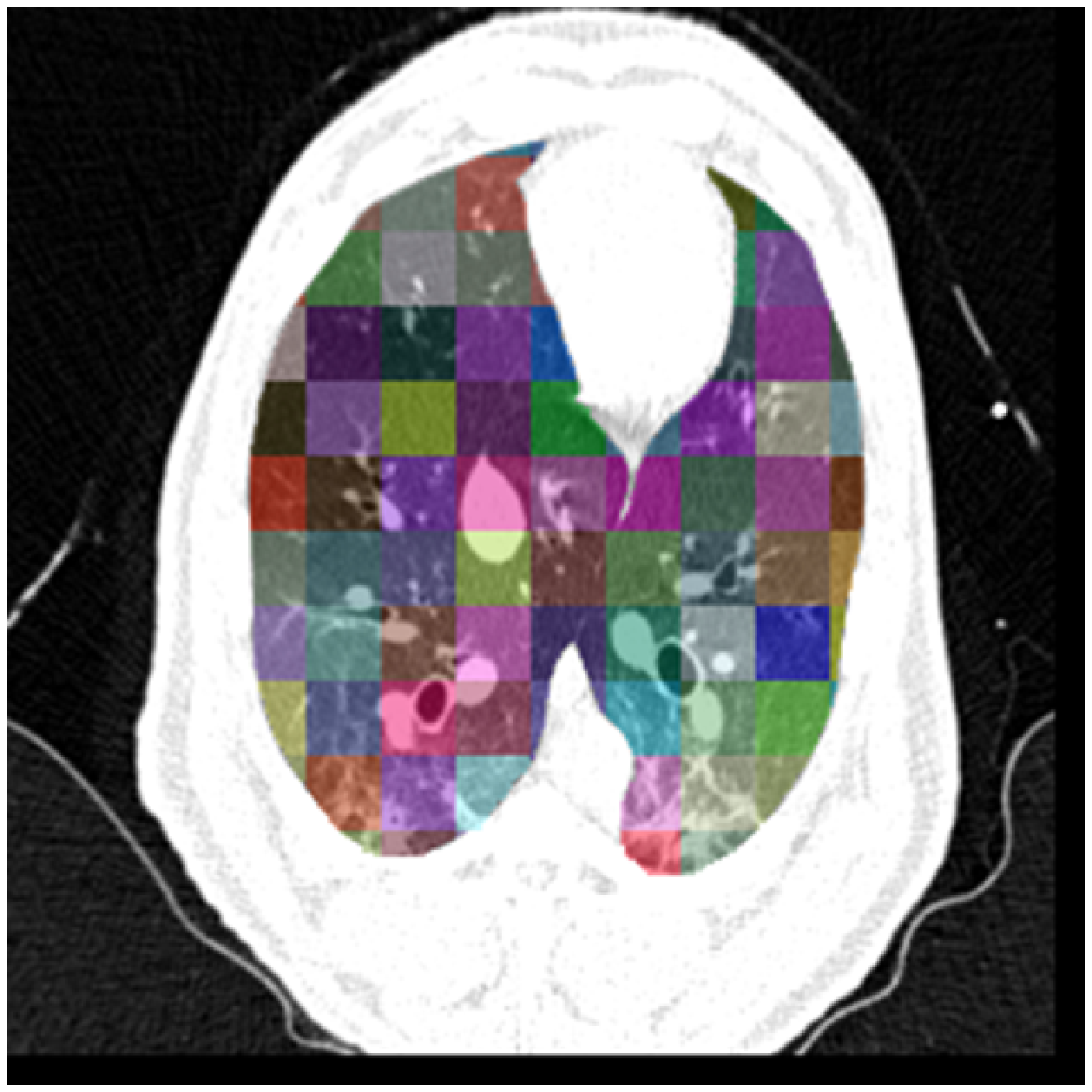}\\
    (a)\\
    \includegraphics[width=0.35\textwidth]{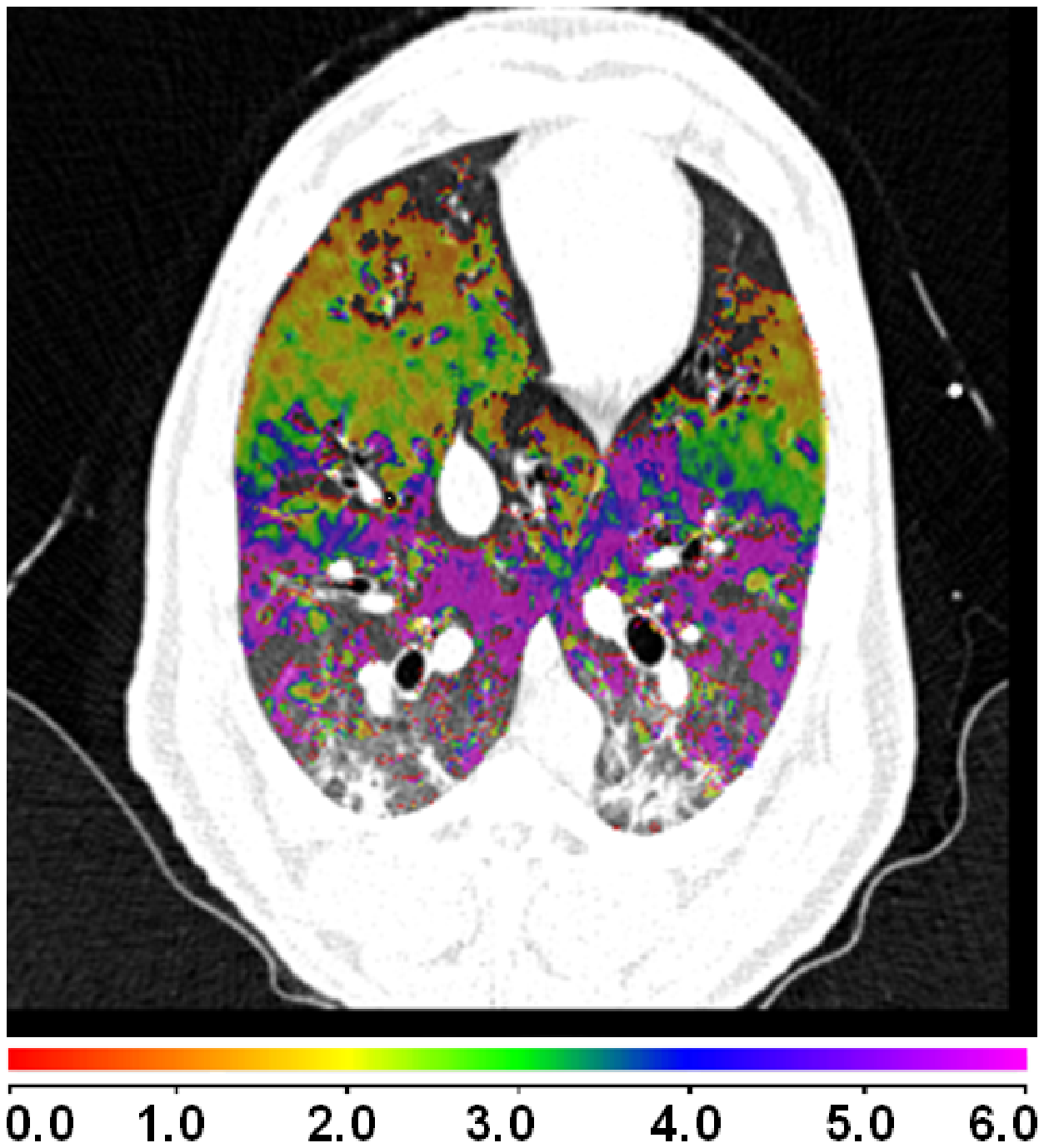}
    &
    \includegraphics[width=0.35\textwidth]{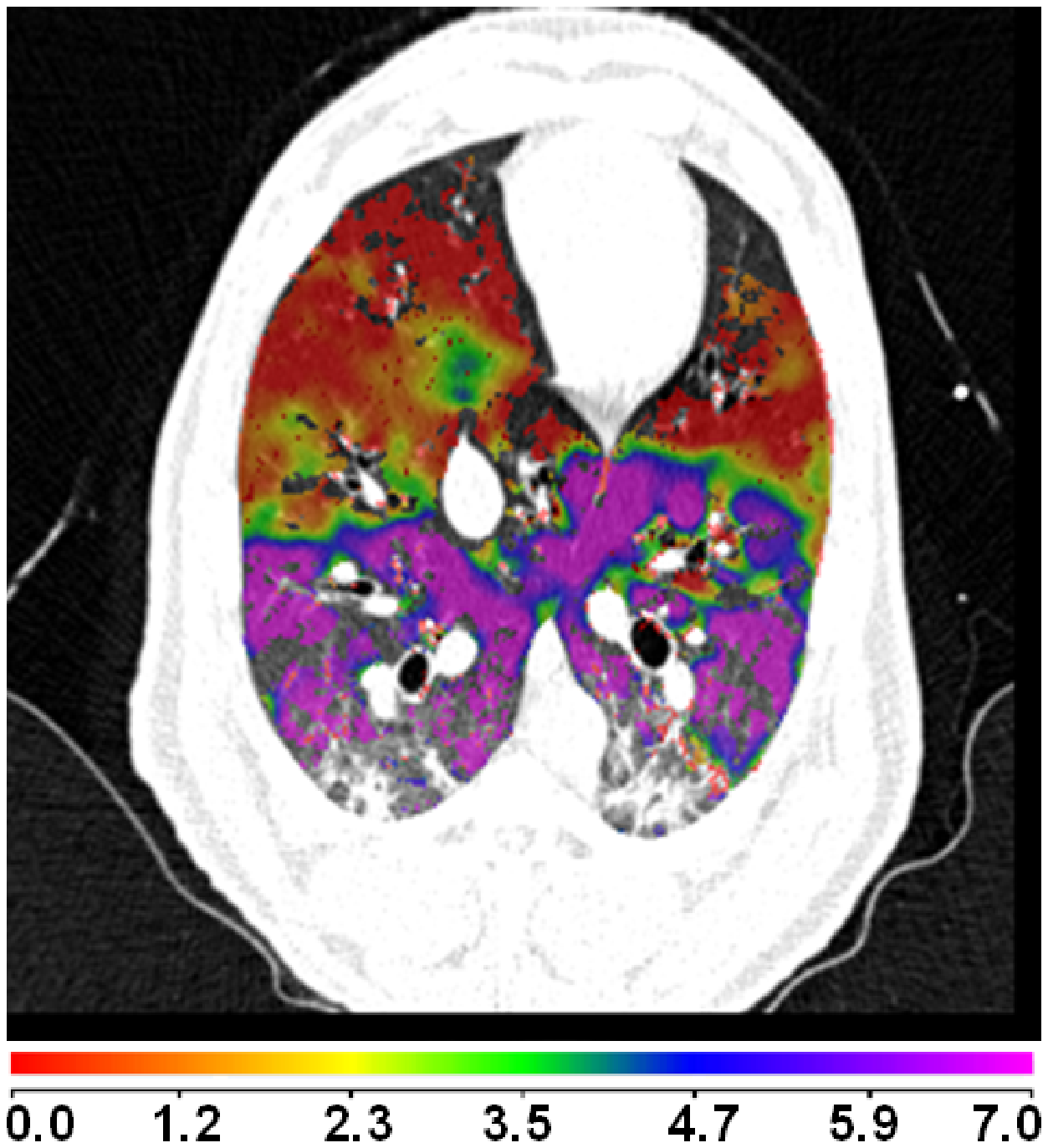}\\
    (b)&(c)\\
    \includegraphics[width=0.35\textwidth]{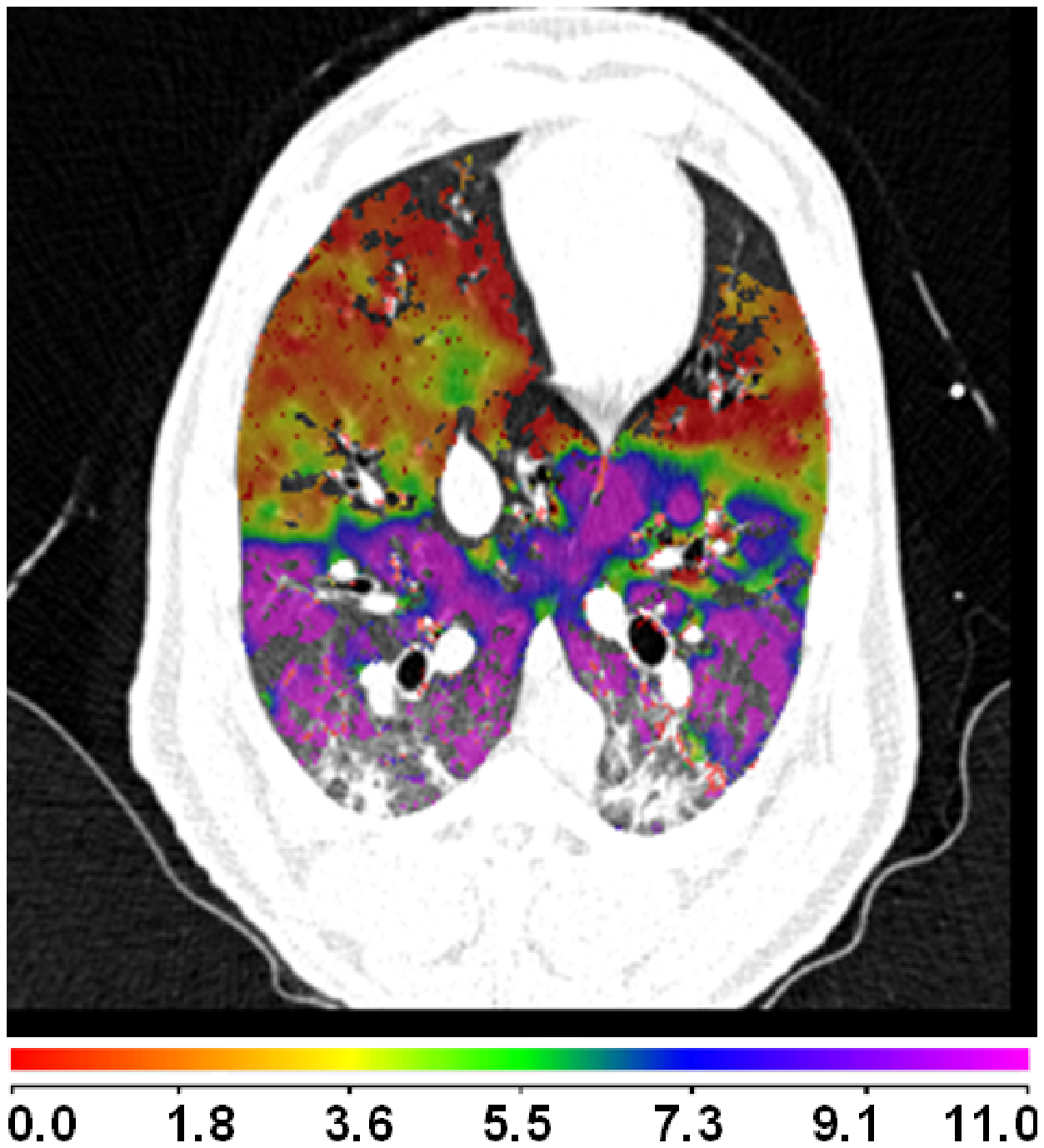}
    &
    \includegraphics[width=0.35\textwidth]{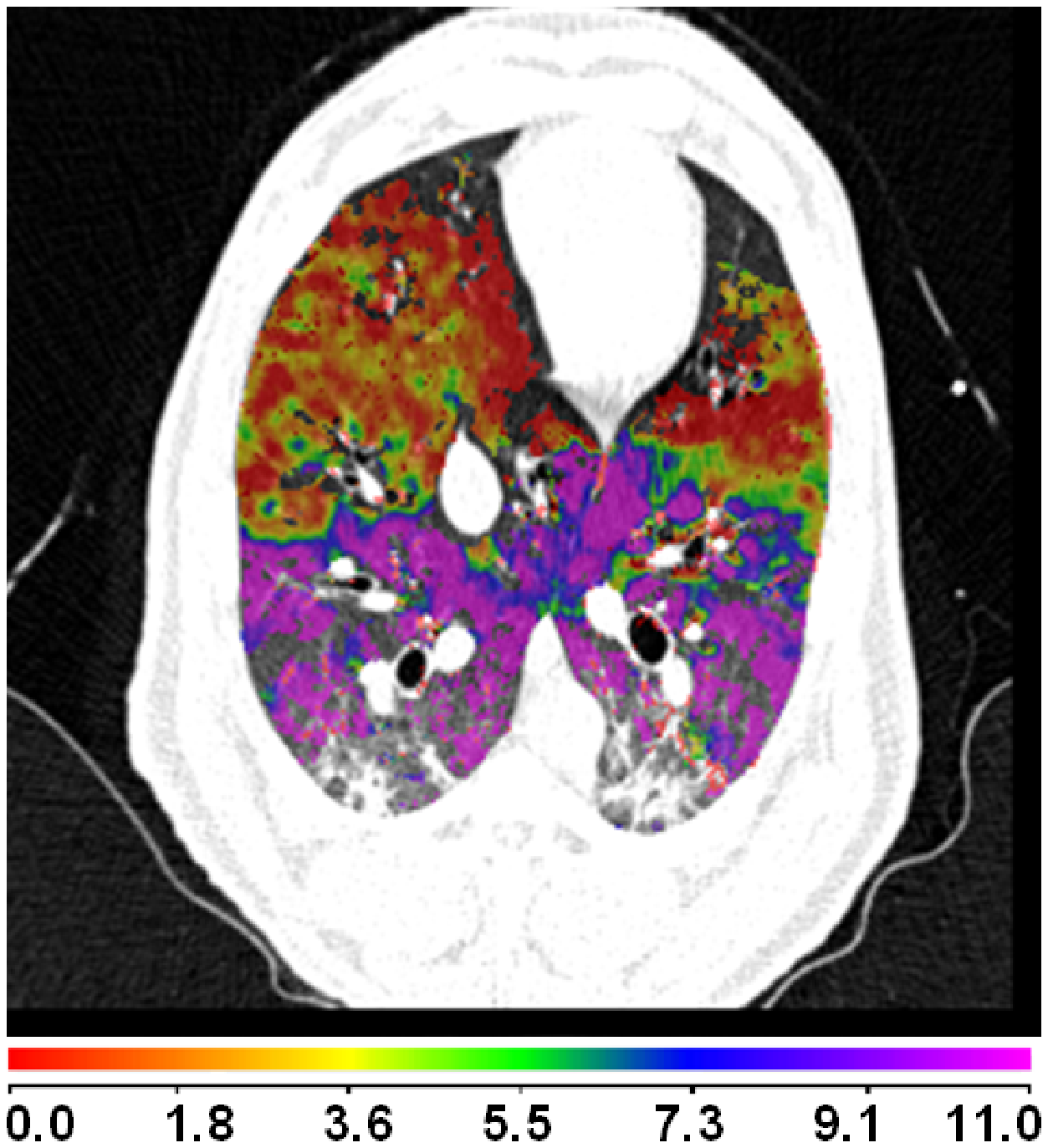}\\
    (d)&(e)\\
  \end{tabular}
  \caption{Comparison of the regional ventilation measures for animal D.
           (a): \ee{} with color coded cubes showing the sample region.
           (b), (c), (d) and (e): color map of the sV, SAJ, SACJ and SAI.
           Note that the color scales are different for (b)-(e), and are set
           based on the range of values from the appropriate plot in Fig.~\ref{fig:regvent-xevent-cuberoi}. The results were similar for the other three animals}
   \label{fig:vent}
\end{figure}

\begin{figure}[p]
  \centering
  \begin{tabular}{ccc}
    \includegraphics[width=0.28\textwidth]{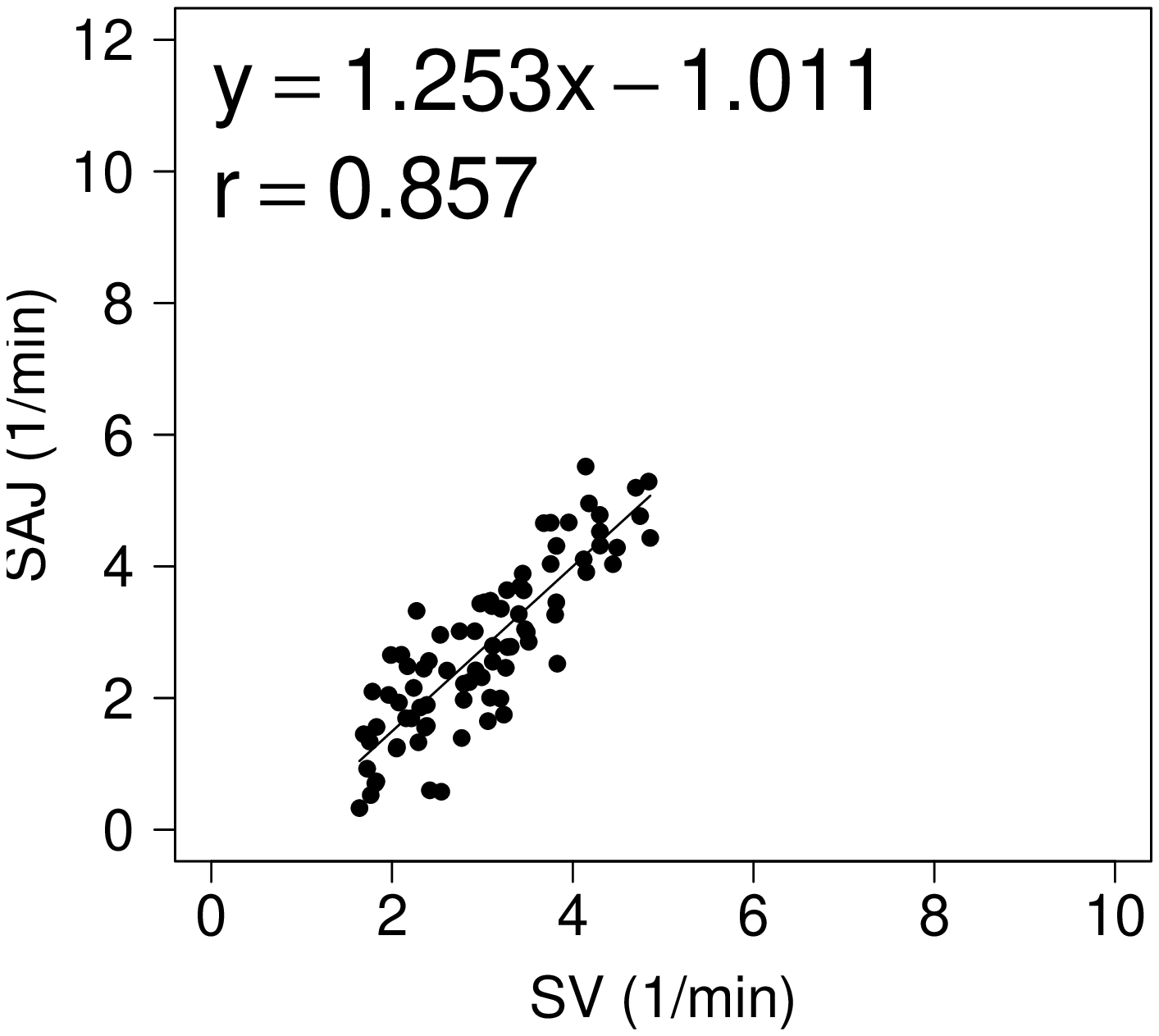}
    &
    \includegraphics[width=0.28\textwidth]{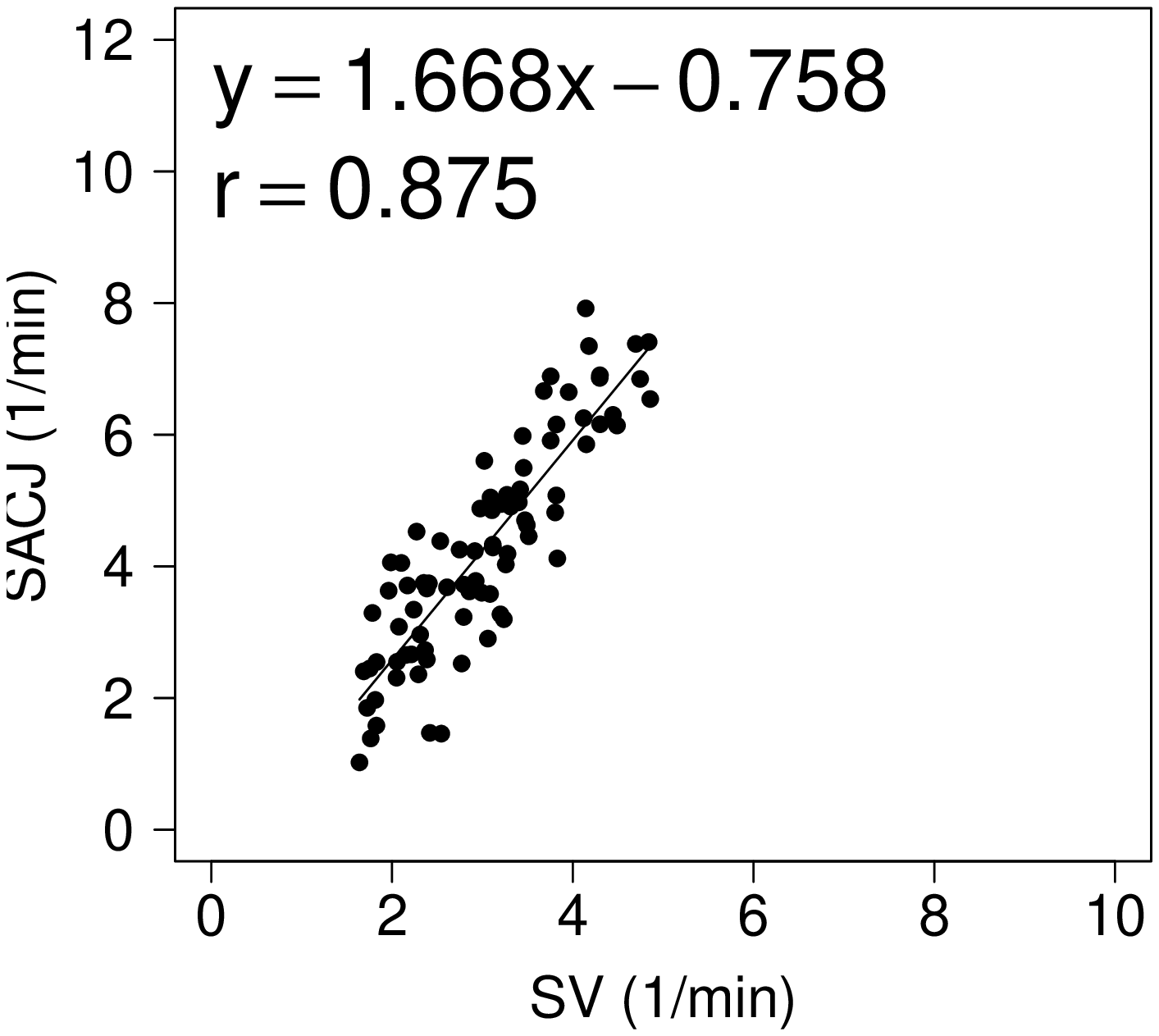}
    &
    \includegraphics[width=0.28\textwidth]{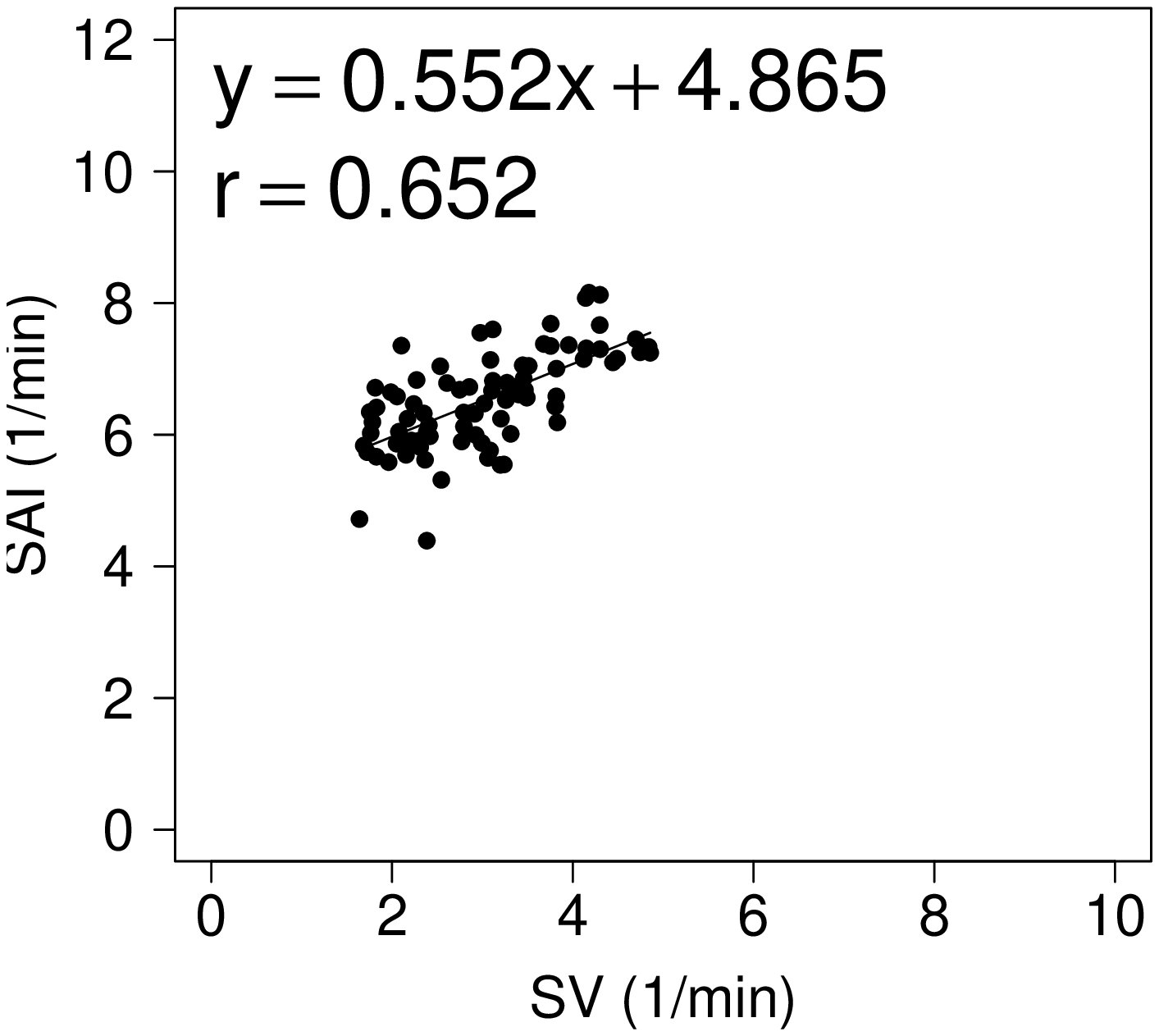}\\
    ~&Animal A&~\\
    \includegraphics[width=0.28\textwidth]{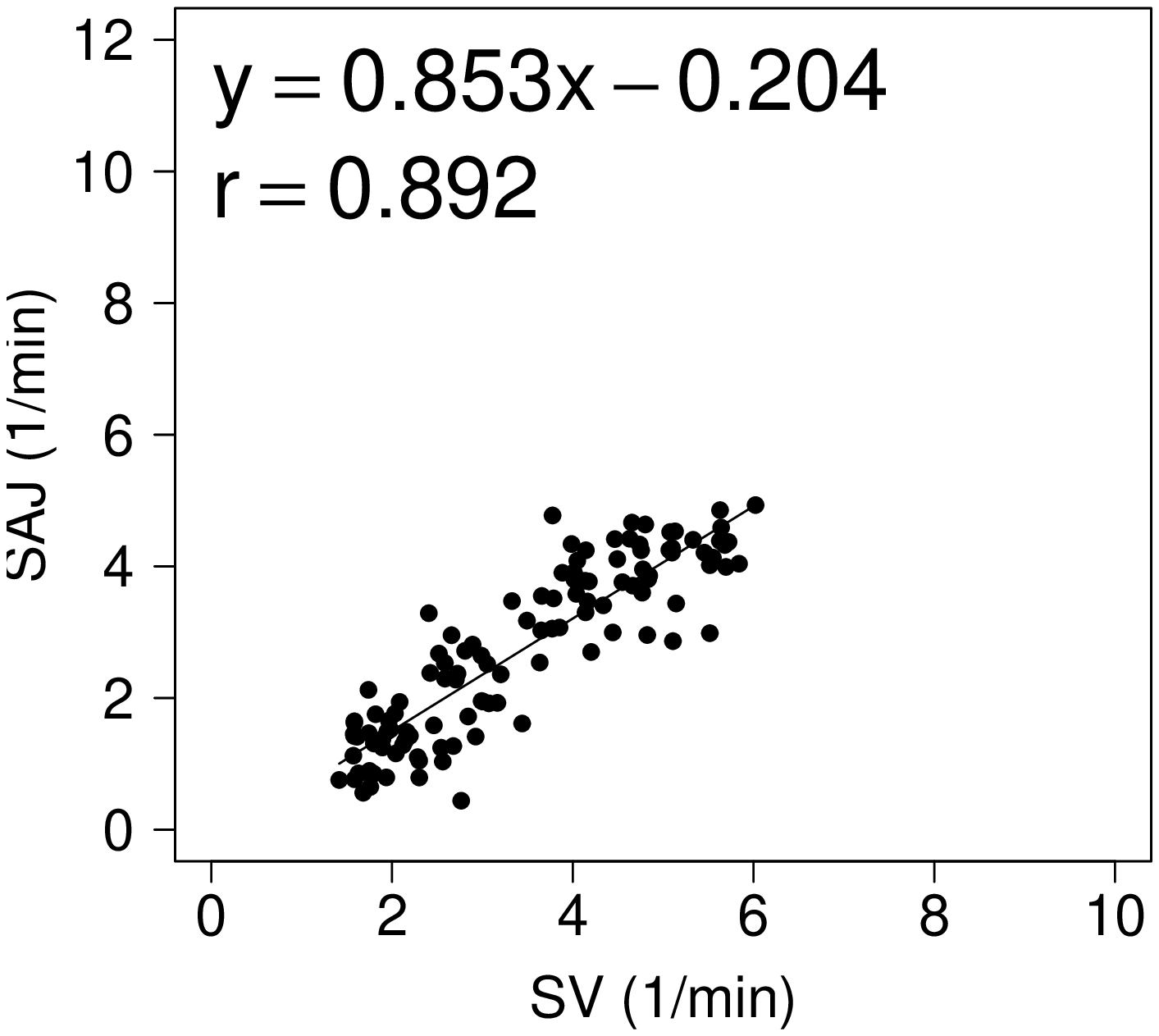}
    &
    \includegraphics[width=0.28\textwidth]{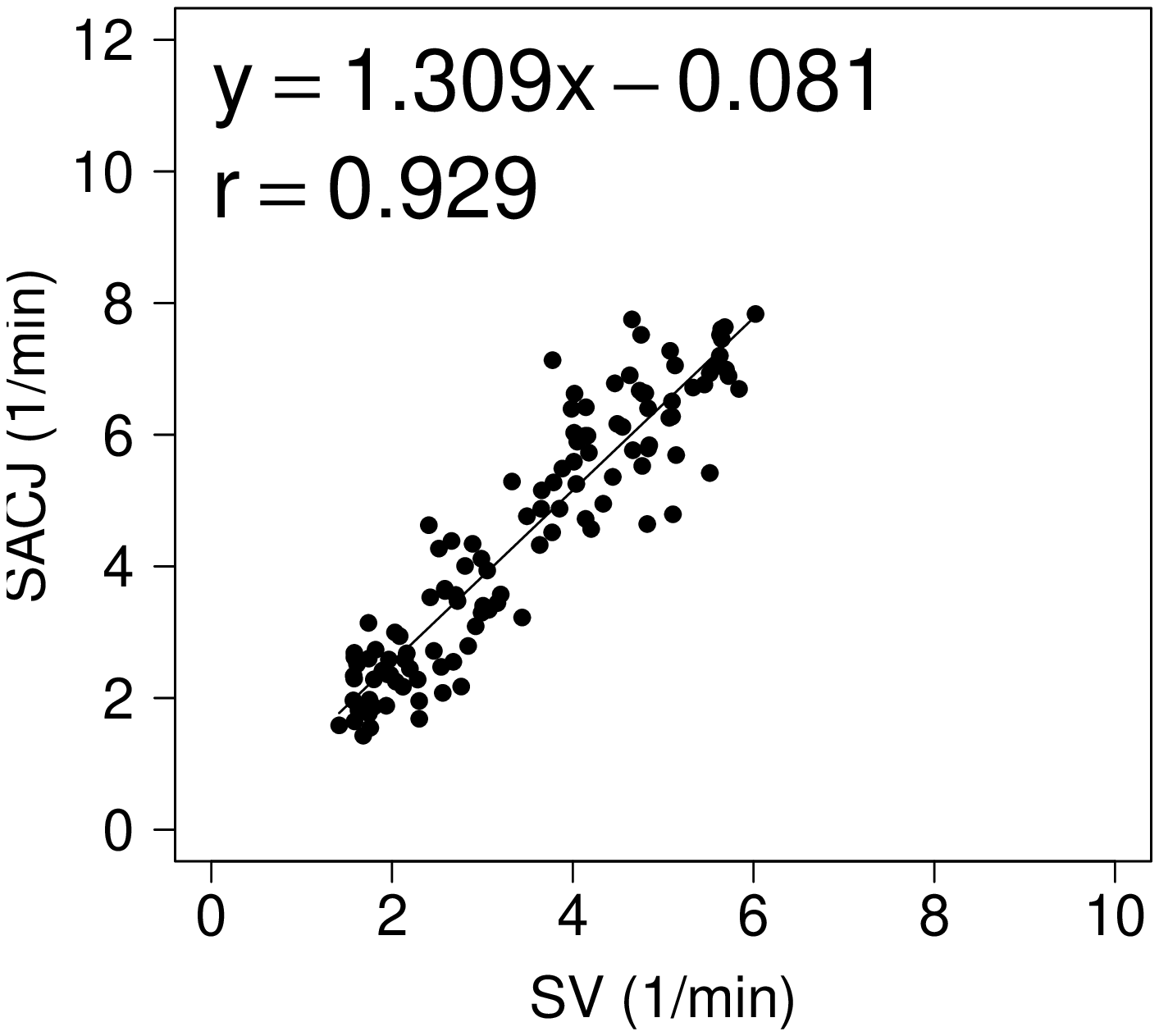}
    &
    \includegraphics[width=0.28\textwidth]{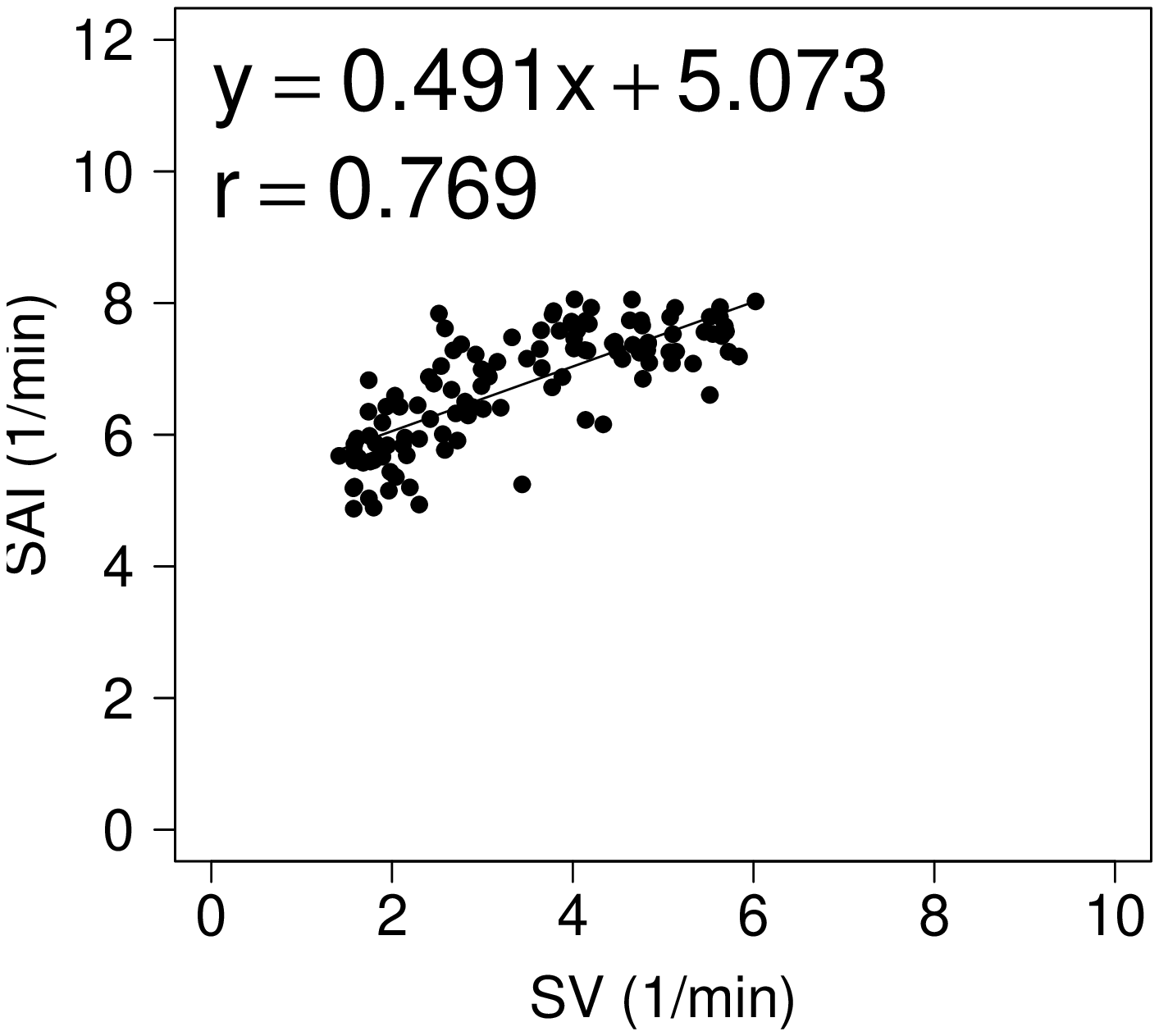}\\
    ~&Animal B&~\\
    \includegraphics[width=0.28\textwidth]{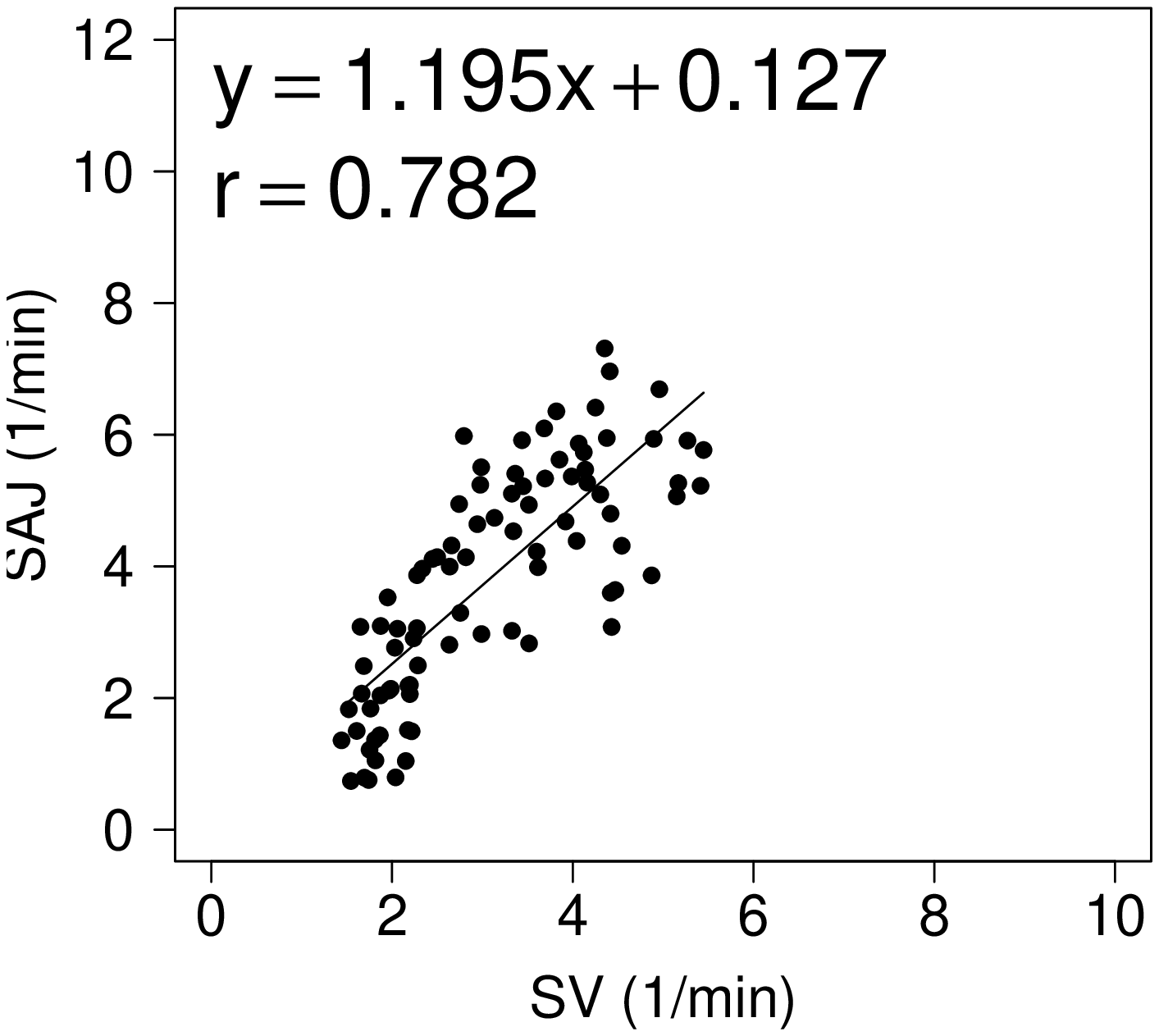}
    &
    \includegraphics[width=0.28\textwidth]{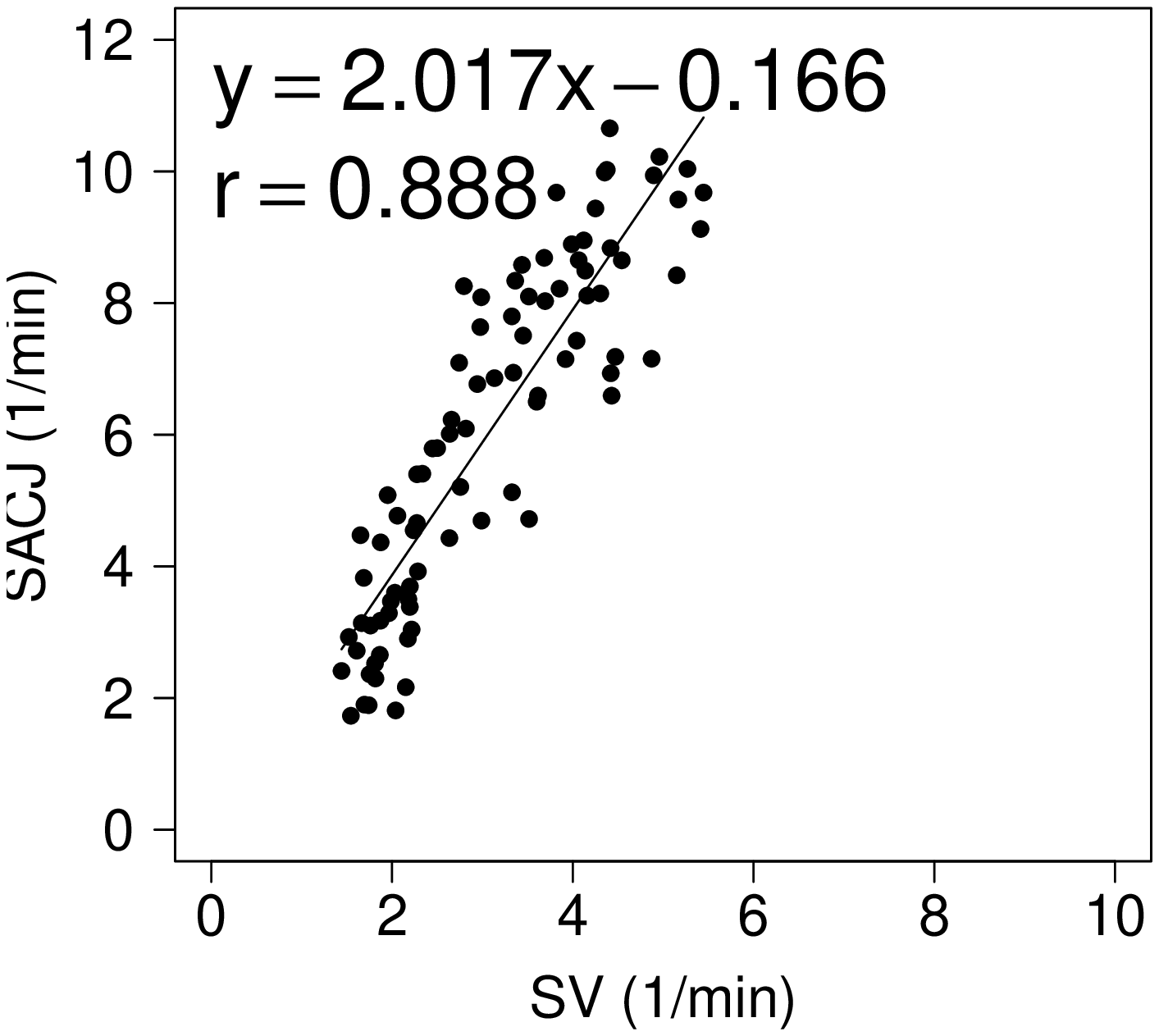}
    &
    \includegraphics[width=0.28\textwidth]{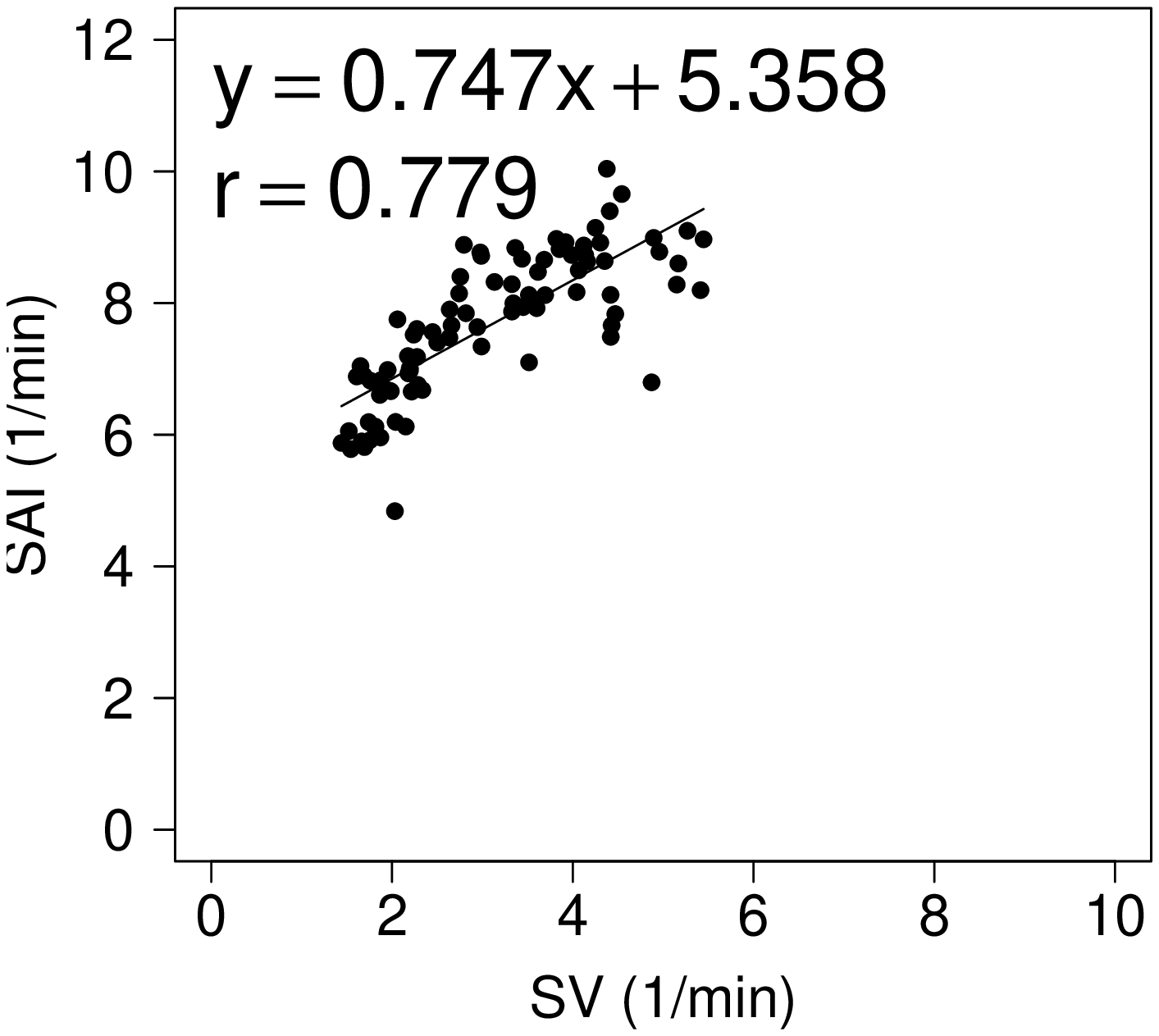}\\
    ~&Animal C&~\\
    \includegraphics[width=0.28\textwidth]{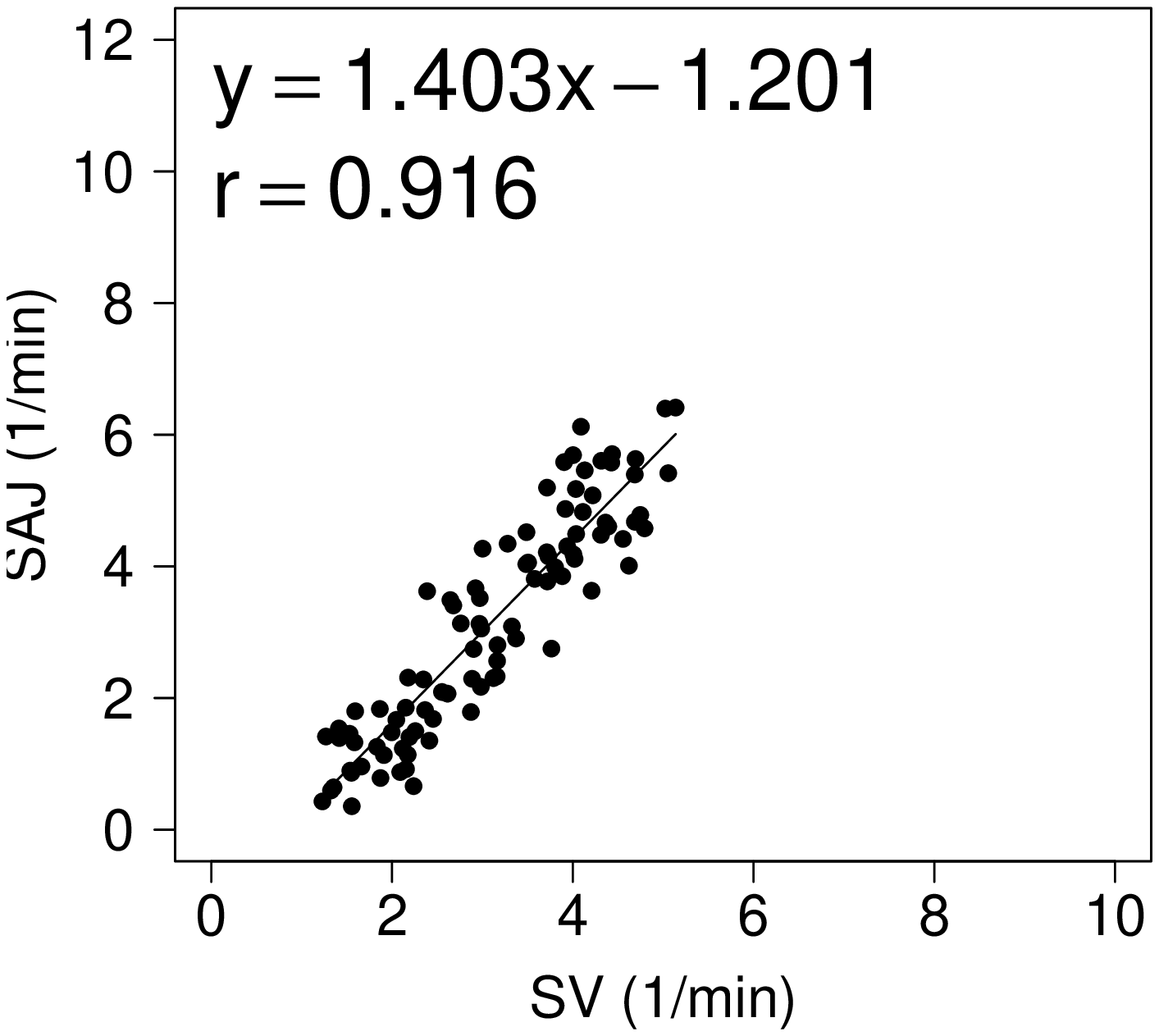}
    &
    \includegraphics[width=0.28\textwidth]{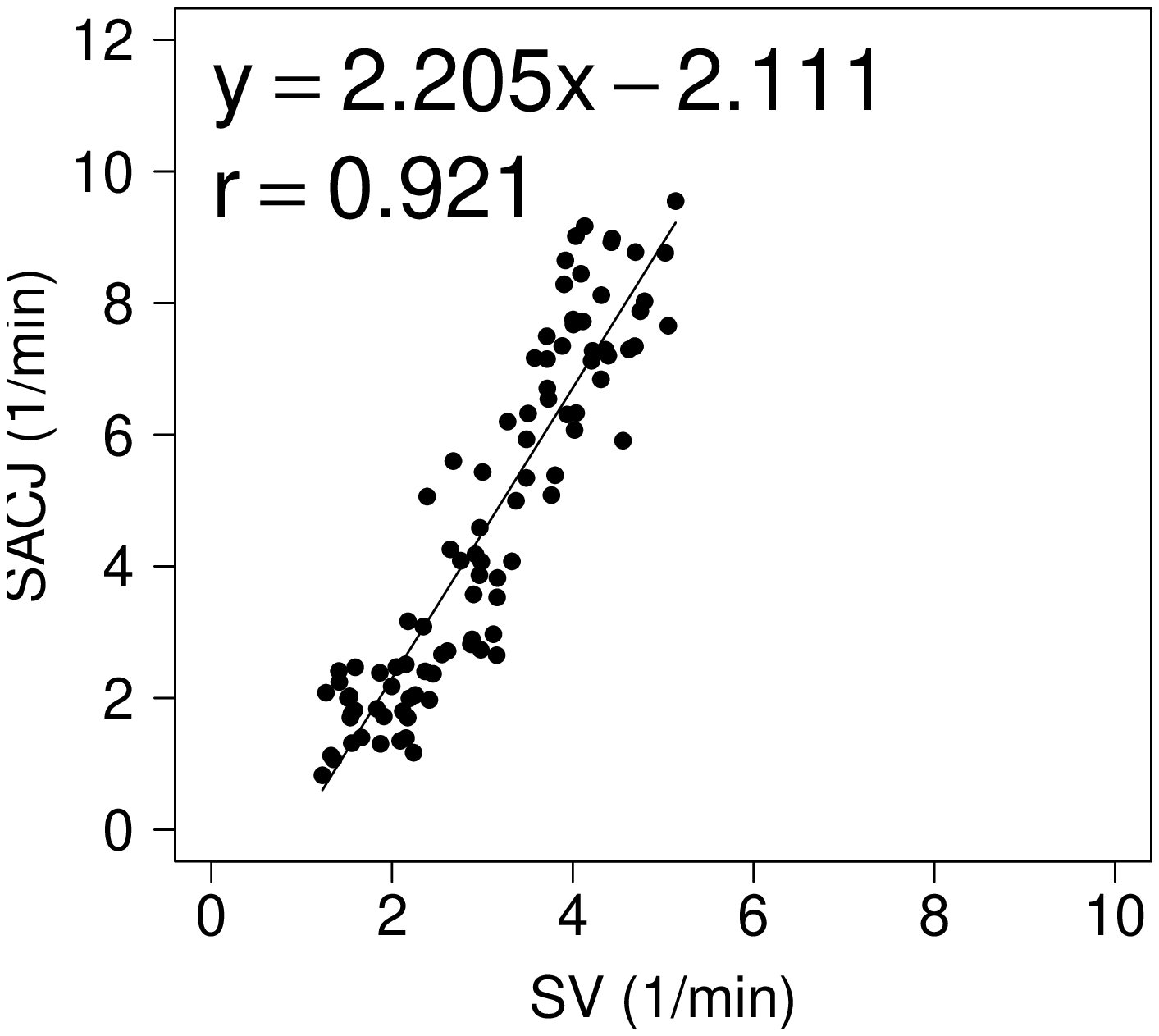}
    &
    \includegraphics[width=0.28\textwidth]{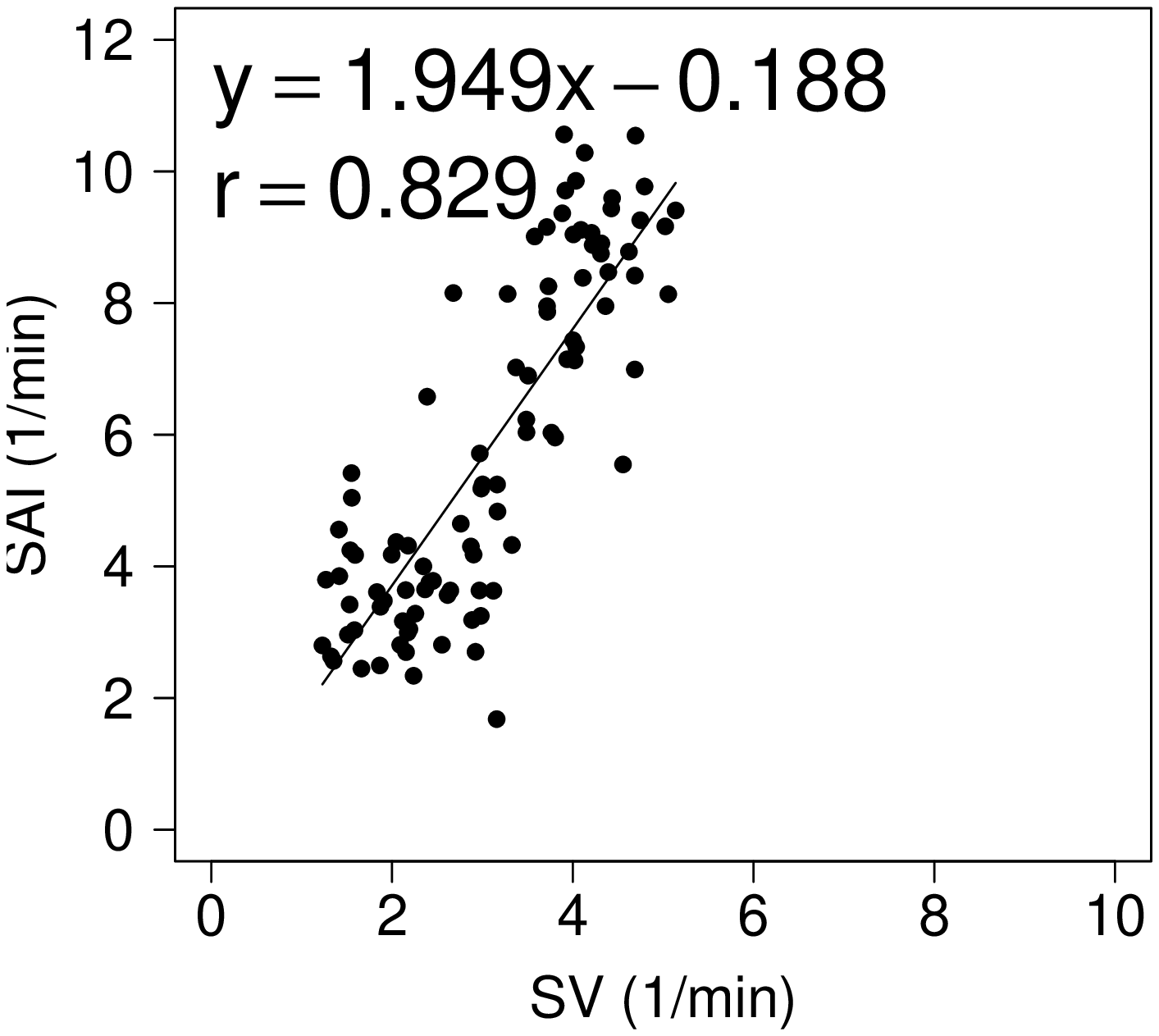}\\
    ~&Animal D&~\\
  \end{tabular}
  \caption{Small cube ROIs with size 20 mm $\times$ 20 mm $\times$ 20
    mm results for registration estimated ventilation measures
    compared to the Xe-CT estimated ventilation sV in scatter plot
    with linear regression in four animals. The first column is the
    SAJ vs.\ sV. The second column is the SACJ vs.\ sV. The third column
    is the SAI vs.\  sV.}
  \label{fig:regvent-xevent-cuberoi}
\end{figure}

Figure~\ref{fig:vent}(a) shows a comparison between the
registration-derived indices of ventilation and the Xe-CT estimated
sV in cube-shaped regions of interest for animal D. The
corresponding Xe-CT regions in the \ee{} are divided into about 100
cubes. Figure~\ref{fig:vent}(b) is the Xe-CT estimate of sV.
Figure~\ref{fig:vent}(c), (d), (e) are the corresponding
registration ventilation measures SAJ, SACJ and SAI\@. The regions
with edema are excluded from the comparison.
Figure~\ref{fig:vent}(b) to (d) all show noticeable similar gradient
in the ventral-dorsal direction. Notice that the color scales are
different in each map and are set based on the range of values from
the appropriate plot in Fig.~\ref{fig:regvent-xevent-cuberoi}.

Figure~\ref{fig:regvent-xevent-cuberoi} shows scatter plots
comparing the registration ventilation measures and the Xe-CT
ventilation sV in all four animals. The SACJ
column shows the strongest correlation with the sV (average
$r^2=0.82$). The SAJ, which is directly related to Jacobian as
$\mathrm{SAJ}=\mathrm{J}-1$, also shows good correlation with the sV
(average $r^2=0.75$). The intensity-based measure SAI shows the
lowest correlation with the sV (average $r^2=0.58$).

Table~\ref{tbl:pvalue-cuberoi-sacj} shows the results of comparing
the $r$ values from SACJ vs.\ sV and SAI vs.\ sV. All four animals show
that the correlation coefficient from SACJ vs.\ sV is significantly
stronger than it from SAI vs.\ sV. Similarly,
table~\ref{tbl:pvalue-cuberoi-saj} shows the results of comparing
the r values from SAJ vs.\ sV and SAI vs.\ sV. The registration
ventilation measure SAJ also shows a significantly stronger
correlation with sV than SAI\@. Comparing the r values from SACJ vs.\ sV and SAJ vs.\ sV, only animal B and C show that the SACJ has significantly stronger correlation with sV than SAJ.

\begin{figure}[p]
  \centering
  \begin{tabular}{ccc}
    \includegraphics[width=0.28\textwidth]{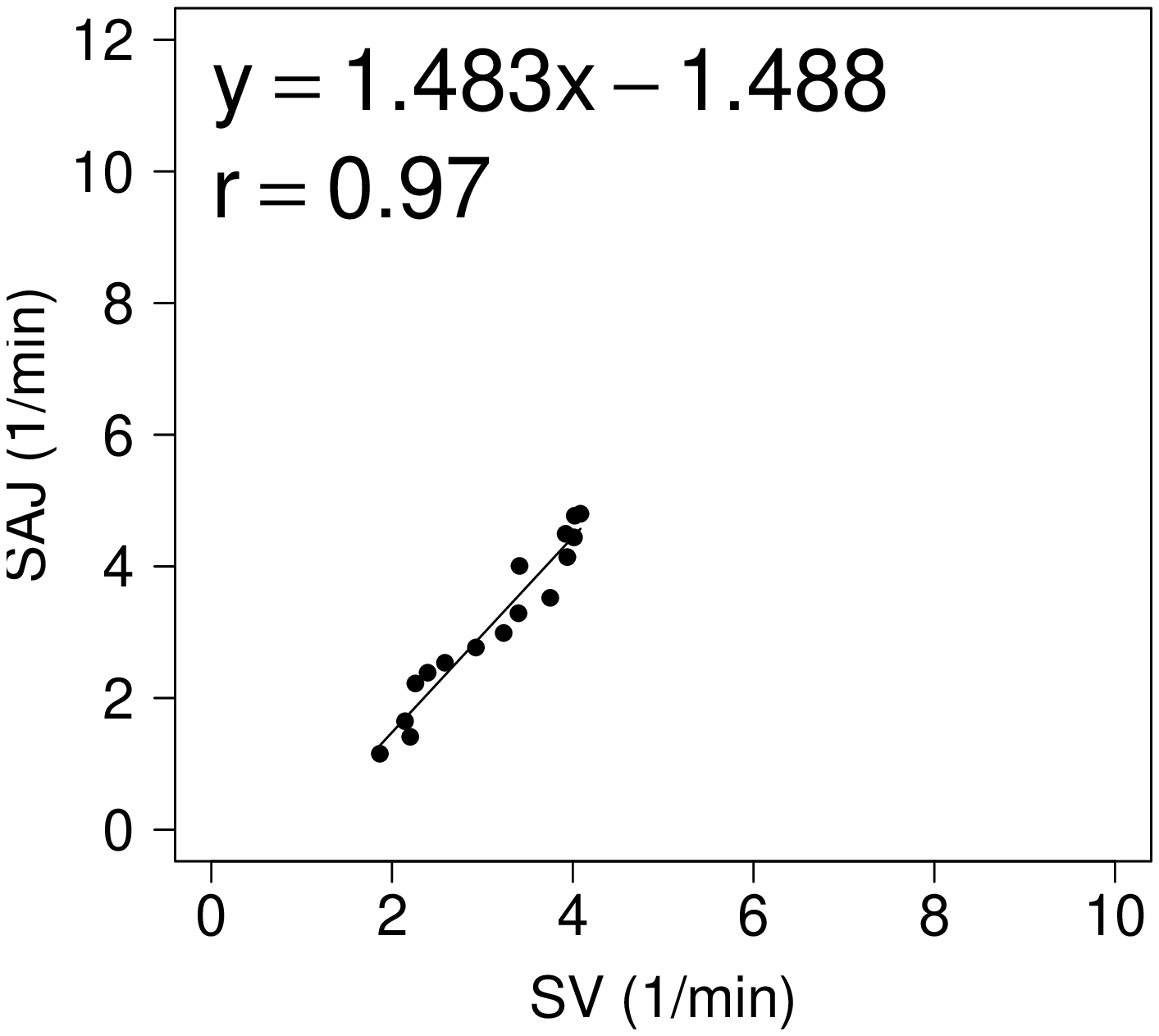}
    &
    \includegraphics[width=0.28\textwidth]{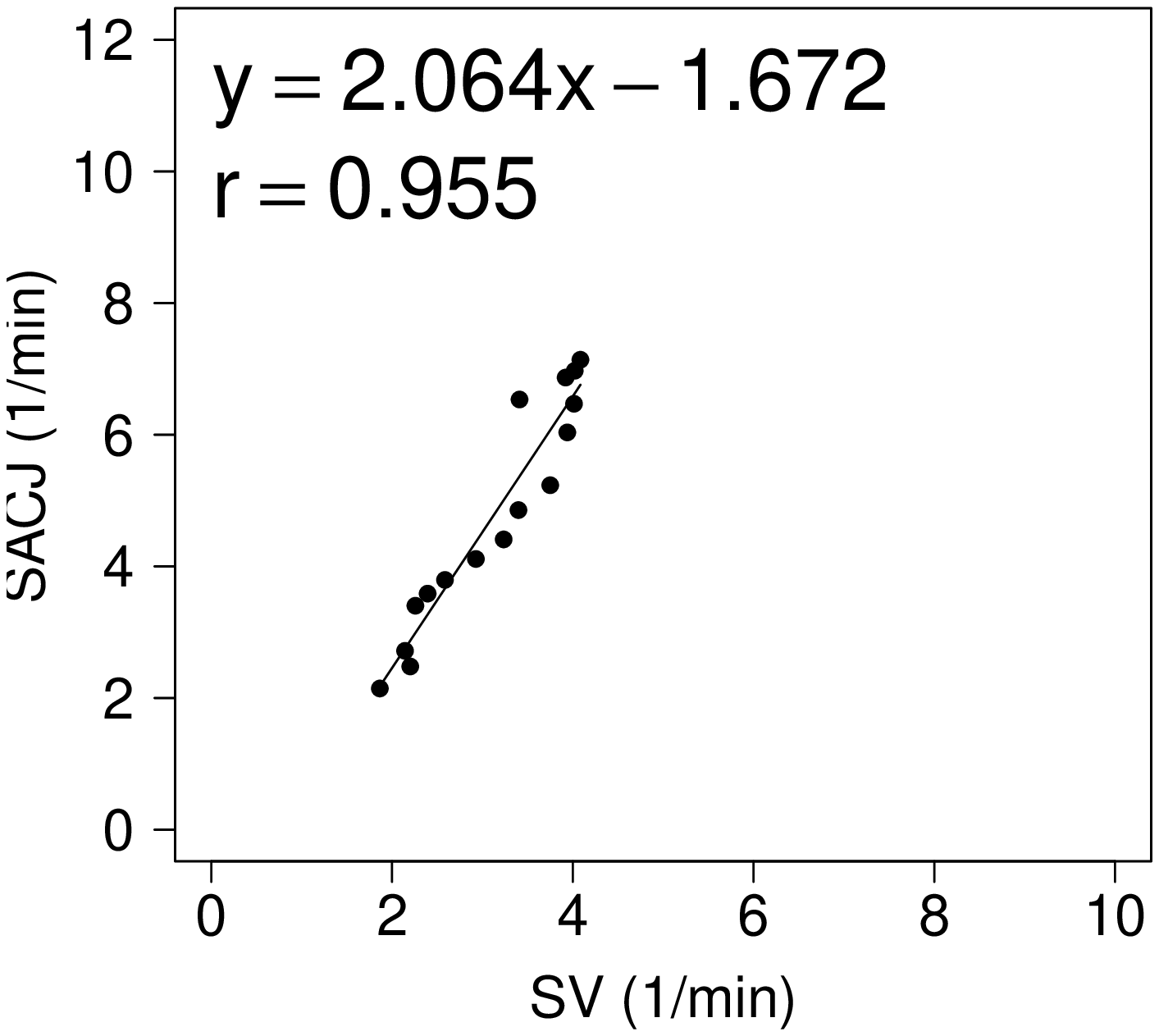}
    &
    \includegraphics[width=0.28\textwidth]{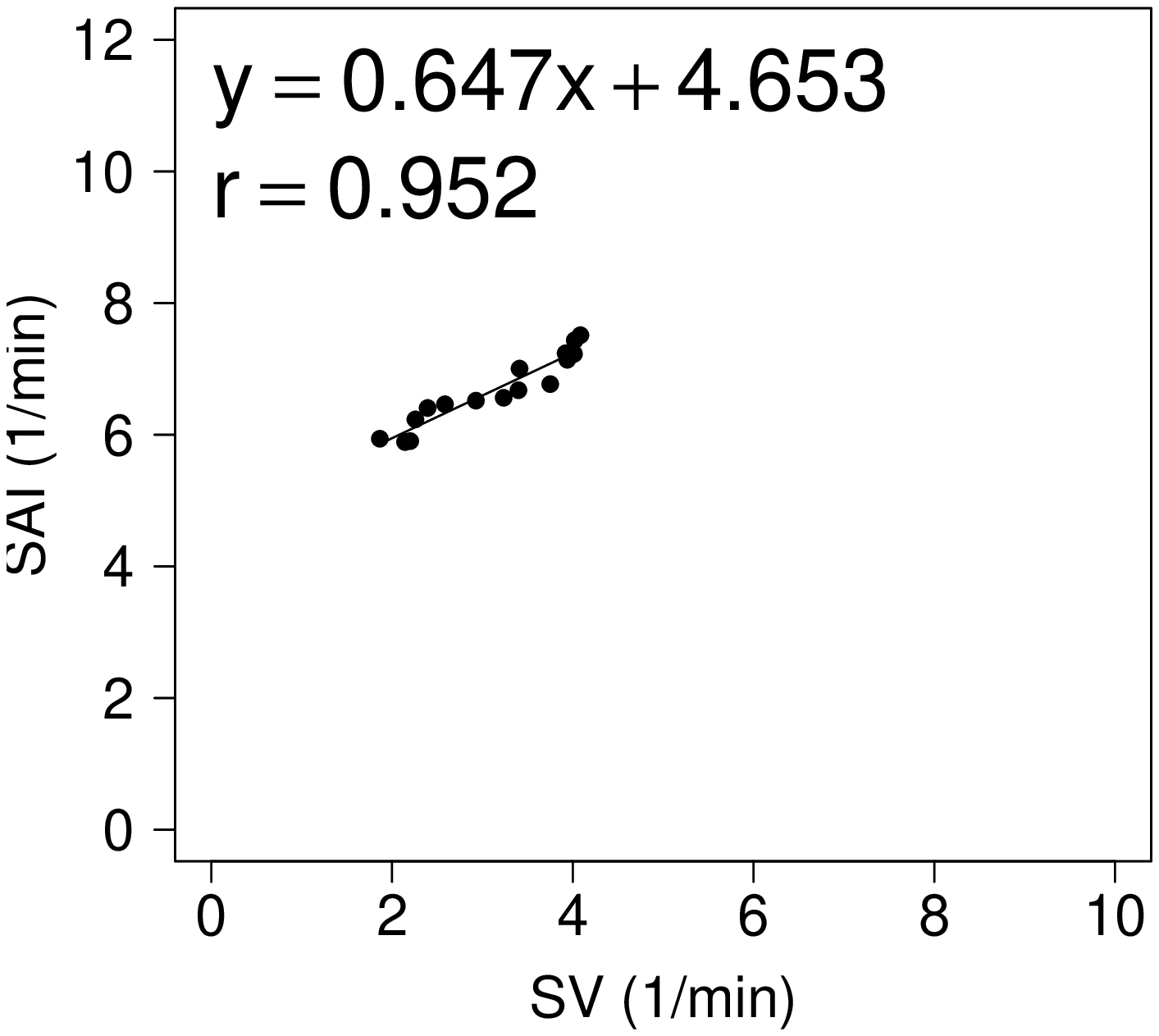}\\
    ~&Animal A&~\\
    \includegraphics[width=0.28\textwidth]{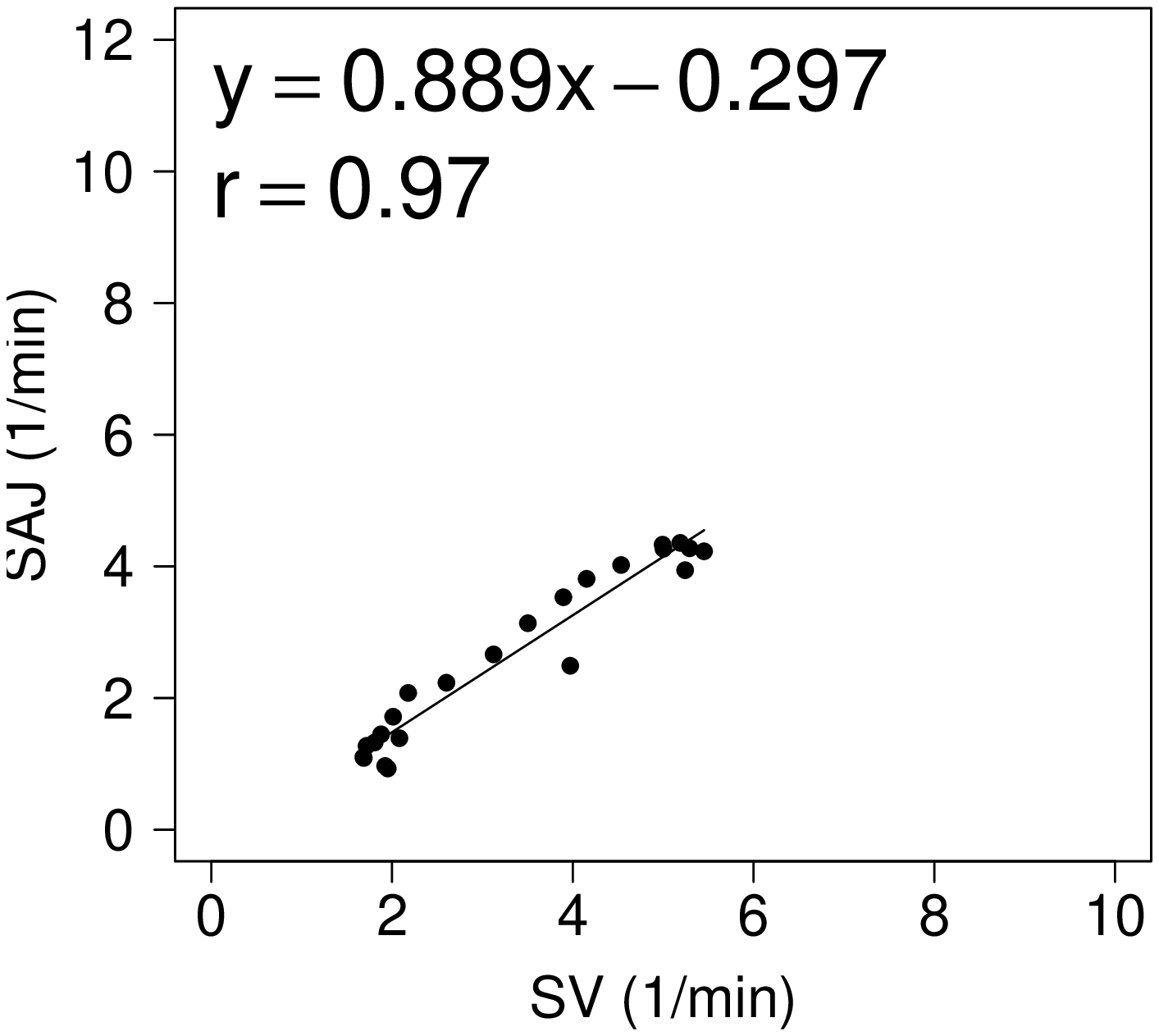}
    &
    \includegraphics[width=0.28\textwidth]{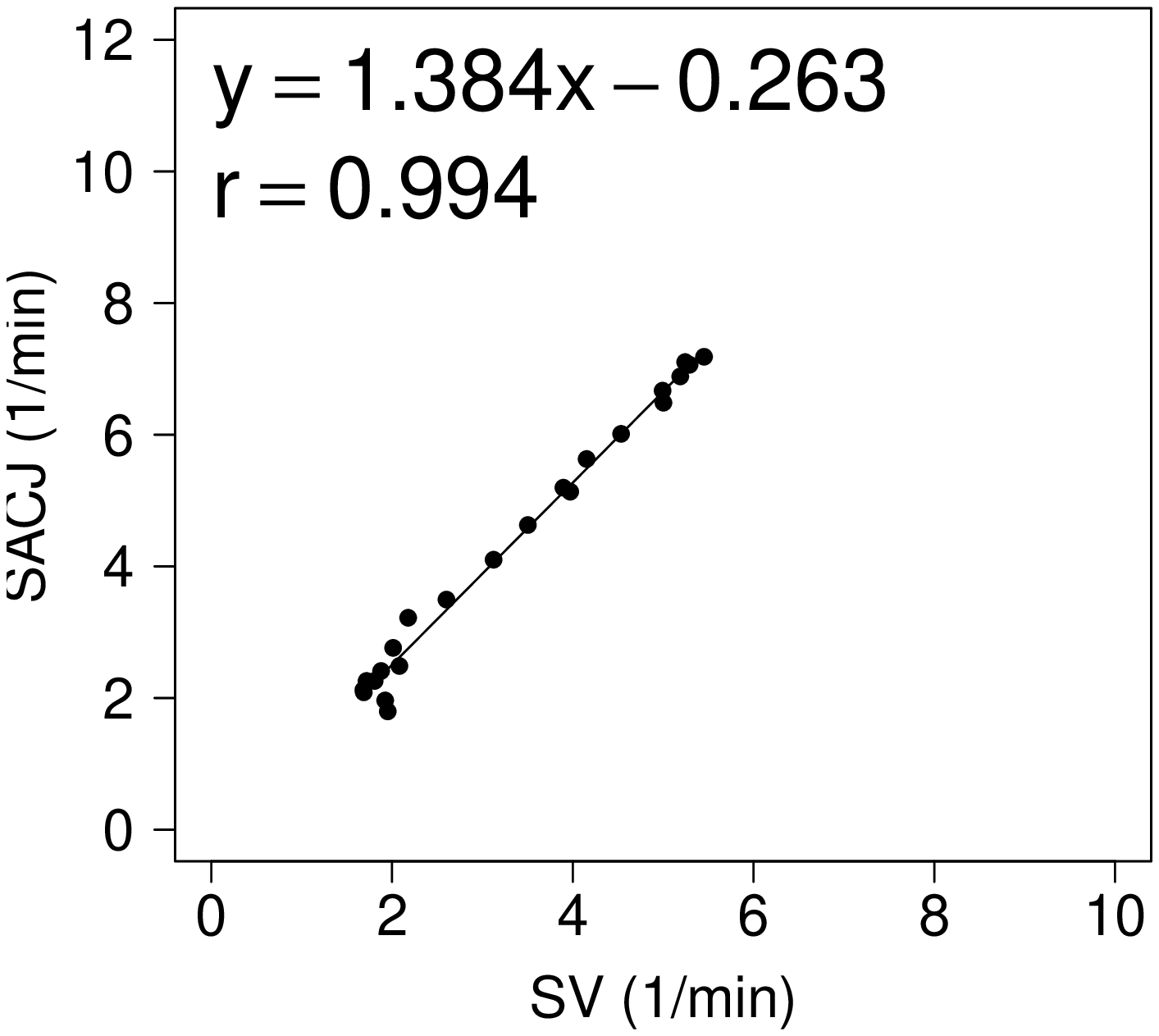}
    &
    \includegraphics[width=0.28\textwidth]{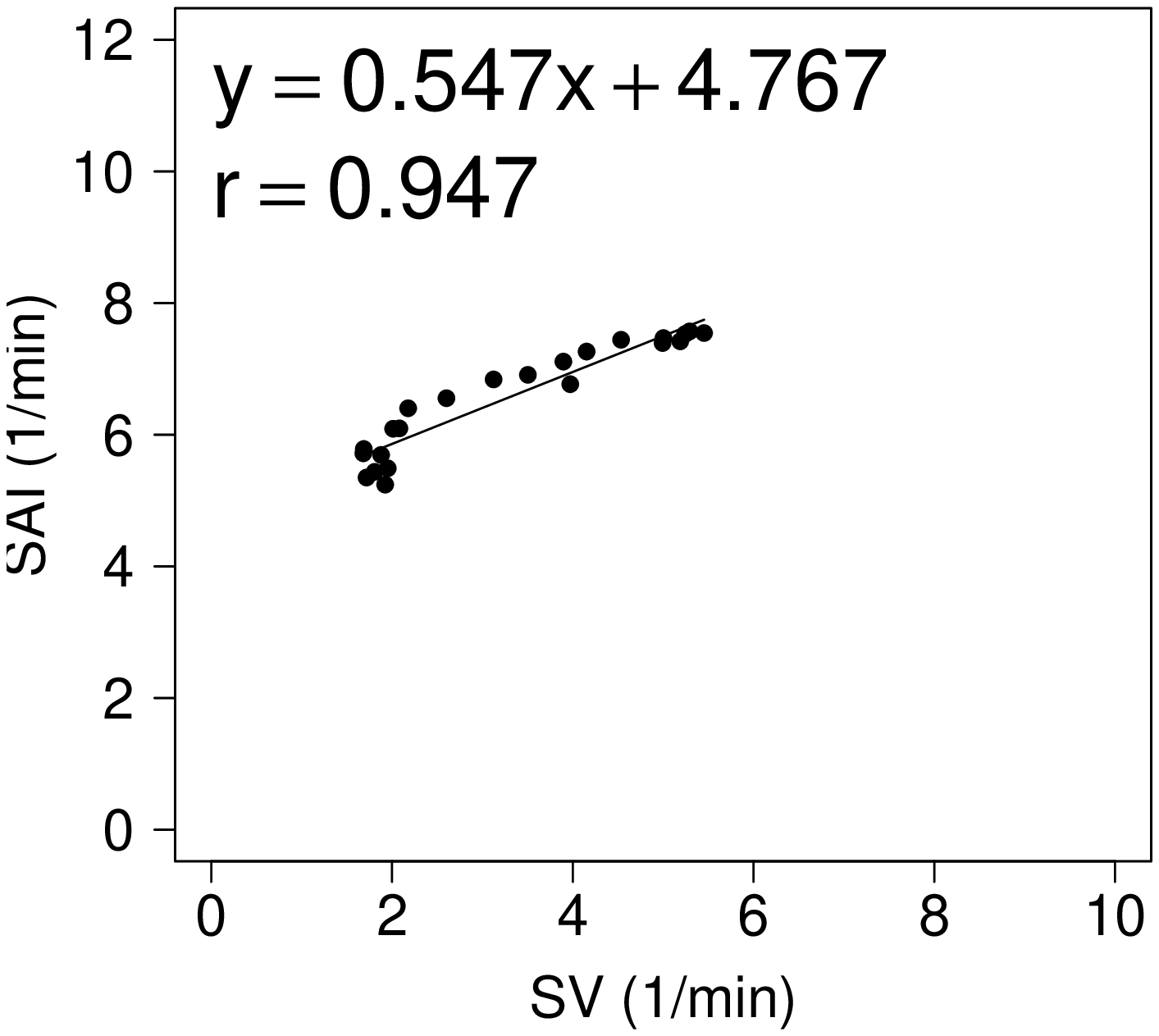}\\
    ~&Animal B&~\\
    \includegraphics[width=0.28\textwidth]{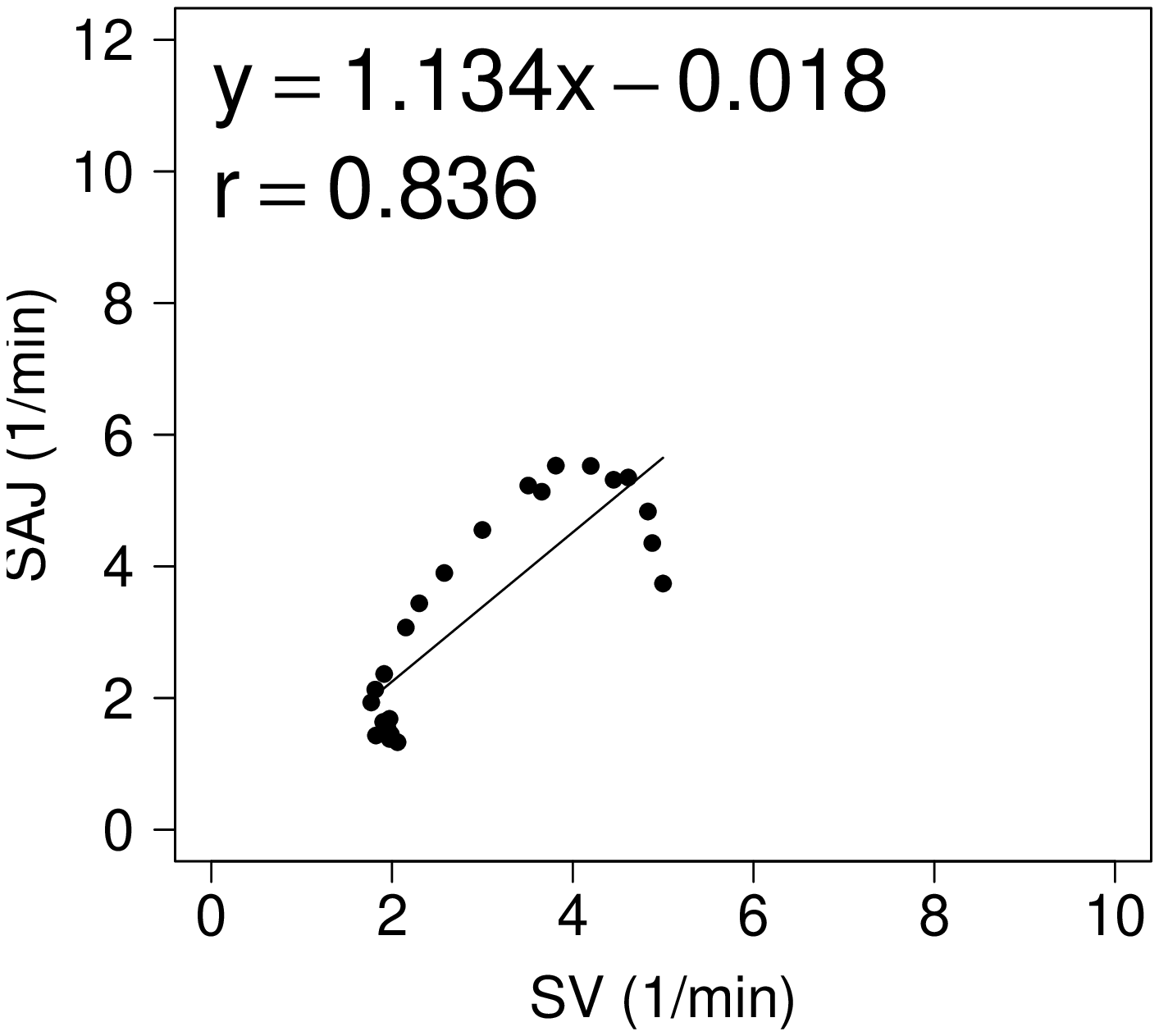}
    &
    \includegraphics[width=0.28\textwidth]{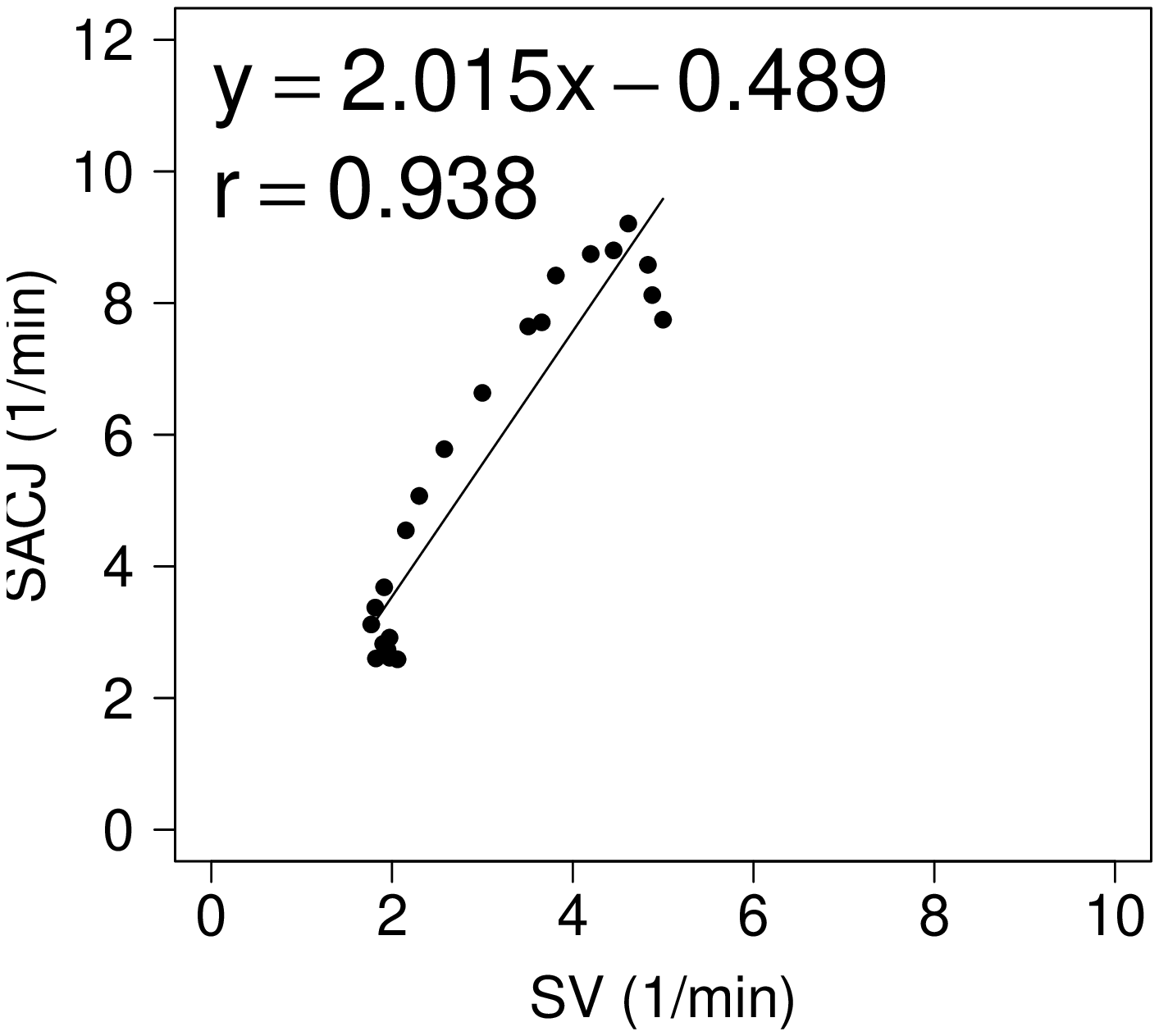}
    &
    \includegraphics[width=0.28\textwidth]{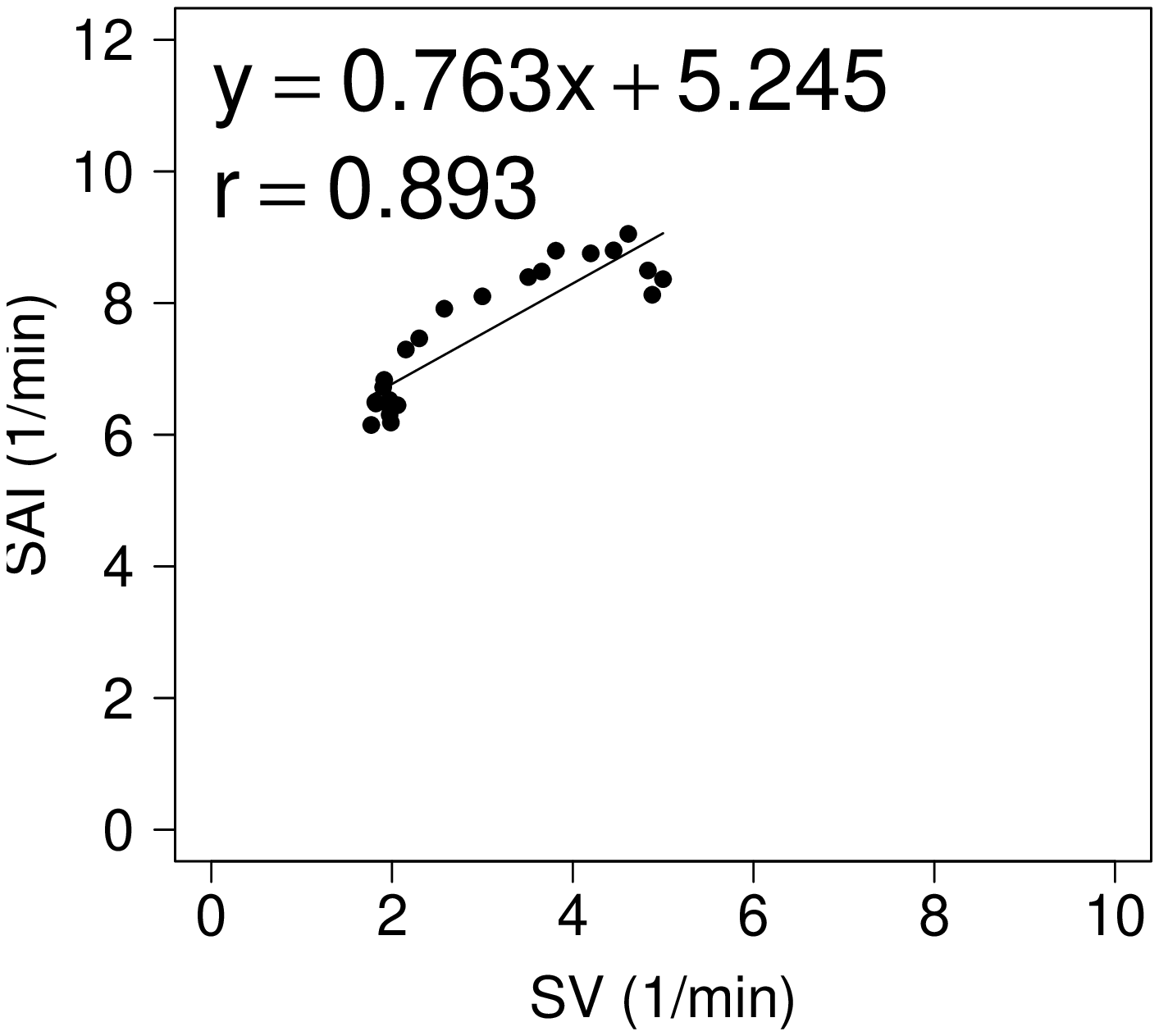}\\
    ~&Animal C&~\\
    \includegraphics[width=0.28\textwidth]{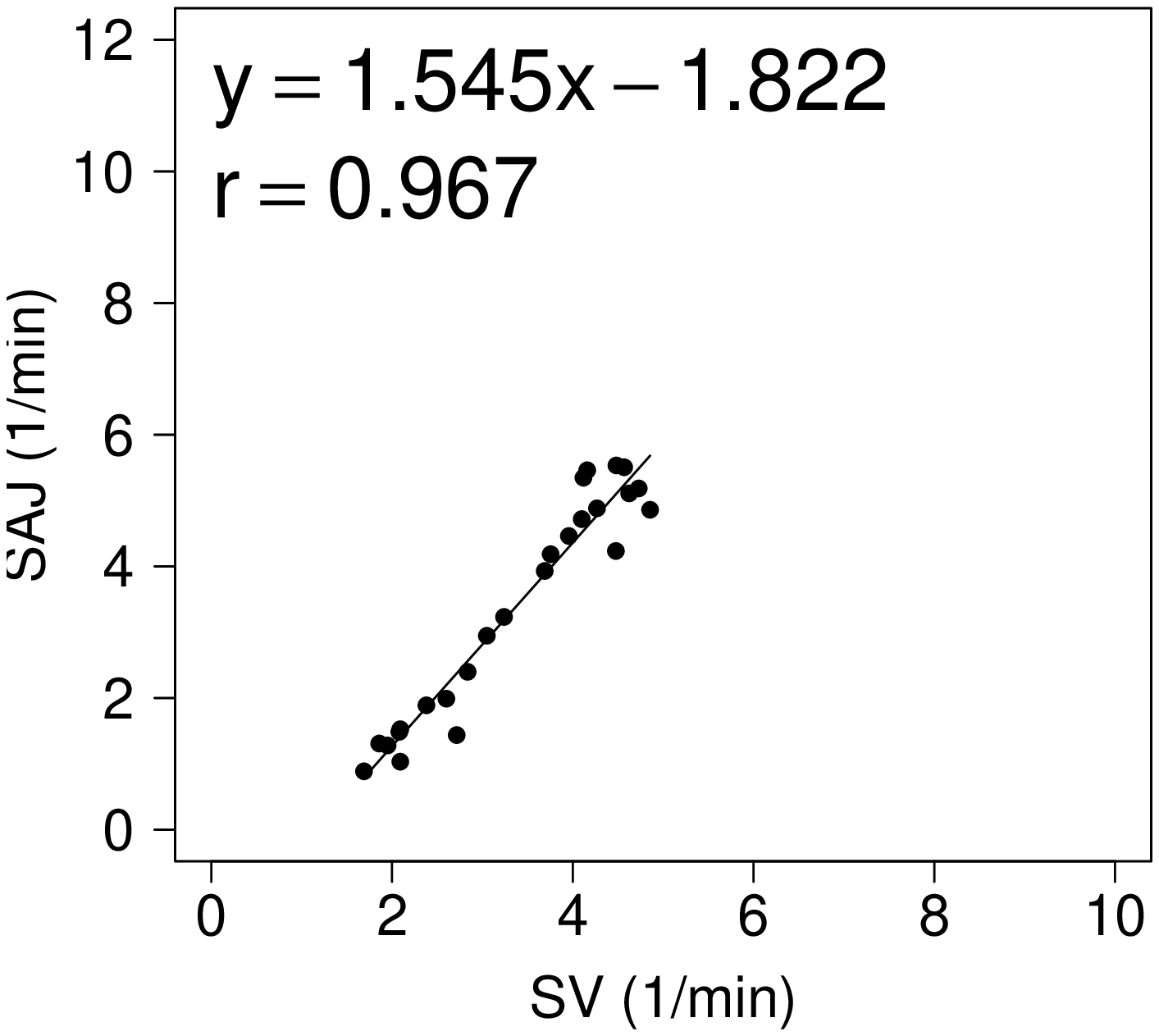}
    &
    \includegraphics[width=0.28\textwidth]{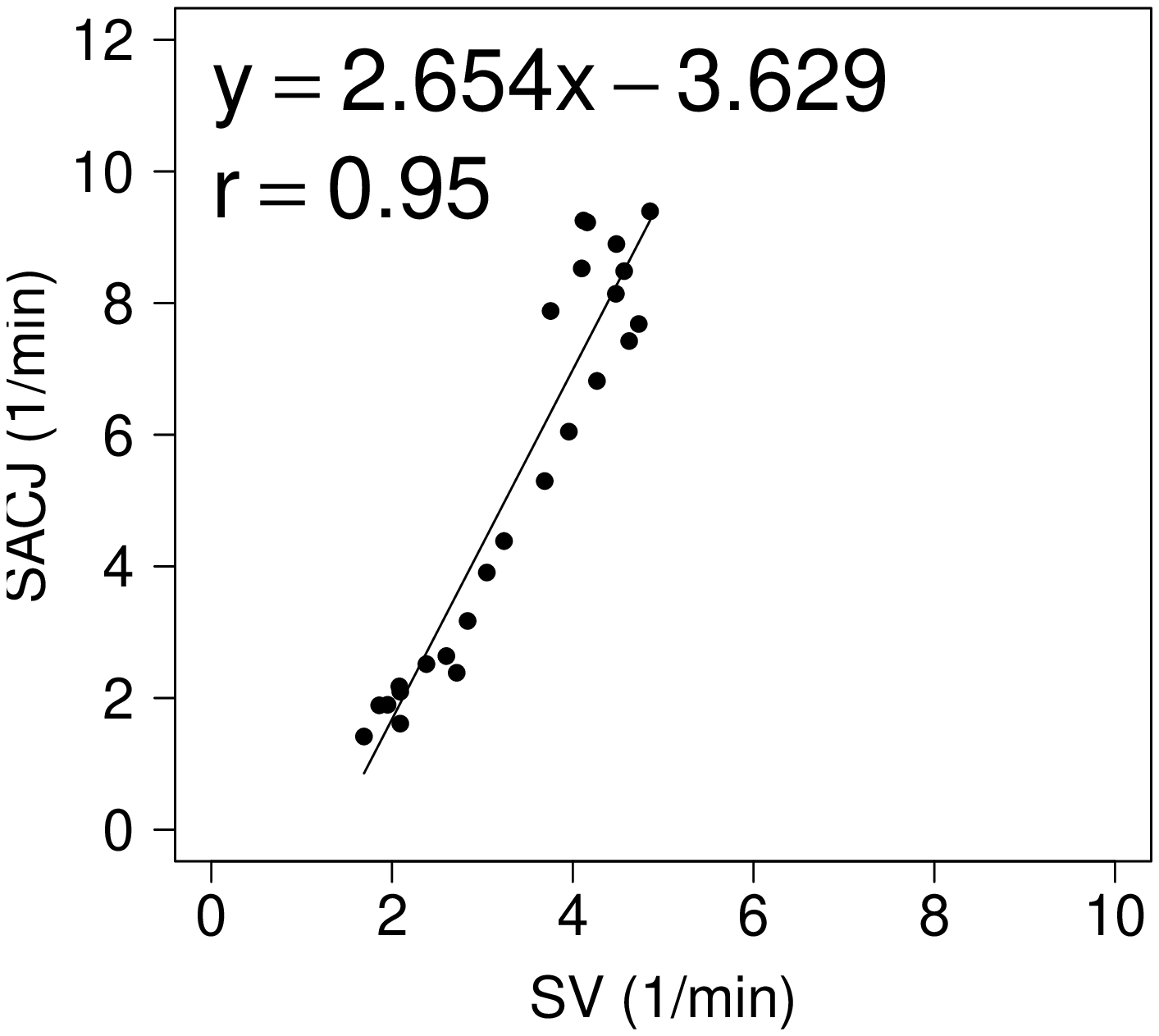}
    &
    \includegraphics[width=0.28\textwidth]{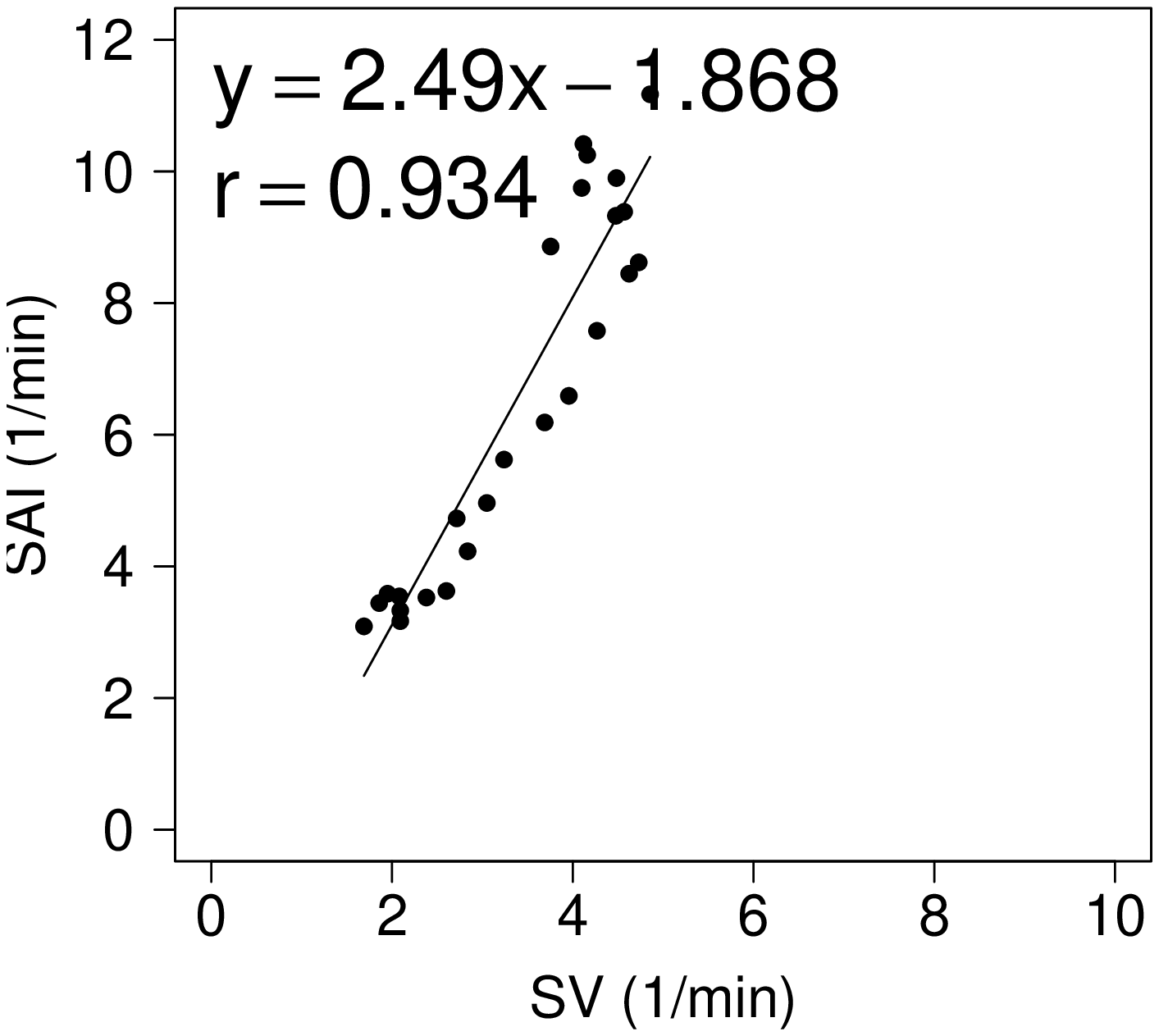}\\
    ~&Animal D&~\\
  \end{tabular}

  \caption{Large slab ROIs with size 150 mm $\times$ 8 mm $\times$ 40
    mm results for registration estimated ventilation measures
    compared to the Xe-CT estimated ventilation sV in scatter plot
    with linear regression in four animals. The first column is the
    SAJ vs.\ sV. The second column is the SACJ vs.\ sV. The third column
    is the SAI vs.\  sV.}
  \label{fig:regvent-xevent-slabroi}
\end{figure}

To analyze the effect of the size of the region of interest, the
corresponding region of Xe-CT image \eeinit{} in the \ee{} is
divided into about 30 slabs along the ventral-dorsal direction with
size of 150 mm $\times$ 8 mm $\times$ 40 mm as similarly in our
previous work~\cite{Reinhardt2008752,ding2009}.
Figure~\ref{fig:regvent-xevent-slabroi} shows the scatter plots
between the registration ventilation measures and the Xe-CT
ventilation sV similar as Fig.~\ref{fig:regvent-xevent-slabroi} but
in larger ROIs. The SACJ column shows the strongest correlation with
the sV (average $r^2=0.92$). Both the SAJ and SAI show good
correlation with sV as well (average $r^2=0.88$ and $r^2=0.87$).
However, though the average $r^2$ value still show the SACJ has the
highest correlation with Xe-CT based sV,
Table~\ref{tbl:pvalue-slabroi-sacj} and~\ref{tbl:pvalue-slabroi-saj}
show that with larger averaging
region as defined slabs, there is no significant difference between
the correlation coefficients from SACJ vs.\ sV and SAI vs.\ sV, between SAJ vs.\ sV and SAI vs.\ sV, or SACJ vs.\ sV and SAJ vs.\ sV as in Table~\ref{tbl:pvalue-slabroi-sacj_vs_saj}.

\begin{figure}[p]
  \centering
  \begin{tabular}{cc}
    \includegraphics[width=0.5\textwidth]{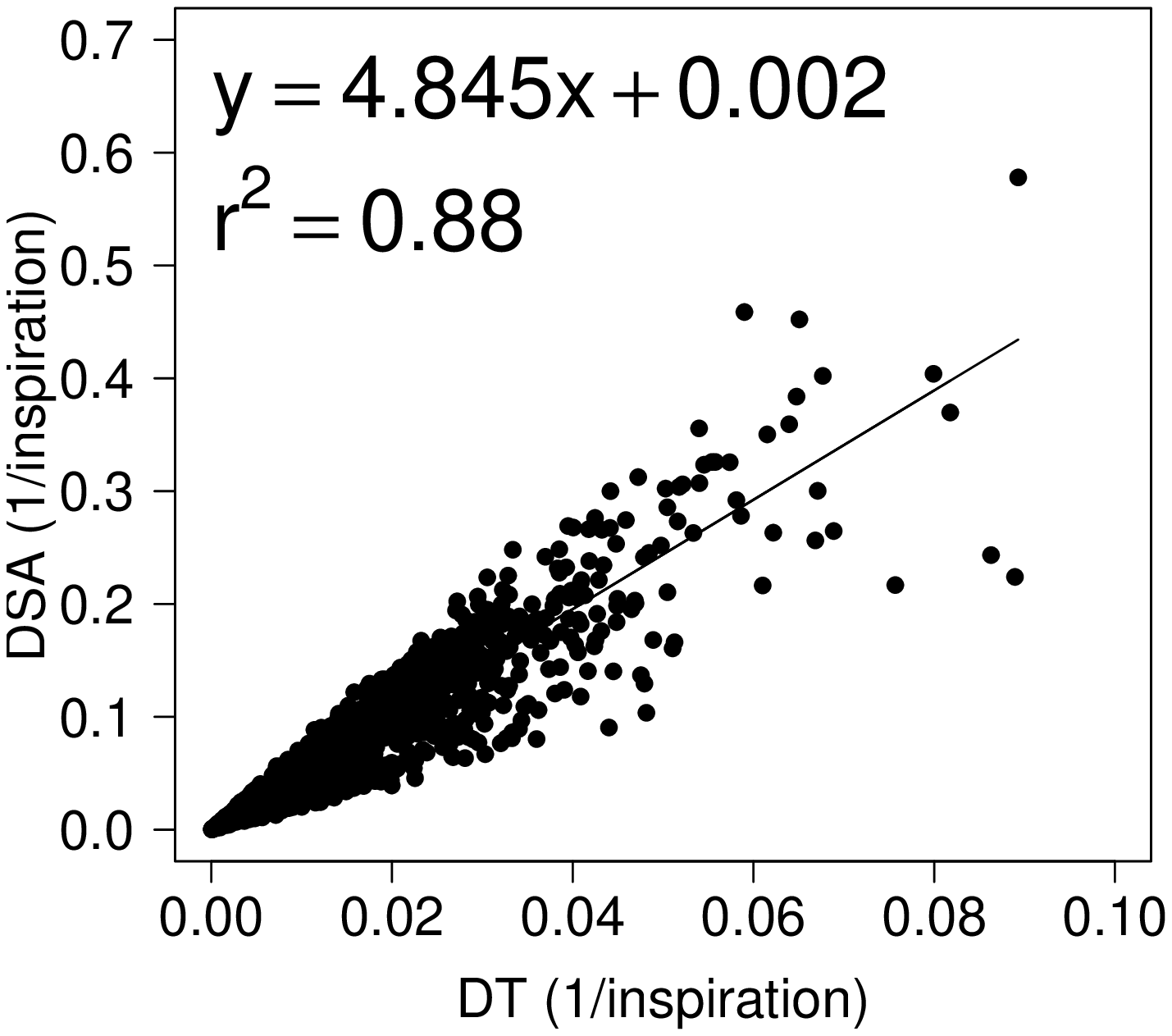}
    &
    \includegraphics[width=0.5\textwidth]{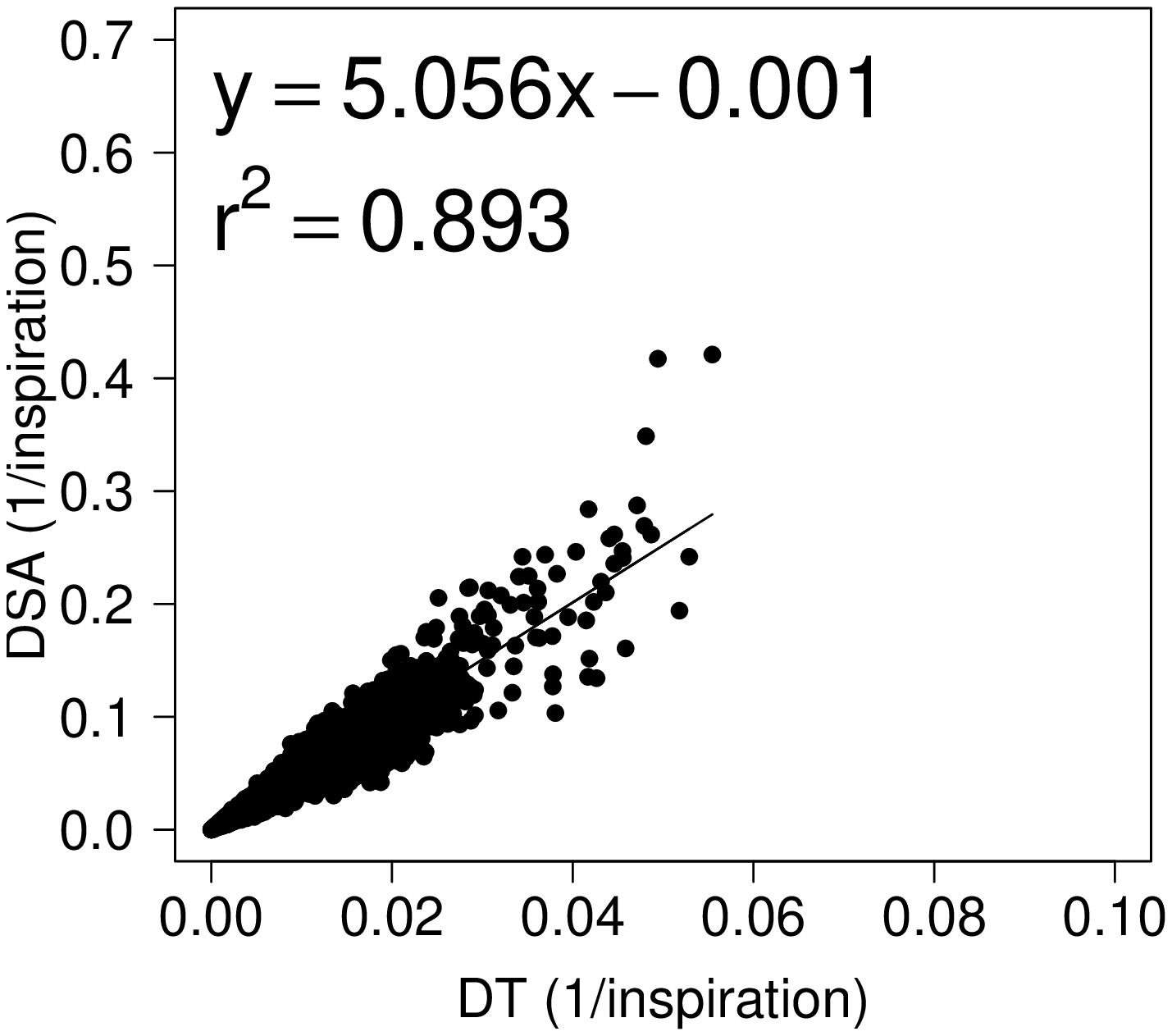}\\
    (a)&(b)\\
    \includegraphics[width=0.5\textwidth]{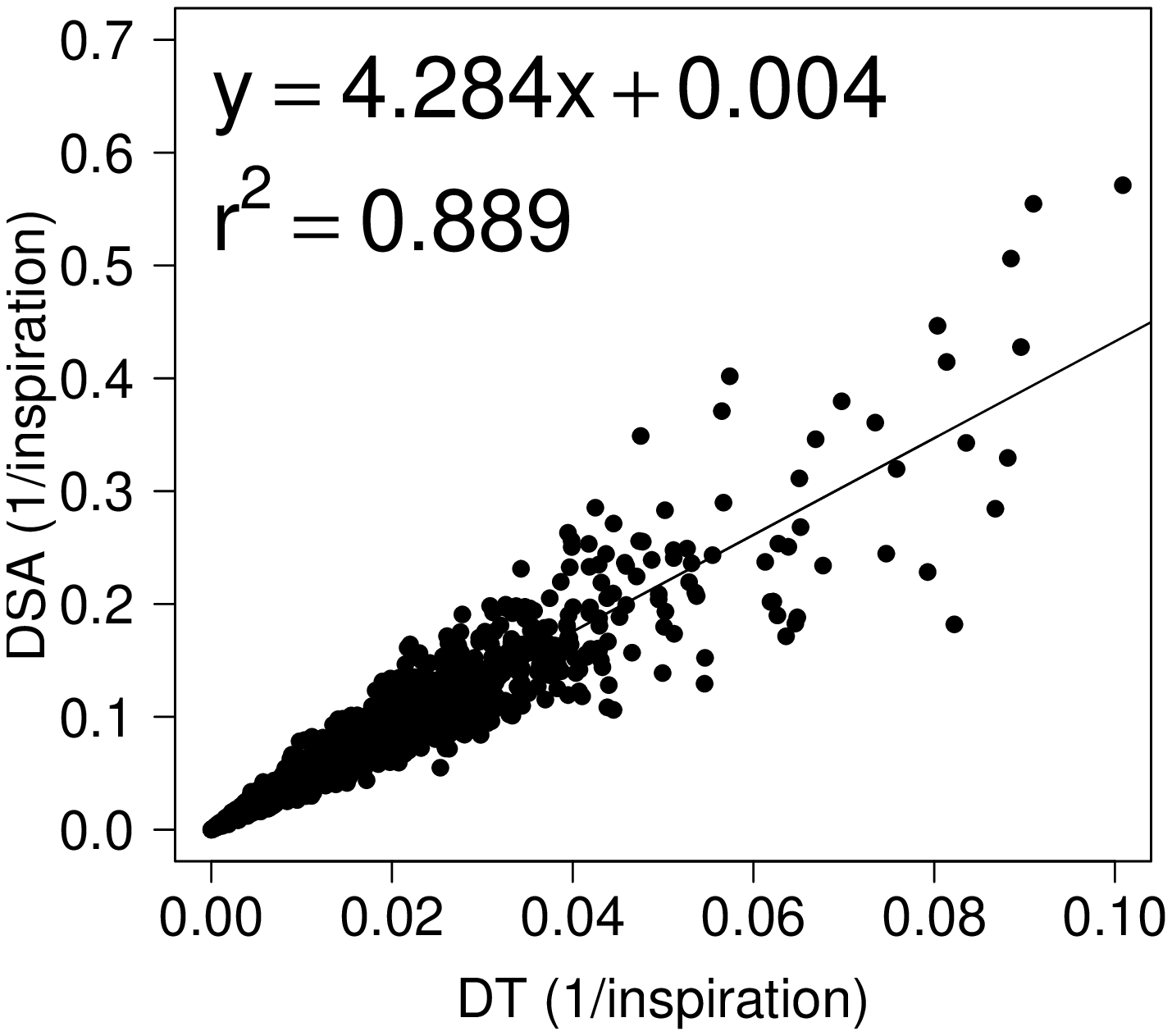}
    &
    \includegraphics[width=0.5\textwidth]{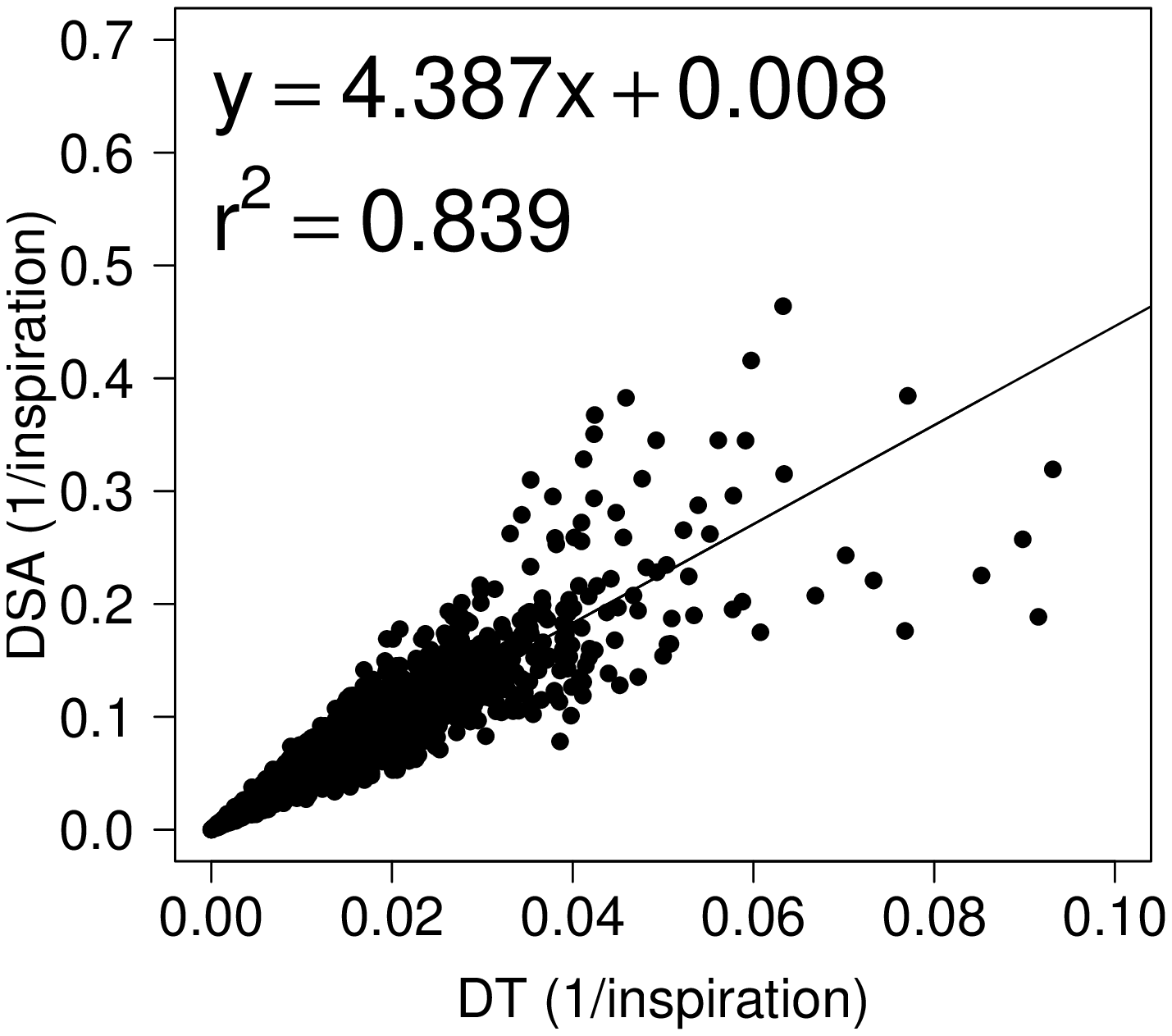}\\
    (c)&(d)\\
  \end{tabular}
  \caption{Linear regression analysis between DSA and DT. (a) to (d): DSA (the absolute difference of the value between the SACJ and SAI) compared to DT (the absolute difference of the tissue volume) in animals A, B, C and D.}
   \label{fig:sair-tissue}
\end{figure}

Figure~\ref{fig:sair-tissue} shows the scatter plots between DSA
(the absolute difference of the value between the SACJ and SAI) and
the DT (the absolute difference of the tissue volume) with linear
regression in all four animals (average $r^2=0.86$). From the
equation (\ref{eq:General_SACJ}) and (\ref{eq:General_SAI}), we know
that the SAI measurement assumes no tissue volume change in a given
region being registered, which may not be valid assumption in all
lung regions. Figure~\ref{fig:sair-tissue} shows that as the tissue
volume change increases, the difference between the measures of
regional ventilation from SACJ and SAI increases linearly in all
four animals. It indicates that the lower correlation of SAI with sV
compared with SACJ with sV may be caused by the tissue volume change
between two volumes.

\begin{table}
\centering

\begin{tabular}{ccccc}
\toprule
\multicolumn{ 1}{c}{Animal} & Correlation  & Correlation with sV & Number of  & \multicolumn{ 1}{c}{SACJ vs.\ SAI} \\

\multicolumn{ 1}{c}{} &  pair &  ($r$ value) &    samples & \multicolumn{ 1}{c}{$p$ value} \\
\midrule
\multicolumn{ 1}{c}{A} &      SACJ vs.\ sV &       0.88 & \multicolumn{ 1}{c}{83} & \multicolumn{ 1}{c}{$p<=0.0001$} \\

\multicolumn{ 1}{c}{} &        SAI vs.\ sV &       0.65 & \multicolumn{ 1}{c}{} & \multicolumn{ 1}{c}{} \\*[10pt]

\multicolumn{ 1}{c}{B} &      SACJ vs.\ sV &       0.93 & \multicolumn{ 1}{c}{119} & \multicolumn{ 1}{c}{$p<=1.18e^{-6}$} \\

\multicolumn{ 1}{c}{} &        SAI vs.\ sV &       0.77 & \multicolumn{ 1}{c}{} & \multicolumn{ 1}{c}{} \\*[10pt]

\multicolumn{ 1}{c}{C} &      SACJ vs.\ sV &       0.89 & \multicolumn{ 1}{c}{86} & \multicolumn{ 1}{c}{$p<=0.0075$} \\

\multicolumn{ 1}{c}{} &        SAI vs.\ sV &       0.78 & \multicolumn{ 1}{c}{} & \multicolumn{ 1}{c}{} \\*[10pt]

\multicolumn{ 1}{c}{D} &      SACJ vs.\ sV &       0.92 & \multicolumn{ 1}{c}{110} & \multicolumn{ 1}{c}{$p<=0.0017$} \\

\multicolumn{ 1}{c}{} &        SAI vs.\ sV &       0.83 & \multicolumn{ 1}{c}{} & \multicolumn{ 1}{c}{} \\
 \bottomrule
\end{tabular}
\caption{Comparison of ventilation measures between SACJ and
  SAI in small cube ROIs with size 20 mm $\times$ 20 mm
  $\times$ 20 mm.}
\label{tbl:pvalue-cuberoi-sacj}
\end{table}

\begin{table}
\centering

\begin{tabular}{ccccc}
\toprule
\multicolumn{ 1}{c}{Animal} & Correlation & Correlation with sV & Number of  & \multicolumn{ 1}{c}{SAJ vs.\ SAI} \\

\multicolumn{ 1}{c}{} &  pair &  ($r$ value) &    samples & \multicolumn{ 1}{c}{$p$ value} \\
\midrule
\multicolumn{ 1}{c}{A} &      SAJ vs.\ sV &       0.86 & \multicolumn{ 1}{c}{83} & \multicolumn{ 1}{c}{$p<=0.0005$} \\

\multicolumn{ 1}{c}{} &        SAI vs.\ sV &       0.65 & \multicolumn{ 1}{c}{} & \multicolumn{ 1}{c}{} \\*[10pt]

\multicolumn{ 1}{c}{B} &      SAJ vs.\ sV &       0.89 & \multicolumn{ 1}{c}{119} & \multicolumn{ 1}{c}{$p<=0.002$} \\

\multicolumn{ 1}{c}{} &        SAI vs.\ sV &       0.77 & \multicolumn{ 1}{c}{} & \multicolumn{ 1}{c}{} \\*[10pt]

\multicolumn{ 1}{c}{C} &      SAJ vs.\ sV &       0.78 & \multicolumn{ 1}{c}{86} & \multicolumn{ 1}{c}{$p<=0.5$} \\

\multicolumn{ 1}{c}{} &        SAI vs.\ sV &       0.78 & \multicolumn{ 1}{c}{} & \multicolumn{ 1}{c}{} \\*[10pt]

\multicolumn{ 1}{c}{D} &      SAJ vs.\ sV &       0.92 & \multicolumn{ 1}{c}{110} & \multicolumn{ 1}{c}{$p<=0.0017$} \\

\multicolumn{ 1}{c}{} &        SAI vs.\ sV &       0.83 & \multicolumn{ 1}{c}{} & \multicolumn{ 1}{c}{} \\
\bottomrule
\end{tabular}
\caption{Comparison of ventilation measures between SAJ and
SAI in small cube ROIs with size 20 mm $\times$ 20 mm
$\times$ 20 mm.}
\label{tbl:pvalue-cuberoi-saj}
\end{table}

\begin{table}
\centering
\begin{tabular}{ccccc}
\toprule
\multicolumn{ 1}{c}{Animal} & Correlation  & Correlation with sV & Number of  & \multicolumn{ 1}{c}{SACJ vs.\ SAJ} \\

\multicolumn{ 1}{c}{} &  pair &  ($r$ value) &    samples & \multicolumn{ 1}{c}{$p$ value} \\
\midrule
\multicolumn{ 1}{c}{A} &      SACJ vs.\ sV &       0.88 & \multicolumn{ 1}{c}{83} & \multicolumn{ 1}{c}{$p<=0.302$} \\

\multicolumn{ 1}{c}{} &        SAJ vs.\ sV &       0.86 & \multicolumn{ 1}{c}{} & \multicolumn{ 1}{c}{} \\*[10pt]

\multicolumn{ 1}{c}{B} &      SACJ vs.\ sV &       0.93 & \multicolumn{ 1}{c}{119} & \multicolumn{ 1}{c}{$p<=0.035$} \\

\multicolumn{ 1}{c}{} &        SAJ vs.\ sV &       0.89 & \multicolumn{ 1}{c}{} & \multicolumn{ 1}{c}{} \\*[10pt]

\multicolumn{ 1}{c}{C} &      SACJ vs.\ sV &       0.89 & \multicolumn{ 1}{c}{86} & \multicolumn{ 1}{c}{$p<=0.007$} \\

\multicolumn{ 1}{c}{} &        SAJ vs.\ sV &       0.78 & \multicolumn{ 1}{c}{} & \multicolumn{ 1}{c}{} \\*[10pt]

\multicolumn{ 1}{c}{D} &      SACJ vs.\ sV &       0.92 & \multicolumn{ 1}{c}{110} & \multicolumn{ 1}{c}{$p<=0.5$} \\

\multicolumn{ 1}{c}{} &        SAJ vs.\ sV &       0.92 & \multicolumn{ 1}{c}{} & \multicolumn{ 1}{c}{} \\
 \bottomrule
\end{tabular}
\caption{Comparison of ventilation measures between SACJ and SAJ in small cube ROIs with size 20 mm $\times$ 20 mm
  $\times$ 20 mm.}
\label{tbl:pvalue-cuberoi-sacj_vs_saj}
\end{table}

\begin{table}
\centering

\begin{tabular}{ccccc}
\toprule
\multicolumn{ 1}{c}{Animal} & Correlation & Correlation with sV & Number of  & \multicolumn{ 1}{c}{SACJ vs.\ SAI} \\

\multicolumn{ 1}{c}{} &  pair &  ($r$ value) &    samples & \multicolumn{ 1}{c}{$p$ value} \\
\midrule
\multicolumn{ 1}{c}{A} &      SACJ vs.\ sV &       0.95 & \multicolumn{ 1}{c}{17} & \multicolumn{ 1}{c}{$p<=0.5$} \\

\multicolumn{ 1}{c}{} &        SAI vs.\ sV &       0.95 & \multicolumn{ 1}{c}{} & \multicolumn{ 1}{c}{} \\*[10pt]

\multicolumn{ 1}{c}{B} &      SACJ vs.\ sV &       0.99 & \multicolumn{ 1}{c}{23} & \multicolumn{ 1}{c}{$p<=0.005$} \\

\multicolumn{ 1}{c}{} &        SAI vs.\ sV &       0.95 & \multicolumn{ 1}{c}{} & \multicolumn{ 1}{c}{} \\*[10pt]

\multicolumn{ 1}{c}{C} &      SACJ vs.\ sV &       0.94 & \multicolumn{ 1}{c}{23} & \multicolumn{ 1}{c}{$p<=0.15$} \\

\multicolumn{ 1}{c}{} &        SAI vs.\ sV &       0.89 & \multicolumn{ 1}{c}{} & \multicolumn{ 1}{c}{} \\*[10pt]

\multicolumn{ 1}{c}{D} &      SACJ vs.\ sV &       0.95 & \multicolumn{ 1}{c}{25} & \multicolumn{ 1}{c}{$p<=0.28$} \\

\multicolumn{ 1}{c}{} &        SAI vs.\ sV &       0.93 & \multicolumn{ 1}{c}{} & \multicolumn{ 1}{c}{} \\
\bottomrule
\end{tabular}
\caption{Comparison of ventilation measures between SACJ and
SAI in large slab ROIs with size 150 mm $\times$ 8 mm
$\times$ 40 mm.}
\label{tbl:pvalue-slabroi-sacj}
\end{table}

\begin{table}
\centering

\begin{tabular}{ccccc}
\toprule
\multicolumn{ 1}{c}{Animal} & Correlation  & Correlation with sV & Number of  & \multicolumn{ 1}{c}{SAJ vs.\ SAI} \\

\multicolumn{ 1}{c}{} &  pair &  ($r$ value) &    samples & \multicolumn{ 1}{c}{$p$ value} \\
\midrule
\multicolumn{ 1}{c}{A} &      SAJ vs.\ sV  &       0.95 & \multicolumn{ 1}{c}{17} & \multicolumn{ 1}{c}{$p<=0.5$} \\

\multicolumn{ 1}{c}{} &        SAI vs.\ sV &       0.95 & \multicolumn{ 1}{c}{} & \multicolumn{ 1}{c}{} \\*[10pt]

\multicolumn{ 1}{c}{B} &      SAJ  vs.\ sV &       0.99 & \multicolumn{ 1}{c}{23} & \multicolumn{ 1}{c}{$p<=0.005$} \\

\multicolumn{ 1}{c}{} &        SAI vs.\ sV &       0.95 & \multicolumn{ 1}{c}{} & \multicolumn{ 1}{c}{} \\*[10pt]

\multicolumn{ 1}{c}{C} &      SAJ  vs.\ sV &       0.94 & \multicolumn{ 1}{c}{23} & \multicolumn{ 1}{c}{$p<=0.16$} \\

\multicolumn{ 1}{c}{} &        SAI vs.\ sV &       0.89 & \multicolumn{ 1}{c}{} & \multicolumn{ 1}{c}{} \\*[10pt]

\multicolumn{ 1}{c}{D} &      SAJ  vs.\ sV &       0.95 & \multicolumn{ 1}{c}{25} & \multicolumn{ 1}{c}{$p<=0.28$} \\

\multicolumn{ 1}{c}{} &        SAI vs.\ sV &       0.93 & \multicolumn{ 1}{c}{} & \multicolumn{ 1}{c}{} \\
\bottomrule
\end{tabular}
\caption{Comparison of ventilation measures between SAJ and
SAI in large slab ROIs with size 150 mm $\times$ 8 mm
$\times$ 40 mm.}
\label{tbl:pvalue-slabroi-saj}
\end{table}

\begin{table}
\centering

\begin{tabular}{ccccc}
\toprule
\multicolumn{ 1}{c}{Animal} & Correlation & Correlation with sV & Number of  & \multicolumn{ 1}{c}{SACJ vs.\ SAI} \\

\multicolumn{ 1}{c}{} &  pair &  ($r$ value) &    samples & \multicolumn{ 1}{c}{$p$ value} \\
\midrule
\multicolumn{ 1}{c}{A} &      SACJ vs.\ sV &       0.95 & \multicolumn{ 1}{c}{17} & \multicolumn{ 1}{c}{$p<=0.5$} \\

\multicolumn{ 1}{c}{} &        SAJ vs.\ sV &       0.95 & \multicolumn{ 1}{c}{} & \multicolumn{ 1}{c}{} \\*[10pt]

\multicolumn{ 1}{c}{B} &      SACJ vs.\ sV &       0.99 & \multicolumn{ 1}{c}{23} & \multicolumn{ 1}{c}{$p<=0.5$} \\

\multicolumn{ 1}{c}{} &        SAJ vs.\ sV &       0.99 & \multicolumn{ 1}{c}{} & \multicolumn{ 1}{c}{} \\*[10pt]

\multicolumn{ 1}{c}{C} &      SACJ vs.\ sV &       0.94 & \multicolumn{ 1}{c}{23} & \multicolumn{ 1}{c}{$p<=0.5$} \\

\multicolumn{ 1}{c}{} &        SAJ vs.\ sV &       0.94 & \multicolumn{ 1}{c}{} & \multicolumn{ 1}{c}{} \\*[10pt]

\multicolumn{ 1}{c}{D} &      SACJ vs.\ sV &       0.95 & \multicolumn{ 1}{c}{25} & \multicolumn{ 1}{c}{$p<=0.5$} \\

\multicolumn{ 1}{c}{} &        SAJ vs.\ sV &       0.95 & \multicolumn{ 1}{c}{} & \multicolumn{ 1}{c}{} \\
\bottomrule
\end{tabular}
\caption{Comparison of ventilation measures between SACJ and
SAJ in large slab ROIs with size 150 mm $\times$ 8 mm
$\times$ 40 mm.}
\label{tbl:pvalue-slabroi-sacj_vs_saj}
\end{table}

\section{Summary and Conclusions}
\label{sec:discussion}   We described three measures to estimate
regional lung tissue ventilation from tissue volume and vesselness
preserving image registration of CT images.  These measures have
been compared with each other, and compared to Xe-CT estimates of
specific ventilation.  We examined the assumption of constant tissue
volume between registered lung regions, and demonstrated that the
difference between two of the registration derived measures (SACJ
and SAI) may be explained by differences in tissue volume between
the lung regions being compared with registration.

The tissue volume and vesselness preserving non-linear registration
algorithm was used to match the \ei{} image to the \ee{} image to
produce the registration deformation field and estimates of regional
ventilation. It was used to register the \eeinit{} image to the
\ee{} image for comparing the three ventilation measures to the
Xe-CT based sV. About 200 anatomical landmarks were identified and
annotated in each data set to evaluate registration accuracy. The
average landmark error is on the order of 1 mm after registration.

The ventilation measures SAJ, SACJ, and SAI were derived using a
simple model of a lung region, containing a mixture of air and
tissue, which deforms during inspiration or exhalation.   The SAJ
measure, which is a linear function of the Jacobian of the
registration displacement field, measures regional ventilation based
on the assumption that the lung region contains only air (i.e., no
tissue volume).   The SACJ is the most general form of the
three measures and is based on model where both the air and tissue
volumes can change during inspiration and exhalation. Finally, the
SAI measure is computed based on intensity change alone, and assumes
that the region may have a tissue volume that is non-zero, but this
volume does not change during inspiration and exhalation.
Thus, the SAJ measure relies solely on the volume change information
computed from the Jacobian of the deformation field and the SAI
measure relies solely on the change in region intensity as measured
by the CT\@. The SACJ measure uses a more complete model of local
ventilation, and combines the geometric information from the
Jacobian with the density information calculated from the change in
region intensity.

The three registration-based ventilation measures and the Xe-CT sV
measurement were averaged and compared in cubic-shaped regions of
interest. In 20 mm $\times$ 20 mm $\times$ 20 mm ROIs, the SACJ
shows significantly higher correlation with Xe-CT sV than the
SAI in all four animals. By studying the difference between
the SACJ and SAI measures and the tissue volume difference
estimated by the CT intensity change, we showed that the difference
between SACJ and SAI may be explained by the constant tissue volume
assumption implicit in the SAI model~(\ref{eq:SAI-assumption}).
From Fig.~\ref{fig:sair-tissue}, we see that the difference between
the SACJ and SAI measures is approximately linearly-related to the
estimated tissue volume change.
Tables~\ref{tbl:pvalue-cuberoi-sacj} and~\ref{tbl:pvalue-cuberoi-saj}
show that the both the SACJ and
SAJ have significantly better correlation with sV than the SAI\@.
This is consistent with the findings by Kabus et
al.~\cite{kabus2008workshop} who showed that the Jacobian-based
measure of ventilation has less error than the intensity-based ventilation
measure, using the segmented total lung volume as a global
comparison. Though all the regional ventilation measures and Xe-CT
based sV from the linear regression analysis in
Fig.~\ref{fig:regvent-xevent-slabroi} show high correlations,
Tables~\ref{tbl:pvalue-slabroi-sacj} and~\ref{tbl:pvalue-slabroi-saj}
show that there is no significant
difference in the correlation with sV between the Jacobian-based
measures and intensity-based measure. This result indicates that the
validation methods using global comparison such as segmented total
volume may not be able to distinguish the Jacobian-based measure and
the intensity-based measure.

The comparison of the ventilation measures was limited to the
resolution of 20 mm $\times$ 20 mm $\times$ ROIs. As the size of the
ROIs decreases, the correlation between the ventilation measures
with Xe-CT based sV decreases.  This may be due to the
underlying noise of the Xe-CT measurement of ventilation or the
decreased sensitivity of registration based measure to local
ventilation heterogeneity which is relative to the case.
Additional Xe-CT image analysis work including using
multi-compartment models, thinner slice, and inter-phase
registration to improve sV measurement are required to reduce the
noise in Xe-CT based sV measurement.

To compare with the intensity-based ventilation measure used in
previous work in Simon~\cite{Simon2000}, Guerrero et
al.~\cite{Guerrero2005630}, and Fuld et al.~\cite{Fuld04012008}, we
followed the assumption that $HU_{air}$ is -1000 HU and
$HU_{tissue}$ is 0 HU (equaling water~\cite{Hoffman1985}) in this
work. The ventilation measures were calculated under the assumption
that $HU_{air}$ is -1000 HU and $HU_{tissue}$ is 55 HU, which are
the values used by Yin et al.\ in~\cite{yin:4213}. Our analysis shows
that the correlation coefficients between any two estimates (SAJ-sV,
SACJ-sV or SAI-sV) change less than $1\%$ with two different
$HU_{tissue}$ values. However, it would be important to have
sensitivity analysis in the future to compare different ventilation
measures against intensity changes.

The image registration algorithm used to find the transformation
from \ei{} to \ee{} for measurement of regional ventilation produces
accurate registrations by minimizing the tissue volume and
vesselness measure difference between the template image and the
target image. It would be interesting to compare different image
registration algorithms and their effects on the registration-based
ventilation measures. For example, if two registration algorithms
achieve the similar landmark accuracy, the one does not preserve
tissue volume change may show even larger difference in the SACJ and
SAI measures than the results using TVP as described above.

In conclusion, with the same deformation field by the same image
registration algorithm, a significant difference between the
Jacobian/-corrected Jacobian-based ventilation measures and the
intensity-based ventilation measure is found in a regional level
using Xe-CT based ventilation measure sV. The ventilation measure by
corrected Jacobian SACJ gives best correlation with Xe-CT based sV
and the correlation is significantly higher than from the
ventilation by intensity SAI indicating the ventilation measure by
corrected Jacobian SACJ may be a better measure of regional lung
ventilation from image registration of 4DCT images.

\section{Acknowledgments}
The authors would like to thank Ms.\ K.\ Murphy and Dr.\ B. van
Ginneken for providing the software iX for generating and annotating
landmarks. This work was supported in part by NIH grant HL079406 and
EB004126.

\vspace{2in}


\begin{thebibliography}{10}
\bibitem{Keall2002}
P.~J. Keall, V.~R. Kini, S.~S. Vedam, and R.~Mohan, ``Potential
radiotherapy
  improvements with respiratory gating,'' \emph{Australasian physical \&
  engineering sciences in medicine}, vol.~25, no.~1, pp. 1 -- 6, 2002.

\bibitem{low:1254}
D.~A. Low, M.~Nystrom, E.~Kalinin, P.~Parikh, J.~F. Dempsey, J.~D.
Bradley,
  S.~Mutic, S.~H. Wahab, T.~Islam, G.~Christensen, D.~G. Politte, and B.~R.
  Whiting, ``A method for the reconstruction of four-dimensional synchronized
  {CT} scans acquired during free breathing,'' \emph{Medical Physics}, vol.~30,
  no.~6, pp. 1254--1263, 2003.

\bibitem{pan:627}
T.~Pan, ``Comparison of helical and cine acquisitions for {4D-CT}
imaging with
  multislice {CT},'' \emph{Medical Physics}, vol.~32, no.~2, pp. 627--634,
  2005.

\bibitem{Reinhardt2008752}
J.~M. Reinhardt, K.~Ding, K.~Cao, G.~E. Christensen, E.~A. Hoffman,
and S.~V.
  Bodas, ``Registration-based estimates of local lung tissue expansion compared
  to xenon {CT} measures of specific ventilation,'' \emph{Medical Image
  Analysis}, vol.~12, no.~6, pp. 752 -- 763, 2008, special issue on information
  processing in medical imaging 2007.

\bibitem{ding2008b}
K.~Ding, K.~Cao, G.~E. Christensen, M.~L. Raghavan, E.~A. Hoffman,
and J.~M.
  Reinhardt, ``{R}egistration-based lung tissue mechanics assessment during
  tidal breathing,'' in \emph{First International Workshop on Pulmonary Image
  Analysis}, M.~Brown, M.~de~Bruijne, B.~van Ginneken, A.~Kiraly, J.-M.
  Kuhnigk, C.~Lorenz, K.~Mori, and J.~M. Reinhardt, Eds., New York, 2008,
  p.~63.

\bibitem{ding2009}
K.~Ding, K.~Cao, G.~E. Christensen, E.~A. Hoffman, and J.~M.
Reinhardt,
  ``Registration-based regional lung mechanical analysis: {R}etrospectively
  reconstructed dynamic imaging versus static breath-hold image acquisition,''
  X.~P. Hu and A.~V. Clough, Eds., vol. 7262, no.~1.\hskip 1em plus 0.5em minus
  0.4em\relax SPIE, 2009, p. 72620D.

\bibitem{ding2009b}
K.~Ding, Y.~Yin, K.~Cao, G.~E. Christensen, C.-L. Lin, E.~A.
Hoffman, and J.~M.
  Reinhardt, ``Evaluation of lobar biomechanics during respiration using image
  registration,'' in \emph{Proc.\ of International Conference on Medical Image
  Computing and Computer-Assisted Intervention 2009}, vol. 5761, 2009, pp.
  739--746.

\bibitem{Yaremko2007562}
B.~P. Yaremko, T.~M. Guerrero, J.~Noyola-Martinez, R.~Guerra, D.~G.
Lege, L.~T.
  Nguyen, P.~A. Balter, J.~D. Cox, and R.~Komaki, ``Reduction of normal lung
  irradiation in locally advanced non-small-cell lung cancer patients, using
  ventilation images for functional avoidance,'' \emph{International Journal of
  Radiation Oncology*Biology*Physics}, vol.~68, no.~2, pp. 562 -- 571, 2007.
  
\bibitem{Yamamoto2010}
T.~Yamamoto, S.~Kabus, J.~von Berg, C.~Lorenz, and P.~J. Keall, ``Impact of
  four-dimensional computed tomography pulmonary ventilation imaging-based
  functional avoidance for lung cancer radiotherapy,'' {\em Int. J. Radiation
  Oncology Biol. Phys.}, vol.~2, pp.~1--10, 2010.
  
\bibitem{ding:1261}
K.~Ding, J.~E. Bayouth, J.~M. Buatti, G.~E. Christensen, and J.~M.
Reinhardt,
  ``4dct-based measurement of changes in pulmonary function following a course
  of radiation therapy,'' \emph{Medical Physics}, vol.~37, no.~3, pp.
  1261--1272, 2010.

\bibitem{Guerrero2005630}
T.~Guerrero, K.~Sanders, J.~Noyola-Martinez, E.~Castillo, Y.~Zhang,
R.~Tapia,
  R.~Guerra, Y.~Borghero, and R.~Komaki, ``Quantification of regional
  ventilation from treatment planning {CT},'' \emph{International Journal of
  Radiation Oncology*Biology*Physics}, vol.~62, no.~3, pp. 630 -- 634, 2005.

\bibitem{guerrero2006a}
T.~Guerrero, K.~Sanders, E.~Castillo, Y.~Zhang, L.~Bidaut, and
T.~P.~R. Komaki,
  ``Dynamic ventilation imaging from four-dimensional computed tomography,''
  \emph{Phys Med Biol.}, vol.~51, no.~4, pp. 777--791, Feb. 21 2006.

\bibitem{christensen2007a}
G.~E. Christensen, J.~H. Song, W.~Lu, I.~E. Naqa, and D.~A. Low,
``Tracking
  lung tissue motion and expansion/compression with inverse consistent image
  registration and spirometry.'' \emph{Med Physics}, vol.~34, no.~6, pp.
  2155--2165, June 2007.

\bibitem{Castillo2010}
R.~Castillo, E.~Castillo, J.~Martinez, and T.~Guerrero, ``Ventilation from
  four-dimensional computed tomography: density versus jacobian methods,'' {\em
  Physics in Medicine and Biology}, vol.~55(16), pp.~4661--4685, 2010.
  
\bibitem{Marcucci02012001}
C.~Marcucci, D.~Nyhan, and B.~A. Simon, ``{Distribution of pulmonary
  ventilation using {Xe}-enhanced computed tomography in prone and supine
  dogs},'' \emph{J Appl Physiol}, vol.~90, no.~2, pp. 421--430, 2001.

\bibitem{tajik2002}
J.~K. Tajik, D.~Chon, C.-H. Won, B.~Q. Tran, and E.~A. Hoffman,
``Subsecond
  multisection {CT} of regional pulmonary ventilation,'' \emph{Academic
  Radiology}, vol.~9, pp. 130--146, 2002.

\bibitem{Chon200565}
D.~Chon, B.~A. Simon, K.~C. Beck, H.~Shikata, O.~I. Saba, C.~Won,
and E.~A.
  Hoffman, ``Differences in regional wash-in and wash-out time constants for
  xenon-{CT} ventilation studies,'' \emph{Respiratory Physiology \&
  Neurobiology}, vol. 148, no. 1-2, pp. 65 -- 83, 2005.

\bibitem{guo2008}
J.~Guo, M.~K. Fuld, S.~K. Alford, J.~M. Reinhardt, and E.~A.
Hoffman,
  ``Pulmonary analysis software suite 9.0: Integrating quantitative measures of
  function with structural analyses,'' in \emph{First International Workshop on
  Pulmonary Image Analysis}, New York, 2008, pp. 283--292.

\bibitem{cao:762309}
K.~Cao, K.~Ding, G.~E. Christensen, and J.~M. Reinhardt, ``Tissue
volume and
  vesselness measure preserving nonrigid registration of lung ct images,''
  B.~M. Dawant and D.~R. Haynor, Eds., vol. 7623, no.~1.\hskip 1em plus 0.5em
  minus 0.4em\relax SPIE, 2010, p. 762309.

\bibitem{caowbir2010}
K.~Cao, K.~Ding, G.~E. Christensen, M.~L. Raghavan, R.~E. Amelon,
and J.~M.
  Reinhardt, ``{U}nifying {V}ascular {I}nformation in {I}ntensity-{B}ased
  {N}onrigid {L}ung {CT} {R}egistration,'' in \emph{Biomedical Image
  Registration}, B.~Fischer, B.~Dawant, and C.~Lorenz, Eds., L\"{u}beck, 2010,
  pp. 1--12.

\bibitem{Yin2009}
Y.~Yin, E.~A. Hoffman, and C.-L. Lin, ``Local tissue-weight-based
nonrigid
  registration of lung images with application to regional ventilation,'' X.~P.
  Hu and A.~V. Clough, Eds., vol. 7262, no.~1.\hskip 1em plus 0.5em minus
  0.4em\relax SPIE, 2009, p. 72620C.

\bibitem{yin:4213}
Y.~Yin, E.~A. Hoffman, and C.-L. Lin, ``Mass preserving nonrigid
registration of {CT} lung images using cubic
  {B}-spline,'' \emph{Medical Physics}, vol.~36, no.~9, pp. 4213--4222, 2009.

\bibitem{cao2009a}
K.~Cao, G.~E. Christensen, K.~Ding, and J.~M. Reinhardt,
  ``{I}ntensity-and-{L}andmark-{D}riven, {I}nverse {C}onsistent, {B}-{S}pline
  {R}egistration and {A}nalysis for {L}ung {I}magery,'' in \emph{Second
  International Workshop on Pulmonary Image Analysis}, M.~Brown, M.~de~Bruijne,
  B.~van Ginneken, A.~Kiraly, J.-M. Kuhnigk, C.~Lorenz, J.~R. McClelland,
  K.~Mori, A.~Reeves, and J.~M. Reinhardt, Eds., London, UK, 2009, p. 137.

\bibitem{Yin20102159}
Y.~Yin, J.~Choi, E.~A. Hoffman, M.~H. Tawhai, and C.-L. Lin,
``Simulation of
  pulmonary air flow with a subject-specific boundary condition,''
  \emph{Journal of Biomechanics}, vol.~43, no.~11, pp. 2159 -- 2163, 2010.

\bibitem{Hoffman1985}
E.~A. Hoffman and E.~L. Ritman, ``{Effect of body orientation on
regional lung
  expansion in dog and sloth},'' \emph{J Appl Physiol}, vol.~59, no.~2, pp.
  481--491, 1985.

\bibitem{Frangi98multiscalevessel}
A.~F. Frangi, W.~J. Niessen, K.~L. Vincken, and M.~A. Viergever,
``Multiscale
  vessel enhancement filtering,'' in \emph{MICCAI}, vol. 1496, 1998, pp.
  130--137.

\bibitem{shikata2004}
H.~Shikata, E.~A. Hoffman, and M.~Sonka, ``Automated segmentation of
pulmonary
  vascular tree from {3D} {CT} images,'' A.~A. Amini and A.~Manduca, Eds., vol.
  5369, no.~1.\hskip 1em plus 0.5em minus 0.4em\relax SPIE, 2004, pp. 107--116.

\bibitem{Shikata2009}
H.~Shikata, G.~Mc{L}ennan, E.~A. Hoffman, and M.~Sonka,
``Segmentation of
  pulmonary vascular trees from thoracic 3{D} {CT} images,''
  \emph{International Journal of Biomedical Imaging}, vol. 2009, no.~1, p.
  36240.

\bibitem{joshi2010a}
V.~Joshi, J.~M. Reinhardt, and M.~D. Abramoff, ``{A}utomated
measurement of
  retinal blood vessel tortuosity,'' in \emph{Proc. SPIE Conf. Medical
  Imaging}, N.~Karssemeijer and R.~M. Summers, Eds., vol. 7624, 2010, p. 76243.

\bibitem{lbfgsb}
R.~H. Byrd, P.~Lu, J.~Nocedal, and C.~Zhu, ``{A limited memory
algorithm for
  bound constrained optimization},'' \emph{SIAM J. Sci. Comput.}, vol.~16,
  no.~5, pp. 1190--1208, 1995.

\bibitem{Choi00injectivityconditions}
Y.~Choi and S.~Lee, ``Injectivity conditions of 2{D} and 3{D}
uniform cubic
  b-spline functions,'' \emph{Graphical Models}, vol.~62, no.~6, pp. 411--427,
  2000.

\bibitem{Simon2000}
B.~A. Simon, ``Non-invasive imaging of regional lung function using
{X-Ray}
  computed tomography,'' \emph{Journal of Clinical Monitoring and Computing},
  vol.~16, no.~5, pp. 433 -- 442, 2000.


\bibitem{Fuld04012008}
M.~K. Fuld, R.~B. Easley, O.~I. Saba, D.~Chon, J.~M. Reinhardt,
E.~A. Hoffman,
  and B.~A. Simon, ``{CT-measured regional specific volume change reflects
  regional ventilation in supine sheep},'' \emph{J Appl Physiol}, vol. 104,
  no.~4, pp. 1177--1184, 2008.

\bibitem{murphy2008}
K.~Murphy, B.~van Ginneken, J.~Pluim, S.~Klein, and M.~Staring,
  ``Semi-automatic reference standard construction for quantitative evaluation
  of lung {CT} registration,'' in \emph{Proc.\ of International Conference on
  Medical Image Computing and Computer-Assisted Intervention 2008}, vol. 5242,
  2008, pp. 1006--1013.

\bibitem{papoulis1990}
A.~Papoulis, Ed., \emph{Probability and Statistics}.\hskip 1em plus
0.5em minus
  0.4em\relax Englewood Cliffs, NJ: Prentence Hall, 1990.

\bibitem{kabus2008workshop}
S.~Kabus, J.~{von Berg}, T.~Yamamoto, R.~Opfer, and P.~J. Keall,
``Lung
  ventilation estimation based on {4D-CT} imaging,'' in \emph{First
  International Workshop on Pulmonary Image Analysis}, New York, 2008, pp.
  73--81.

\end{thebibliography}
\end{document}